\documentclass{jfm}

\usepackage{xcolor}					
\usepackage{xfrac}					

\usepackage{subfigure}              
\usepackage{overpic}
\usepackage{bm}

\usepackage{amsmath}

\usepackage{tikz}

\newcommand{\red}[1]{\textcolor{black}{#1}}
\newcommand{\blue}[1]{\textcolor{black}{#1}}
\newcommand{\green}[1]{\textcolor{black}{#1}}
\newcommand{\cyan}[1]{\textcolor{black}{#1}}

\newcommand{\white}[1]{\textcolor{white}{#1}}

\shorttitle{Airfoil secondary tones and acoustic feedback loop}
\shortauthor{T. R. Ricciardi, W. R. Wolf and K. Taira}

\title{\red{Transition, intermittency and phase interference effects in airfoil secondary tones and acoustic feedback loop}}

\author{Tulio R. Ricciardi\aff{1}, William R. Wolf\aff{1}
	\corresp{\email{wolf@fem.unicamp.br}}
	\and Kunihiko Taira\aff{2}}

\affiliation{\aff{1}School of Mechanical Engineering, University of Campinas,
	Campinas, SP 13083-860, Brazil
	\aff{2}Department of Mechanical and Aerospace Engineering, University of California, Los Angeles, CA 90095, USA}

\begin{document}
\maketitle

\begin{abstract}
	A large eddy simulation is performed to study secondary tones generated by a NACA0012 airfoil at angle of attack of $\alpha = 3^{\circ}$ with freestream Mach number of $M_{\infty} = 0.3$ and Reynolds number of $Re = 5 \times 10^4$. Laminar separation bubbles are observed over the suction side and near the trailing edge, on the pressure side.
	Flow visualization and spectral analysis are employed to investigate vortex shedding aft of the suction side separation bubble. Vortex interaction results in merging or bursting such that coherent structures or turbulent packets are advected towards the trailing edge leading to different levels of noise emission. Despite the intermittent occurrence of laminar-turbulent transition, the noise spectrum depicts a main tone with multiple equidistant secondary tones.
	To understand the role of flow instabilities on the tones, the linearized Navier-Stokes equations are examined in its operator form through bi-global stability and resolvent analyses, and by time evolution of disturbances using a matrix-free method. These linear global analyses reveal amplification of disturbances over the suction side separation bubble. Non-normality of the linear operator leads to further transient amplification due to modal interaction among eigenvectors.
	\blue{Two-point, one time autocovariance calculations of pressure along the spanwise direction elucidate aspects of the acoustic feedback loop mechanism in the non-linear solutions. This feedback process is self-sustained by acoustic waves radiated from the trailing edge, which reach the most sensitive flow location between the leading edge and the separation bubble, as identified by the resolvent analysis. Leading edge disturbances arising from secondary diffraction and phase interference among the most unstable frequencies computed in the eigenspectrum are also shown to have an important role in the feedback loop.}
\end{abstract}


\section{Introduction}

	Trailing-edge noise is an overriding concern for design of quiet air vehicles. At low and moderate Reynolds numbers, tonal noise becomes an important component of the acoustic spectrum. Several studies on trailing-edge aeroacoustics were conducted starting in the 1970s to examine the tonal noise generation by airfoils \citep{Paterson1973,Tam1974,Fink1975,Longhouse1977,Arbey1983}. Noise measurements were performed by \citet{Paterson1973} for symmetric NACA airfoils at various angles of attack over a Reynolds number range between $10^{5}$ and $10^{6}$. Their results showed the existence of multiple tones in a ladder-like structure in terms of frequency and freestream velocity. These authors also found strong two-point correlation of surface pressure along the spanwise direction, which indicated that the flow phenomenon associated with tonal noise generation could be modeled as two-dimensional.

	\citet{Tam1974} suggested that the ladder-like structure is due to a self-excited feedback loop between disturbances in the boundary layer and the airfoil wake. \citet{Fink1975} assumed that the discrete tonal frequencies are related to the laminar boundary layer on the pressure side. In order to elucidate aspects of airfoil noise,
	\citet{Arbey1983} performed experiments in an open wind tunnel for different NACA airfoils at $\alpha = 0^{\circ}$ for $1 \times 10^5 \le Re \le 7 \times 10^5$.
	The aforementioned studies showed that the noise spectrum has a broadband component with a main tonal peak plus a set of equidistant secondary tones due to a feedback mechanism closing at the point of maximum flow velocity along the airfoil.
	The broadband component was assumed to appear due to scattering of Tollmien-Schlichting (TS) instabilities. For airfoils at incidence, \citet{Lowson1994} found that the presence of secondary tones was related to a separation bubble developed on the airfoil pressure side. In this case, TS instabilities developing along the laminar boundary layer would lead to acoustic scattering on the trailing edge and acoustic waves would then propagate upstream closing the feedback loop, with the separation bubble acting as an amplifier of acoustic disturbances.

	\citet{Nash1999} performed experimental studies of airfoil noise for a NACA0012 profile up to a Reynolds number of $1.45\times10^{6}$ and several angles of attack. A closed-section wind tunnel, with and without acoustic-absorbing lining on its walls, was used in the experiments and results from the hard-wall tunnel revealed multiple frequency peaks. However, the authors argued that these tonal peaks were correlated to resonant frequencies of the wind tunnel. Thus, they carried out measurements with lined walls simulating anechoic conditions and a single dominant tone was observed instead of several peaks. Furthermore, no ladder-like structure of tonal frequency was observed, in disagreement with previous studies of \citet{Paterson1973}, \citet{Fink1975} and \citet{Arbey1983}. It is important to mention that secondary tones were often observed in experiments conducted in open-jet facilities.

	More recently, \citet{Plogmann2013} also found multiple tones in their experiments (NACA0012, $3.1 \times 10^5 \le Re \le 1.5 \times 10^{6}$, $0^{\circ} \le \alpha \le 9^{\circ}$) and demonstrated that tripping the pressure side boundary layer leads to turbulent flow, eliminating the separation bubble and the secondary tones. These authors emphasize that the feedback loop is extremely sensitive to small variations in the flow conditions what, in turn, lead to changes in the tonal components. \blue{This occurs particularly at higher Reynolds numbers, where the flow is more prone to transition}. The dependency of angle of attack and Reynolds number on tonal noise emission is highlighted by \citet{Probsting2015_regimes}. These authors performed experiments for a NACA0012 airfoil for $0.3\times10^{5} \le Re \le 2.3\times10^{5}$ and effective angles of attack $0^{\circ} \le \alpha \le 6.3^{\circ}$. Tripping devices were applied separately on each side of the airfoil to identify their respective role in the noise generation and it was found that suction-side (pressure-side) events dominate at lower (higher) Reynolds numbers. Moreover, it was observed that, at low angles of attack, interactions between the two sides of the airfoil become increasingly important.

	As discussed above, early experiments report the presence of a separation bubble on the airfoil pressure side. Most of these investigations were conducted at higher Reynolds numbers, where the flow was likely turbulent on the suction side. \blue{These observations are in agreement with the} large eddy simulations of a NACA0012 at $\alpha = 5^{\circ}$ and $Re = 4 \times 10^{5}$ performed by \citet{Wolf2012} and \citet{Ricciardi2019}. Flow visualization and proper orthogonal decomposition were used in these references to identify coherent structures shed from the pressure side near the trailing edge. Such flow structures were found responsible for the intense tonal noise generation despite the fact that a turbulent boundary layer developed on the airfoil suction side. In this case, the pressure side boundary layer was laminar due to the favorable pressure gradient. In agreement with the experiments of \citet{Plogmann2013}, \citet{Wolf2012} showed that the tonal component vanishes when both boundary layers are tripped.

	At lower Reynolds numbers, a laminar separation bubble (LSB) exists on the airfoil suction side and is responsible for the overall flow dynamics and noise generation. In the context of airfoil flows, this flow feature has been studied for different purposes. For instance, direct numerical simulations were performed by \citet{jones2008} for a NACA0012 airfoil at Reynolds number $Re = 5 \times 10^4$, $M_\infty = 0.4$ and $\alpha = 5^{\circ}$. It was shown that, despite being absolutely stable by means of linear stability analysis, turbulence is self-sustained even with the absence of forcing. Later, experimental investigations by \citet{Probsting2015_bubble} with a NACA0012 airfoil at moderate Reynolds numbers $0.65 \times 10^5 \le Re \le 4.5 \times 10^5$ and $\alpha = 2^{\circ}$ present intermittent laminar-turbulent transition that affects the advection of coherent structures from the LSB towards the trailing edge.
	Further analyses were also presented by \citet{yarusevich2016_coherent,yarusevych2018_transition, yarusevich2019_merging}, \citet{yarusevych2018_spanwise} and \cite{Probsting2021} to study the impact of acoustic excitation, three-dimensional effects and shedding/merging of vortices from the suction side LSB.

	The first numerical simulations on airfoil secondary tones were conducted by \citet{Desquesnes2007} considering two-dimensional flows over a NACA0012 airfoil for a Reynolds number $Re = 1\times10^{5}$ at an angle of attack $\alpha = 5^{\circ}$, and for $Re = 2\times10^{5}$ at $\alpha = 2^{\circ}$. In agreement with most experimental observations, multiple tonal peaks were observed.
	The previous authors also performed a local stability analysis \green{assuming parallel flow} and showed that the main tone frequency radiated to the far-field was close to that most amplified along the pressure side boundary layer. An assessment of the linear dynamics of wavepackets driving the feedback loop mechanism and the flow receptivity to acoustic forcing was investigated by time-marching the linearized Navier-Stokes equations by \citet{jones2010} for a NACA0012 airfoil at $Re = 5\times10^4$, $M_\infty = 0.4$ and $\alpha = 5^{\circ}$. \citet{FosasdePando2014} applied bi-global stability analysis to investigate the dynamics coupling the boundary layers on both airfoil sides and the wake. The authors studied the 2D flow over a NACA0012 airfoil for $Re = 2\times10^5$, $M_\infty = 0.4$ and $\alpha = 2^{\circ}$ and found multiple frequencies in the eigenspectrum related not only to the main tonal peak but also to the secondary tones. Prior to this work, the stability analyses were \green{limited a parallel flow assumption, and only a discussion of the dominant tonal frequency was presented}. By means of adjoint and resolvent analyses, \citet{FosasCTR,FosasdePando2017} also identified sensitive regions of the flow which are prone to close the feedback loop mechanism. \green{Hence, linear stability theory has proven to be an important methodology to investigate the generation of airfoil secondary tones and the feedback loop mechanism.}
	
	Another important observation from \citet{Desquesnes2007} regards the amplitude modulation of velocity fluctuations computed near the trailing edge. These authors discuss that such modulation is caused by interference of vortical structures from both sides of the airfoil which, combined with the feedback loop mechanism, would lead to the presence of multiple tones. Following the discussion on modulation of flow structures, \citet{Probsting2014} employed particle image velocimetry to study airfoils at Reynolds numbers $1\times10^{5} \le Re \le 2.7\times10^{5}$ and $2^{\circ} \le \alpha \le 4^{\circ}$. They investigated the mechanisms associated with tonal noise generation and the interference effects between suction and pressure sides of the airfoil. For $Re \approx 1.5\times10^{5}$, at the lowest angle of attack, the authors showed that the amplitude modulation discussed by \citet{Desquesnes2007} was related to destructive interference of boundary layer instabilities occurring on both sides of the trailing edge.
	More recently, \citet{Ricciardi2019_tones} \blue{performed 2D simulations} of a NACA0012 at $Re = 1.0\times10^{5}$ at $\alpha = 3^{\circ}$ and showed that the multiple tones are related to modulation of the vortical structures developing on the suction side. In this case, the instantaneous main frequency alternates in time due to phase modulation of the flow structures shed by the laminar separation bubble. As a consequence, multiple equidistant frequencies must appear in the Fourier transform to reconstruct the modulated signals. However, as pointed by \citet{Probsting2014}, flow transition on the suction side occurs in experiments and their conclusions may be different from those observed in 2D numerical simulations. 
	\green{In this regard, the simulation of \cite{Sanjose2018} for a thin-cambered airfoil at $Re = 1.5 \times 10^5$ exhibits flow transition where intermittency played a key role in the flow dynamics and noise emission for the configuration investigated. Thus, experimental \citep{Probsting2014} and numerical \citep{Sanjose2018,Nguyen2021} findings contradict previous assumptions that 2D simulations are sufficient to explain all the mechanisms of secondary tones.}

\red{	
	In this study, a large eddy simulation (LES) is performed to investigate a NACA0012 airfoil at $\alpha = 3^{\circ}$ immersed in a freestream flow with Mach number $M_{\infty} = 0.3$ and Reynolds number $Re = 5 \times 10^4$. \red{This condition is chosen based on experimental results from \cite{Probsting2015_regimes} and \cite{Probsting2015_bubble}. These authors show that, for this flow configuration, suction side events are responsible for the noise generation without a contribution from the pressure side. Moreover, the previous references show that the flow behavior at this particular angle of attack is somehow independent of the Reynolds number up to a point where the relevant vortex dynamics switches to the pressure side.} For the current flow setup, some key aspects are investigated including the dynamics of the suction side separation bubble and its vortex shedding, flow intermittency and its impact on amplitude modulation. \blue{Linear stability theory is applied to investigate the presence of sensitive regions and amplification mechanisms within the flow.}
}	
		
\red{
	The present work is organized as follows: section \ref{ssec:meanflow} presents the laminar separation bubbles which are observed over the suction side and near the trailing edge, on the pressure side. Then, flow visualization is employed in section \ref{ssec:dynamics} to investigate vortex shedding aft of the suction side separation bubble. It is shown that vortex interaction results in merging or bursting such that coherent structures or turbulent packets are advected towards the trailing edge, leading to different levels of noise emission. Despite the intermittent occurrence of laminar-turbulent transition, the noise spectrum presents multiple equidistant secondary tones as discussed in section \ref{ssec:intermittency}.
	To understand the role of flow instabilities on tonal noise generation, in section \ref{ssec:linear}, the linearized Navier-Stokes equations are examined in its modal form using bi-global stability and resolvent analyses, and by the time evolution of disturbances which results in periodic wavepackets.
	Aspects of the acoustic feedback loop mechanism in both the linear and non-linear solutions are examined in section \ref{ssec:feedback} and we highlight the role of leading edge disturbances from secondary diffraction in the closure of the acoustic feedback loop mechanism. The intermittent transition which leads to amplitude modulation of noise generation is analyzed in section \ref{ssec:merging} in terms of phase interference of the dominant frequencies from hydrodynamic \blue{instabilities}. Then, the main findings and conclusions are presented in section \ref{ssec:conclusions}.
}

\clearpage
 	\section{Theoretical and Numerical Approaches}
\label{sec:chapter2}

\subsection{Large eddy simulation}

Large eddy simulation is performed to solve the compressible Navier-Stokes equations in general curvilinear coordinates. The spatial discretization of the governing equations employs a sixth-order accurate compact scheme for derivatives and interpolations on a staggered grid \citep{Nagarajan2003}. The time integration is carried out by a hybrid implicit-explicit method. The implicit second-order scheme of \citet{Beam1978} is applied in the near-wall region to overcome the stiffness problem due to a fine boundary layer grid, while a third-order Runge-Kutta scheme is used for time advancement of the equations in flow regions away from the solid boundary. For the communication across the different methods, information is exchanged at overlapping points. \green{An explicit sub-grid scale model is not applied.} However, outside the boundary layer, a sixth-order compact filter \citep{Lele1992} is applied to control high-wavenumber numerical instabilities arising from grid stretching and interpolation between staggered grids. The transfer function associated with such filters has been shown to provide an approximation to sub-grid scale models \citep{Mathew_etal_2003}.

No-slip adiabatic wall boundary conditions are enforced along the airfoil surface and characteristic plus sponge boundary conditions are applied in the far-field locations to minimize wave reflections \green{\citep{Wolf2011}}. Periodic boundary conditions are used in the spanwise direction.
Length scales, velocity components, density, pressure and temperature are non-dimensionalized as $\boldsymbol{x} = \boldsymbol{x}^*/L^*$, $\boldsymbol{u}=\boldsymbol{u}^*/a_{\infty}^*$, $\rho=\rho^*/{\rho}_{\infty}^*$, $p = p^*/{\rho}_{\infty}^*{a_{\infty}^*}^2$ and $T=T^*/\left[(\gamma-1)T_{\infty}^*\right]$, respectively. Here, $L^*$ is the airfoil chord, $a_{\infty}^*$ is the freestream speed of sound, $\rho_{\infty}^*$ is the freestream density, $T_{\infty}^*$ is the freestream temperature and $\gamma$ is the ratio of specific heats. Variables with superscript $^{*}$ are given in dimensional units.
Herein, time and frequency (Strouhal number) are presented non-dimensionalized by freestream velocity as $t=t^* \, U_{\infty}^* / L^*$ and $St=f^* \, L^* / U_{\infty}^*$, respectively. 
\cyan{The present numerical methodology has been extensively validated for various 2D and 3D simulations of compressible airfoil flows at different configurations \citep{Wolf2012, Wolf:DU96, Wolf2013, Brener2019}.}

\begin{figure}
	\centering
	\begin{overpic}[width=0.9\textwidth]{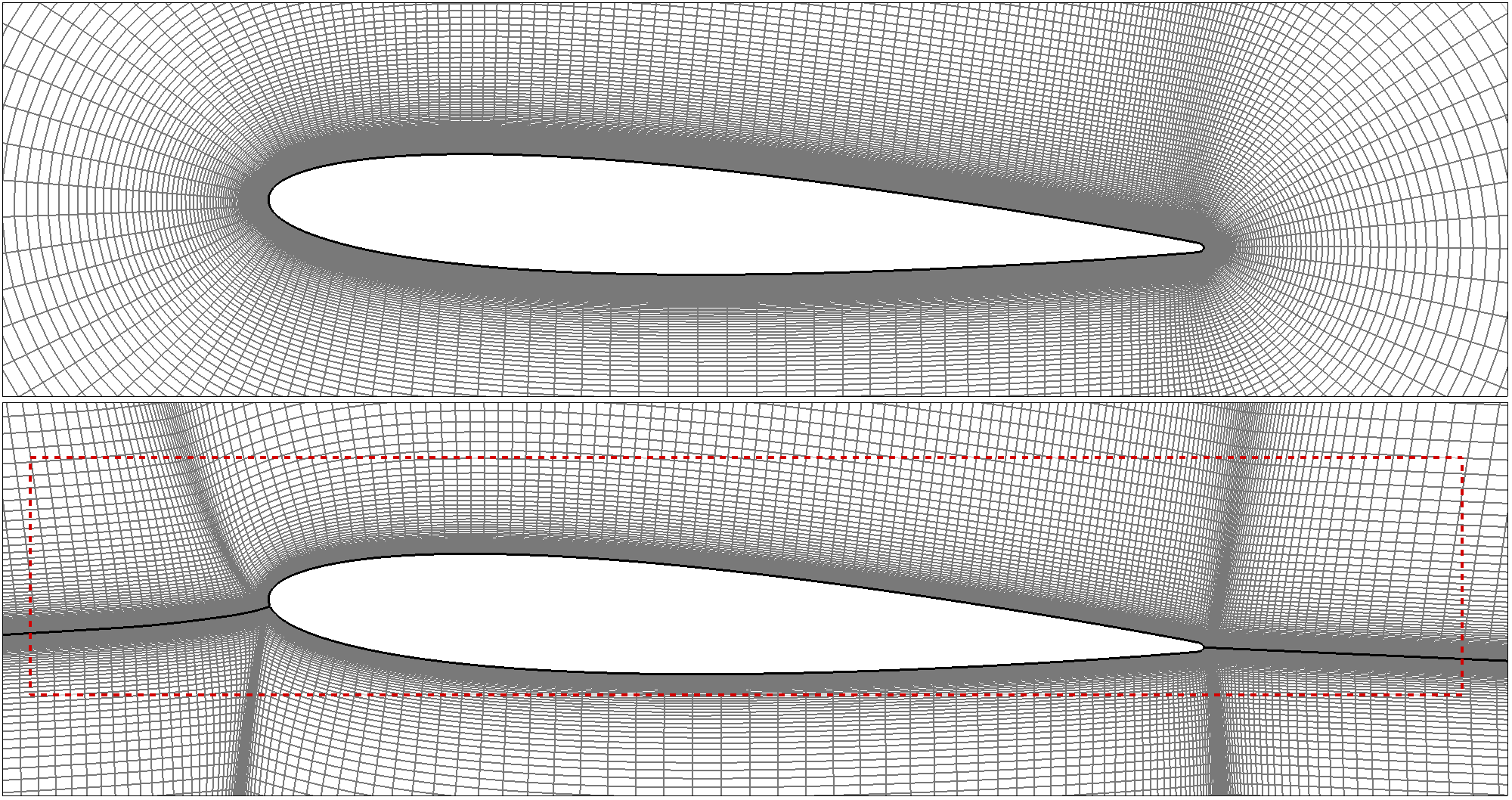}
		\put(0.2,28.1){\fcolorbox{black}{white}{\parbox{0.14\linewidth}{(a) LES mesh}}}
		\put(0.2,1.8){\fcolorbox{black}{white}{\parbox{0.27\linewidth}{(b) Linear analyses mesh}}}
	\end{overpic} 
	\caption{Computational grids near the airfoil for the (a) O-mesh employed in the LES and (b) the H-mesh used in the linear stability analysis. \red{The grids are shown for every 3 points in both directions.}}
	\label{fig:grid}
\end{figure}

The O-grid employed for the current LES is shown in gray lines for every 3 points in figure \ref{fig:grid}(a).
For a smooth O-grid generation, the original airfoil trailing edge is truncated at 98\% of the chord and it has a curvature radius $r/L^* = 0.4\%$. The reference length scale $L^{*}$ is the unit chord from the original NACA0012 airfoil.
The leading edge is placed at $(x, y) = (0, 0)$ and the airfoil is pivoted about this point.
The total wingspan has $0.4 L^*$, being larger compared to previously reported studies of airfoil flows for this range of Reynolds number \citep{jones2008,Ducoin2016}.
\green{The domain extends 37 chord-lengths outwards and a circular sponge, given by $A[(r - r_0)/\Delta r]^{n}$, occupies the last $\Delta r = 10$ chords. In this function, $r$ is the radial distance, $r_0$ is the sponge starting position, and its coefficients are $A = 20$ and $n=4$.}

The mesh has a distribution of points in the streamwise, wall-normal and spanwise directions given by $n_x=660$, $n_y=600$ and $n_z=192$, respectively, which results in approximately $76 \times 10^6$ grid points. The ratio of grid points along the suction side relative to the pressure side is approximately 5:3.
The wall-normal distance of the first grid point is $\Delta n = 0.0001$ and the stretching ratio is 1.5\%.
\cyan{As discussed by \cite{Desquesnes2007}, \citet{Jones2011} and \cite{FosasdePando2014}, the important mechanisms for tonal noise generation arise from two-dimensional flow instabilities in the laminar region of the flow. Motivated by this observation, a grid refinement study in terms of mean and fluctuation properties was conducted for 2D simulations and shown by the present authors in \citet{Ricciardi_Scitech2021}. However, in the present study, the flow is three-dimensional and transitions to turbulence on the airfoil suction side, near the trailing edge. Considering only the turbulent flow region, the estimated mesh resolution in terms of wall units is given by $\Delta x^{+} < 10$, $\Delta y^{+} \approx 0.3$ and $\Delta z^{+} < 5$. These values follow the best practices for wall-resolved LES.}
The time step is $\Delta t = 1.5 \times 10^{-4}$ and 75 convective time units are employed for post-processing and analysis of results. \green{The 3D simulation starts from a 2D flow superposed with random noise and more than 30 convective time units are discarded before collecting flow statistics.}

	\subsection{Linear stability and resolvent analyses}
	\label{sec:LinearStabAnalysis}
	
	\green{The mechanisms of generation and amplification of flow instabilities can be explained by linear stability theory. Hence, bi-global linear stability and resolvent analyses are employed in this work to investigate the most unstable frequencies and their sensitivity to disturbances, besides  their role in the tonal noise generation.} Through a Reynolds decomposition, it is possible to split the unsteady flow \blue{$\boldsymbol{q}(\boldsymbol{x},t)$} in a mean base flow $\bar{\boldsymbol{q}}(\boldsymbol{x})$ plus a time-dependent fluctuation component $\boldsymbol{q}'(\boldsymbol{x},t)$. If the fluctuations are sufficiently small, the Navier-Stokes equations can be linearized about the (mean) base flow.
	Although it is possible to consider the turbulent mean flow $\bar{\boldsymbol{q}}$ as the base flow, the linear stability analysis would not hold because such state is not an equilibrium point of the NS equations.
	Nonetheless, the use of a time-averaged base flow may provide some insights as a model \citep{Taira_AIAAJ2017,Taira_AIAAJ2020}.
	With addition of a time-dependent external forcing $\boldsymbol{f}'(\boldsymbol{x},t)$, the linearized Navier-Stokes equations (LNS) can be written as
	\begin{equation}
	\frac{\partial \boldsymbol{q}'}{\partial t} = \boldsymbol{\mathcal{L}} \boldsymbol{q}' + \boldsymbol{\mathcal{B}} \boldsymbol{f}' \mbox{ ,}
	\label{eq:LNS_1}
	\end{equation}
	where $\boldsymbol{\mathcal{B}} = \boldsymbol{\mathcal{B}}(\boldsymbol{x})$ is an operator which serves as a spatial window where forcing is applied. The matrix $\boldsymbol{\mathcal{L}} = \boldsymbol{\mathcal{L}}(\bar{\boldsymbol{q}})$ is the bi-global LNS operator with the spanwise time-averaged flow properties $\bar{\boldsymbol{q}}(\boldsymbol{x})$ as base flow.
	
	The evolution of linear disturbances by equation \ref{eq:LNS_1} can be performed by a direct analysis of the operator $\boldsymbol{\mathcal{L}}$ or time-integrating the disturbances in a linearized version of the CFD code. In the former case, with a transformation
	\begin{equation}
	\bullet'(x,y,z,t) = \int_{-\infty}^{\infty} \int_{-\infty}^{\infty} \hat{\bullet}(x,y,\beta,\omega) e^{\mathrm{i} (\beta z-\omega t)} \, d \beta \, d \omega \mbox{ ,}
	\end{equation}
	where the wavenumber $\beta \in \mathbb{R}$ and the frequency $\omega \in \mathbb{C}$, it is possible to write the forced LNS equations in discrete form as
	\begin{equation}
	-\mathrm{i} \omega \hat{\boldsymbol{q}} = \mathsfbi{L} \hat{\boldsymbol{q}} + \mathsfbi{B} \hat{\boldsymbol{f}} \mbox{ .}
	\end{equation}
	In this case, the linear operator becomes a function of the spanwise wavenumber $\mathsfbi{L} = \mathsfbi{L}(\bar{\boldsymbol{q}},\beta)$ and $\mathsfbi{B}$ is the discrete version of $\boldsymbol{\mathcal{B}}$. If forcing $\hat{\boldsymbol{f}}$ is absent, the LNS equations can be analyzed separately for each wavenumber $\beta$ as an eigenvalue problem (linear stability analysis) as
	\begin{equation}
	\mathsfbi{Q} \bm{\Lambda} = \mathsfbi{L} \mathsfbi{Q} \mbox{ .}
	\label{eq:LSA}
	\end{equation}
	Here, $\mathsfbi{Q}$ holds the eigenvectors of $\mathsfbi{L}$ and the eigenvalues appear in the diagonal matrix $\bm{\Lambda} = -\mathrm{i} \omega \mathsfbi{I}$, where the frequency and growth rate are the real and imaginary parts of $\omega$, respectively. Due to the velocity gradients appearing off-diagonal, the linear operator becomes a non-normal matrix such that $\mathsfbi{L}^{\mathcal{H}} \mathsfbi{L} \ne \mathsfbi{L} \mathsfbi{L}^{\mathcal{H}}$. As a consequence, the eigenvectors are also non-normal. In this equation, the superscript $\mathcal{H}$ denotes the complex conjugate transpose (Hermitian). The matrix $\mathsfbi{L}^{\mathcal{H}}$ is the adjoint operator, being related to regions of sensitivity within the flow. Its eigenvalue decomposition yields
	\begin{equation}
	\mathsfbi{Q}^{\dagger} \bm{\Lambda}^{\mathcal{H}} =
	\mathsfbi{L}^{\mathcal{H}} {\mathsfbi{Q}^{\dagger}} \mbox{ .}
	\label{eq:LSA_adj}
	\end{equation}
	The eigenvalues of the adjoint operator are the complex conjugate of those from direct analysis. The eigenvectors $\mathsfbi{Q}^{\dagger}$ represent the region of flow sensitivity and they are equal to $\mathsfbi{Q}^{-1}$ for normal systems. On the other hand, this is not true for non-normal systems. A thorough review on the significance of adjoint operators is presented by \citet{adjoint_AnnuReview} and references therein.
	
	In case forcing is applied at a frequency $\tilde{\omega}$, equation \ref{eq:LNS_1} yields
	\begin{equation}
	\hat{\boldsymbol{q}} = (-\mathrm{i} \tilde{\omega} \mathsfbi{I} - \mathsfbi{L} )^{-1} \mathsfbi{B} \hat{\boldsymbol{f}} = \mathsfbi{H} \, \mathsfbi{B} \hat{\boldsymbol{f}} \mbox{ ,}
	\label{eq:resolvent_1}
	\end{equation}
	where the matrix $\mathsfbi{H} = \mathsfbi{H}(\bar{\boldsymbol{q}},\beta,\tilde{\omega})$ is the resolvent operator \citep{reddy_henningson_1993,Schmid2007annrev,mckeon_sharma_2010,Taira_AIAAJ2017,Taira_AIAAJ2020}.	
	For non-normal operators, eigenvalue sensitivity and energy amplification are related to the induced $L_2$ norm of operator $\mathsfbi{H}$, \textit{i.e.}, the leading singular value $\sigma$ obtained from singular value decomposition
	\begin{equation}
	\mathsfbi{H} = \mathsfbi{U} \bm{\Sigma} \mathsfbi{V}^{*} \mbox{ .}
	\label{eq:svd_resolvent}
	\end{equation} 
	In this equation, $\mathsfbi{V}$ and $\mathsfbi{U}$ are unitary matrices holding right and left singular vectors and $\bm{\Sigma}$ is a diagonal matrix containing the singular values $\sigma$, such that
	\begin{equation}
	\left\| \mathsfbi{H} \right\| = \max( \sigma ) \; = \sigma_1 \mbox{ .}
	\end{equation}
	The first column of $\mathsfbi{V}$ contains the forcing term that produces the largest response in the flow (first column in matrix $\mathsfbi{U}$) with the amplification ratio given by $\bm{\Sigma}$.
	
	A transformation using matrix $\mathsfbi{W}$ is applied to convert variables $\hat{\boldsymbol{q}}$ and $\hat{\boldsymbol{f}}$ with an appropriate energy norm prior to performing the SVD. In the case of compressible flows, the Chu norm which relates density, velocity and temperature \citep{ChuNorm} is typically used. A spatial window $\mathsfbi{C} = \mathsfbi{C}(\boldsymbol{x})$ is applied to limit the domain of analysis and the system response in Fourier domain $\hat{\boldsymbol{y}}$ is given by
	\begin{equation}
	\hat{\boldsymbol{y}} = \mathsfbi{W} \mathsfbi{C} \hat{\boldsymbol{q}} \mbox{ .}
	\label{eq:resolvent_2}
	\end{equation}
	Combining equations \ref{eq:resolvent_1} and \ref{eq:resolvent_2} leads to the modified resolvent operator
	\begin{equation}
	\mathsfbi{H}_w = \mathsfbi{W} \mathsfbi{C} (-\mathrm{i} \tilde{\omega} \mathsfbi{I} - \mathsfbi{L} )^{-1} \mathsfbi{B} \mathsfbi{W}^{-1} \mbox{ .}
	\label{eq:final_resolvent}
	\end{equation}
	
	The amplification mechanisms of flow disturbances can be identified by using the eigenvalue decomposition of the linearized NS operator, equation \ref{eq:LSA}, in the resolvent, yielding
	\begin{equation}
	\mathsfbi{H}_w = \mathsfbi{W} \mathsfbi{C} (-\mathrm{i} \tilde{\omega} \mathsfbi{I} - \mathsfbi{Q} \bm{\Lambda} \mathsfbi{Q}^{-1} )^{-1} \mathsfbi{B} \mathsfbi{W}^{-1} = \mathsfbi{W} \mathsfbi{C} \mathsfbi{Q} \left( -\mathrm{i} \tilde{\omega} \mathsfbi{I} - \bm{\Lambda} \right)^{-1} \mathsfbi{Q}^{-1} \mathsfbi{B} \mathsfbi{W}^{-1} \mbox{ ,}
	\label{eq:resolvent_eigen}
	\end{equation}
	where $\bm{\Lambda}$ and $\mathsfbi{Q}$ are the eigenvalues and eigenvectors from the solution of equation \ref{eq:LSA}. The bounds of the induced $L_2$ norm are
	\begin{equation}
	\left\| \left( -\mathrm{i} \tilde{\omega} \mathsfbi{I} - \bm{\Lambda} \right)^{-1} \right\|
	\le \left\| \mathsfbi{H}_w \right\|
	\le \left\| \left( -\mathrm{i} \tilde{\omega} \mathsfbi{I} - \bm{\Lambda} \right)^{-1} \right\| \; || \mathsfbi{Q} || \; || \mathsfbi{Q}^{-1} || \mbox{ ,}
	\label{eq:ressonance}
	\end{equation}
	where the lower bound is the case of an operator with orthonormal eigenvectors. In this case, the norm depends only of the resonances, following a $\sfrac{1}{R}$ decay based on the distance $R$ in complex plane with respect to the eigenvalues. For non-normal systems, the norm also depends on the pseudoresonances, measured by the product of eigenvectors $\mathsfbi{Q}$ and their inverse $\mathsfbi{Q}^{-1}$. The weighting $\mathsfbi{W}$ and the two windowing matrices $\mathsfbi{B}$ and $\mathsfbi{C}$ are included in the norm calculation (see \citet{schmidYellowBook,McKeon2018} for more details).
	
	
	To perform the global stability (spectral) and resolvent (pseudospectral) analyses,
	the discrete linear operator is computed using the second-order accurate methodology from \citet{Yiyang2017} and \citet{yeh_taira_2019}.
	The base flow is the time-spanwised averaged solution from the LES, $\bar{\boldsymbol{q}} = [\bar{\rho}, \bar{u}, \bar{v}, \bar{w}, \bar{T}]$.
	For the far-field and airfoil surface, Dirichlet boundary conditions are set for $[\rho' , u' , v' , w' ] = [0, 0, 0, 0]$ and a Neumann boundary condition is set for $T'$ such that the wall-normal derivative $\partial T' / \partial n = 0$. At the outlet boundary, the same Neumann boundary condition is set for all flow variables.
	With these boundary conditions and the mean base flow $\bar{\boldsymbol{q}}$, the linear operator is computed in its discrete form $L(\bar{\boldsymbol{q}},\beta)$ for a prescribed spanwise wavenumber $\beta$.
	
	The mean flow $\bar{\boldsymbol{q}}$ obtained is interpolated from the O-grid in figure \ref{fig:grid}(a) to the H-mesh shown in figure \ref{fig:grid}(b) for the modal analysis. This is necessary to improve spatial accuracy of the linear operator downstream and upstream of the airfoil, where direct and adjoint eigenvectors are supported, respectively \citep{yeh_taira_2019}.
	The present two-dimensional H-mesh has an extent of $x \in [-5,+7], y \in [-5,+5]$ and it is composed of approximately $520 \times 10^3$ grid points.
	The mesh refinement targets a Strouhal number cut-off $St=8.0$. The sponge region comprises a circle with parabolic growth $f = a(r-r_0)^2$ for $r > r_0$. The sponge is centered at $(x, y) = (0.5, 0)$ with $a = 1.0$ and $r_0 = 1.5$.
	The mesh refinement and sponge placement are assessed by convergence of physically meaningful eigenvalues while suppressing spurious eigenvalues. If the mesh is coarse or the sponge is placed too far from the airfoil, it was observed that the meaningful eigenvectors also exhibit spatial support in the same region of the spurious eigenvectors.
	
	For the pseudospectrum calculation, the dashed red rectangle in figure \ref{fig:grid}(b) depicts the region considered for the present resolvent analysis in terms of operators $\mathsfbi{B}$ and $\mathsfbi{C}$ in equation \ref{eq:final_resolvent}.
	This windowing is important not only from a physical point of view, where the noise sources close to the airfoil surface are most important, but it also eliminates spurious eigenvectors, which are mostly observed far away from the body.
	Finally, the SVD in equation \ref{eq:final_resolvent} is performed with the randomized algorithm presented by \citet{JeanRandomized2020}.
	
	\subsection{Mean flow perturbation}
	
	\blue{The properties of the resolvent operator can be studied as an initial value problem where large transient energy amplification is observed, even in stable problems, due to the non-normality \citep{Trefethen1993,mckeon_sharma_2010}.}
	The time evolution of perturbations with respect to the mean flow can be performed with a linearized version of the CFD code according to equation \ref{eq:LNS_1}. However, here we employ an approach that uses the present LES code with the addition of a force term to the right-hand side of the equations \citep{Touber2009,MFP2015,Ranjan2018}. In this approach, known as mean flow perturbation, the Reynolds decomposition $\boldsymbol{q} = \bar{\boldsymbol{q}} + \boldsymbol{q}'$ is applied to the non-linear NS equations which are expressed as
	\begin{equation}
	\frac{\partial (\bar{\boldsymbol{q}} + \boldsymbol{q}')}{\partial t} = \mathcal{N}(\bar{\boldsymbol{q}} + \boldsymbol{q}') \mbox{ .}
	\end{equation}
	
	The turbulent mean flow $\bar{\boldsymbol{q}}$ is not an equilibrium state of the NS equations and its time derivative $\partial_t \bar{\boldsymbol{q}}$
	is non-null. Hence, in order to use it as a base flow, an additional term $\mathcal{R}$ must be added in the numerical procedure to keep the base flow stationary.	
	A Taylor series expansion about the base state $\bar{\boldsymbol{q}}$ yields
	\begin{equation}
	\frac{\partial (\bar{\boldsymbol{q}} + \boldsymbol{q}')}{\partial t} =
	\mathcal{N}(\bar{\boldsymbol{q}} + \boldsymbol{q}') =
	\mathcal{N}(\bar{\boldsymbol{q}}) + \frac{\partial \mathcal{N}(\bar{\boldsymbol{q}})}{\partial \boldsymbol{q}} \boldsymbol{q}' + \frac{1}{2} \frac{\partial^{2} \mathcal{N}(\bar{\boldsymbol{q}})}{\partial^{2} \boldsymbol{q}} \boldsymbol{q}'^{2} + ... + \mathcal{R}  \mbox{ .}
	\end{equation}	
	Grouping together the base flow terms results in
	\begin{equation}
	\frac{\partial \boldsymbol{q}'}{\partial t} + \left( \frac{\partial \bar{\boldsymbol{q}}}{\partial t} - \mathcal{N}(\bar{\boldsymbol{q}}) - \mathcal{R} \right) =
	\frac{\partial \mathcal{N}(\bar{\boldsymbol{q}})}{\partial \boldsymbol{q}} \boldsymbol{q}' + \frac{1}{2} \frac{\partial^{2} \mathcal{N}(\bar{\boldsymbol{q}})}{\partial^{2} \boldsymbol{q}} \boldsymbol{q}'^{2} + ... \mbox{ ,}
	\label{eq:MFP_1}
	\end{equation}
	where the base flow is stationary when
	\begin{equation}
	\left( \frac{\partial \bar{\boldsymbol{q}}}{\partial t} - \mathcal{N}(\bar{\boldsymbol{q}}) - \mathcal{R} \right) = 0 \mbox{ .}
	\label{eq:MFP_2}
	\end{equation}
	
	Considering that the imposed perturbation amplitude $\boldsymbol{q}'$ is sufficiently small leads to a linearized solution of the NS equations as
	\begin{equation}
	\frac{\partial \boldsymbol{q}'}{\partial t} \approx \frac{\partial \mathcal{N}(\bar{\boldsymbol{q}})}{\partial \boldsymbol{q}} \boldsymbol{q}' = \mathsfbi{L} \boldsymbol{q}' \mbox{ ,}
	\label{eq:MFP}
	\end{equation}
	where the Jacobian $\frac{\partial \mathcal{N}(\bar{\boldsymbol{q}})}{\partial \boldsymbol{q}}$
	is the linearized operator $\mathsfbi{L}$. In contrast to the previous approach in section \ref{sec:LinearStabAnalysis}, this methodology is matrix-free and the linear operator is never explicitly computed. Despite this, the method still accounts for the modal interaction that leads to transient energy amplification within a non-normal analysis of the linear operator.	
	The system response is obtained by the time evolution of the disturbances and the initial condition is set as an impulsive excitation that triggers the dominant response. In the current analysis, the grid and numerical methodology employed are the same used in the LES.

\clearpage
 	\section{Results}

The physical mechanisms responsible for the generation of secondary tones in airfoil flows are investigated by post-processing the non-linear results from LES and also employing linear stability theory.
The linear analysis provides insights on the onset and growth of disturbances in the flow besides its receptivity.
On the other hand, LES results allow the investigation of transition, intermittency and phase interference effects in the context of the multiple secondary tones and the acoustic feedback loop.


\subsection{Mean flow}
\label{ssec:meanflow}

\begin{figure}
	\centering
	\begin{tikzpicture}
	\node[anchor=south west,inner sep=0] (image) at (0,0) {\includegraphics[width=0.90\textwidth]{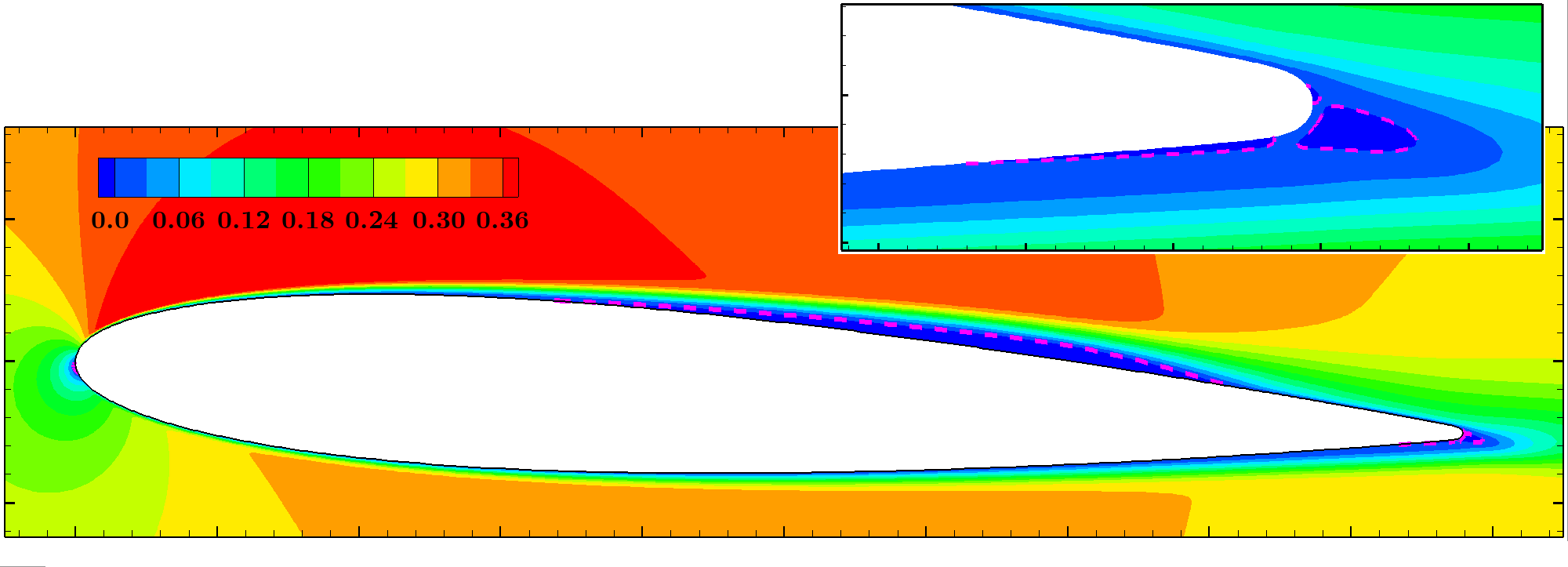}};
	\begin{scope}[x={(image.south east)},y={(image.north west)}]
	\node[] at (0.04,0.615) {\scriptsize $\bar{u}$:};
	\node[] at (0.563,0.6) {\scriptsize $0.92$};
	\node[] at (0.657,0.6) {\scriptsize $0.94$};
	\node[] at (0.751,0.6) {\scriptsize $0.96$};
	\node[] at (0.846,0.6) {\scriptsize $0.98$};						
	\node[] at (0.940,0.6) {\scriptsize $1.00$};
	\node[] at (0.0480,0.02) {\scriptsize $0.0$};
	\node[] at (0.1390,0.02) {\scriptsize $0.1$};
	\node[] at (0.2290,0.02) {\scriptsize $0.2$};
	\node[] at (0.3190,0.02) {\scriptsize $0.3$};						
	\node[] at (0.4100,0.02) {\scriptsize $0.4$};
	\node[] at (0.5000,0.02) {\scriptsize $0.5$};
	\node[] at (0.5900,0.02) {\scriptsize $0.6$};
	\node[] at (0.6810,0.02) {\scriptsize $0.7$};
	\node[] at (0.7710,0.02) {\scriptsize $0.8$};
	\node[] at (0.8620,0.02) {\scriptsize $0.9$};
	\node[] at (0.9520,0.02) {\scriptsize $1.0$};
	\end{scope}
	\end{tikzpicture}
	\caption{Contours of mean streamwise velocity $\bar{u}$ normalized by freestream speed of sound. The magenta dashed lines depict the reversed flow boundaries which include a wide separation bubble on the suction side besides a small bubble on the pressure side, near the trailing edge.}
	\label{fig:mean_flow}
\end{figure}
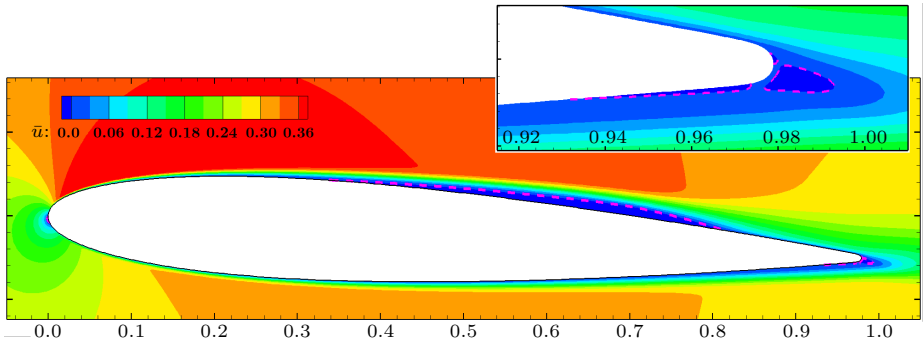

The mean flow is presented with contours of $u$-velocity normalized by the freestream speed of sound in figure \ref{fig:mean_flow}. Important flow features can be observed including a long recirculation bubble that extends over the suction side, from the detachment at $x = 0.34$ until its reattachment at $x = 0.82$. In the magnified view, a small bubble can be observed on the pressure side, followed by a recirculation region at the trailing edge. The blue contours enclosed by magenta lines indicate regions of reversed flow.

\tikzstyle{stuff_fill}=[rectangle,draw,fill=white]

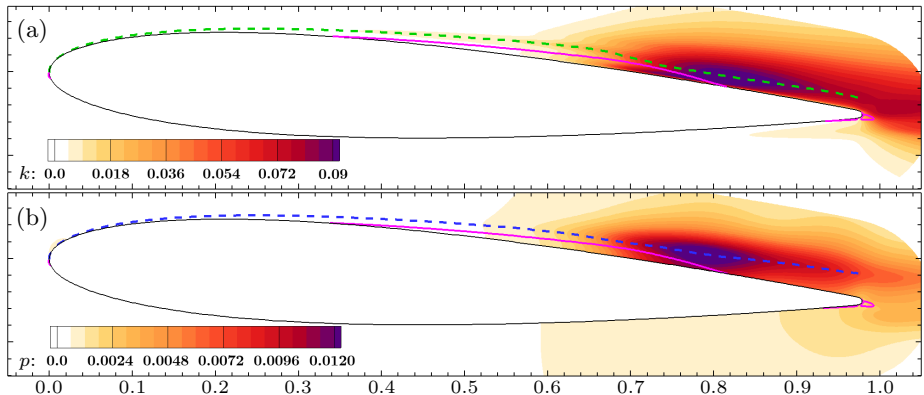
\begin{figure}
	\centering
	\begin{tikzpicture}
	\node[anchor=south west,inner sep=0] (image) at (0,0) {\includegraphics[width=0.90\textwidth]{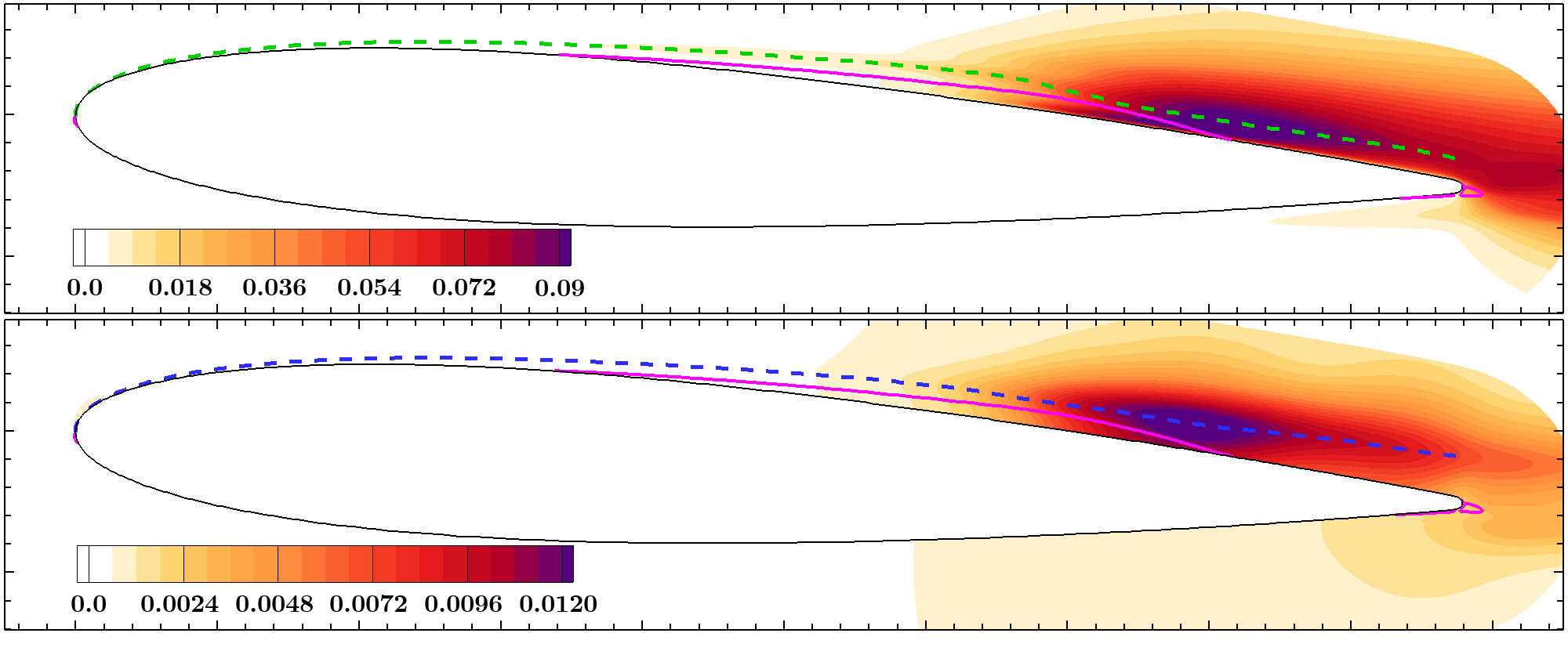}};
	\begin{scope}[x={(image.south east)},y={(image.north west)}]
	\node[] at (0.03,0.93) {(a)};
	\node[] at (0.03,0.45) {(b)};			
	\node[] at (0.025,0.568) {\scriptsize $k$:};
	\node[] at (0.025,0.080) {\scriptsize $p$:};			
	\node[] at (0.0480,0.02) {\scriptsize $0.0$};
	\node[] at (0.1390,0.02) {\scriptsize $0.1$};
	\node[] at (0.2290,0.02) {\scriptsize $0.2$};
	\node[] at (0.3190,0.02) {\scriptsize $0.3$};						
	\node[] at (0.4100,0.02) {\scriptsize $0.4$};
	\node[] at (0.5000,0.02) {\scriptsize $0.5$};
	\node[] at (0.5900,0.02) {\scriptsize $0.6$};
	\node[] at (0.6810,0.02) {\scriptsize $0.7$};
	\node[] at (0.7710,0.02) {\scriptsize $0.8$};
	\node[] at (0.8620,0.02) {\scriptsize $0.9$};
	\node[] at (0.9520,0.02) {\scriptsize $1.0$};
	\end{scope}
	\end{tikzpicture}
	\caption{RMS values of (a) kinetic energy and (b) pressure. The green and blue dashed lines depict locations of maximum fluctuations along the shear layer on the suction side, while the magenta solid line delimits the reversed flow region.}
	\label{fig:k_rms}
\end{figure}

Distributions of root-mean-square (RMS) for kinetic energy $k$ and pressure $p$ are presented in figures \ref{fig:k_rms}(a) and (b), respectively. As it can be seen from the plots, fluctuations start amplifying along the bubble and the highest values appear at the reattachment region, on the suction side. 
The green and blue dashed lines in figures \ref{fig:k_rms}(a) and (b), respectively, depict the locations of maximum $k$ and $p$ along the shear layer forming on the suction side. In the following analyses, these lines will be used \blue{as reference locations for data extraction to track flow disturbances}. Again, the solid magenta line shows the recirculation bubble.

\tikzstyle{stuff_fill}=[rectangle,draw,fill=white]

\begin{figure}
	\centering
	\subfigure[$C_p$]{\includegraphics[width=0.32\textwidth]{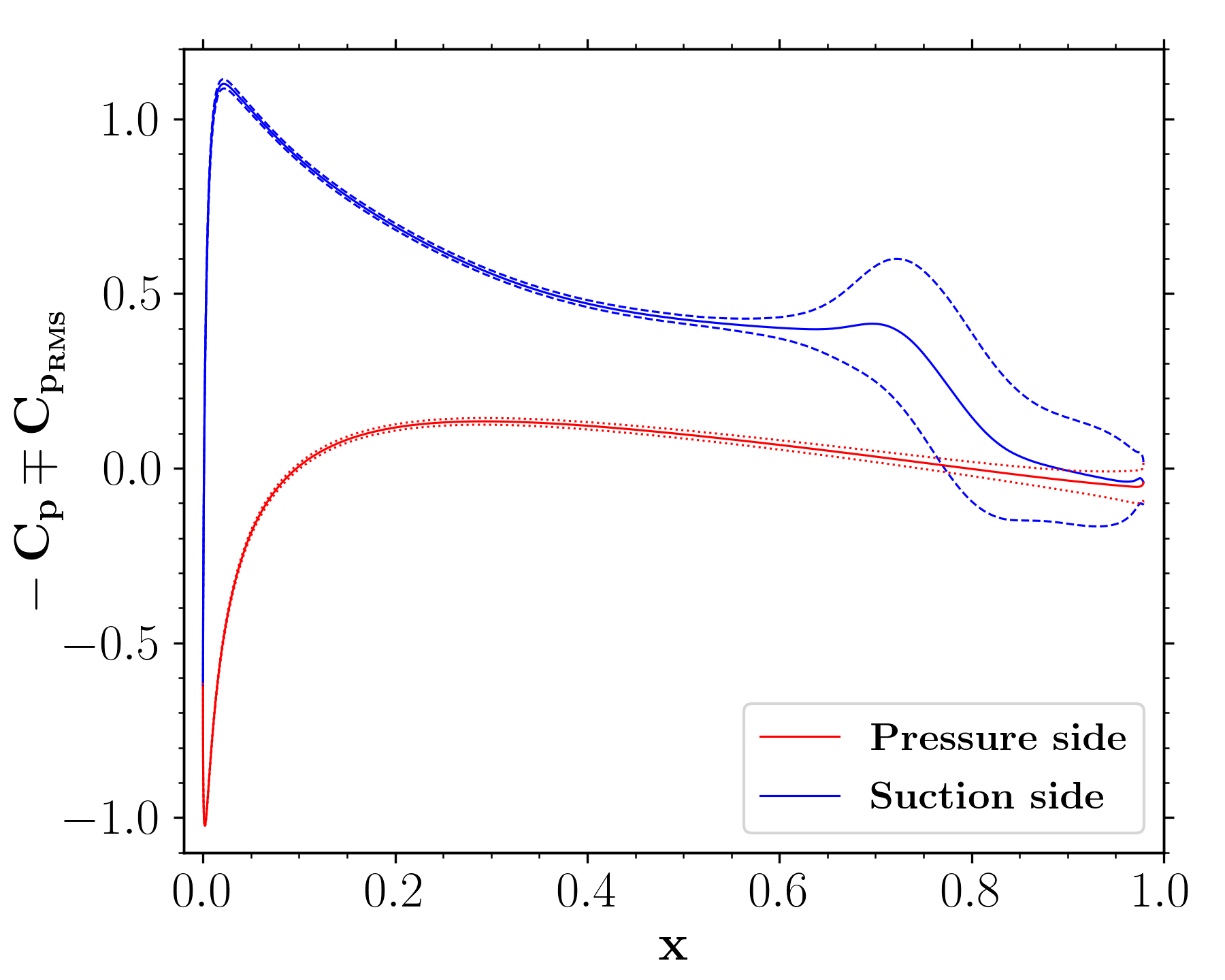}}
	\subfigure[$C_f$]{\includegraphics[width=0.32\textwidth]{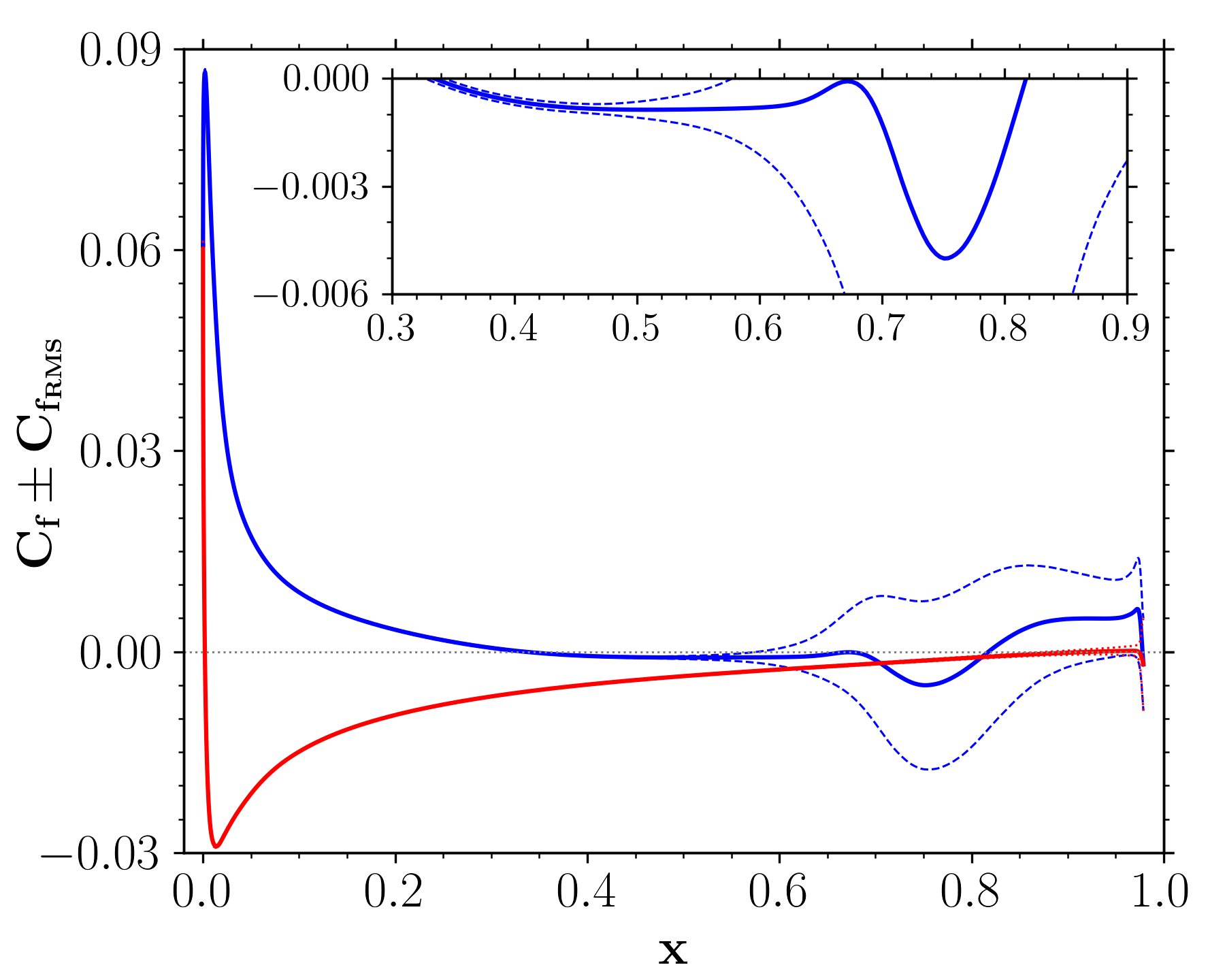}}
	\subfigure[$\bar{u}_t$]{\includegraphics[width=0.32\textwidth]{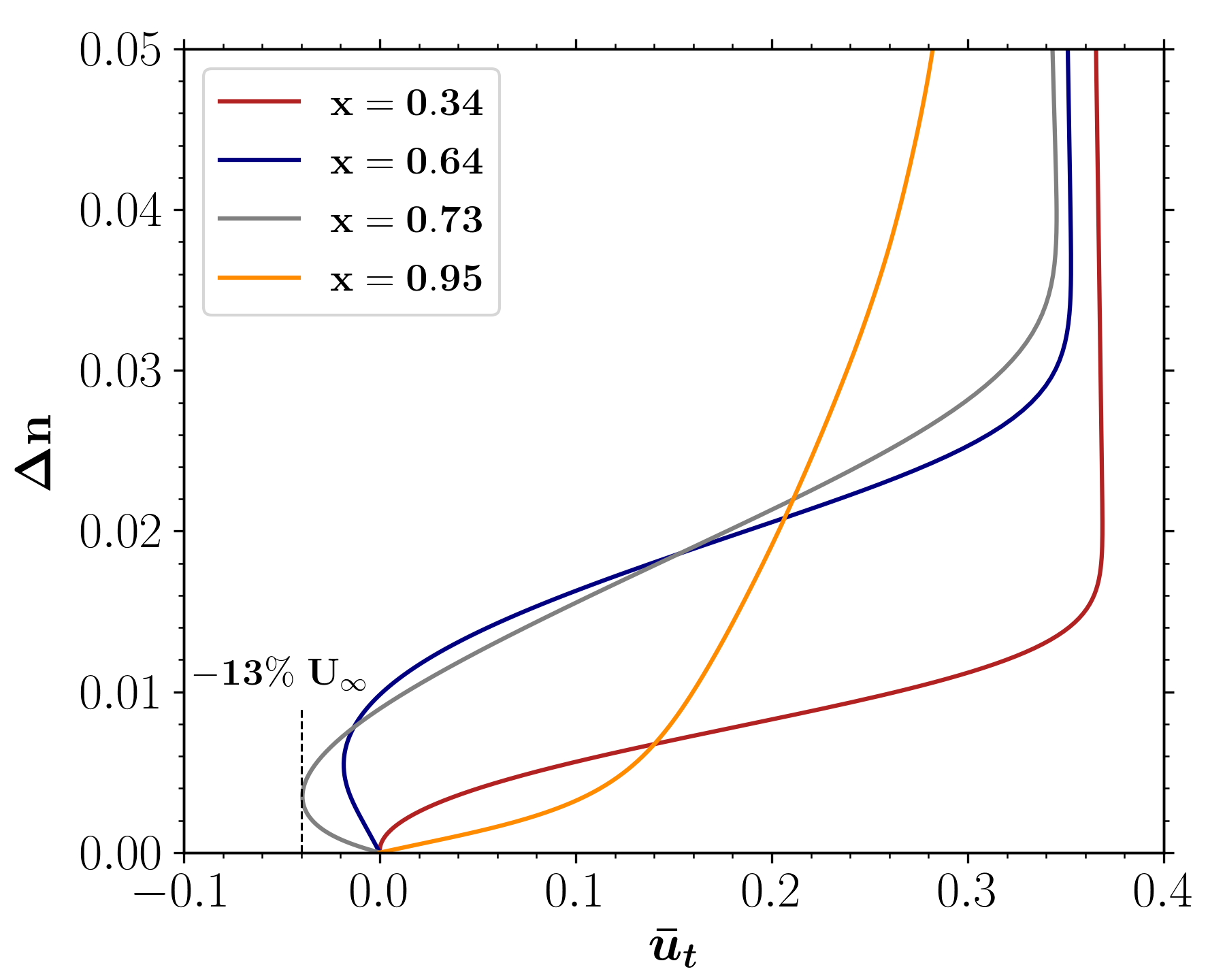}}
	\subfigure[$C_{p_{\mbox{\tiny RMS}}}$]{\includegraphics[width=0.32\textwidth]{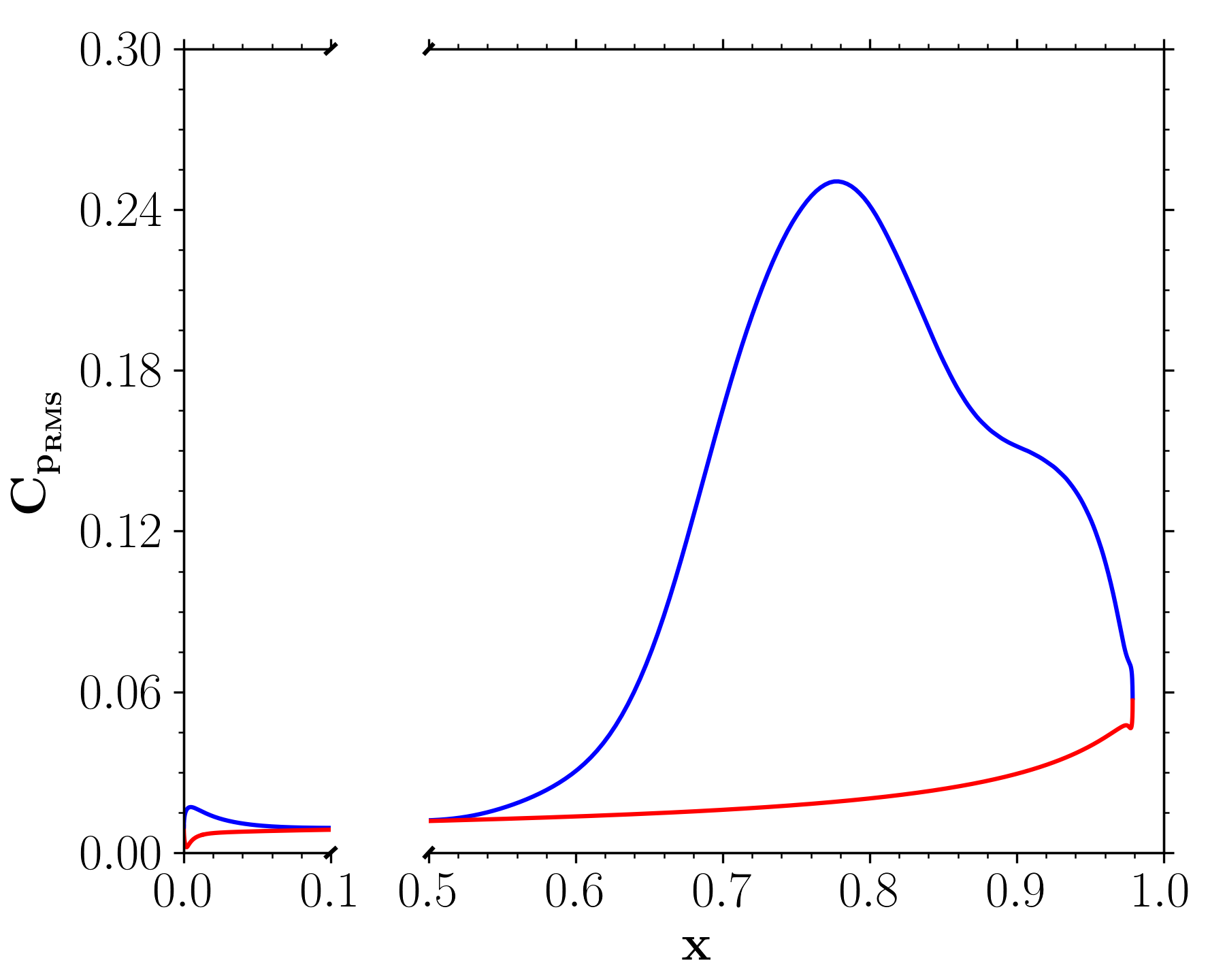}}
	\subfigure[$C_{f_{\mbox{\tiny RMS}}}$]{\includegraphics[width=0.32\textwidth]{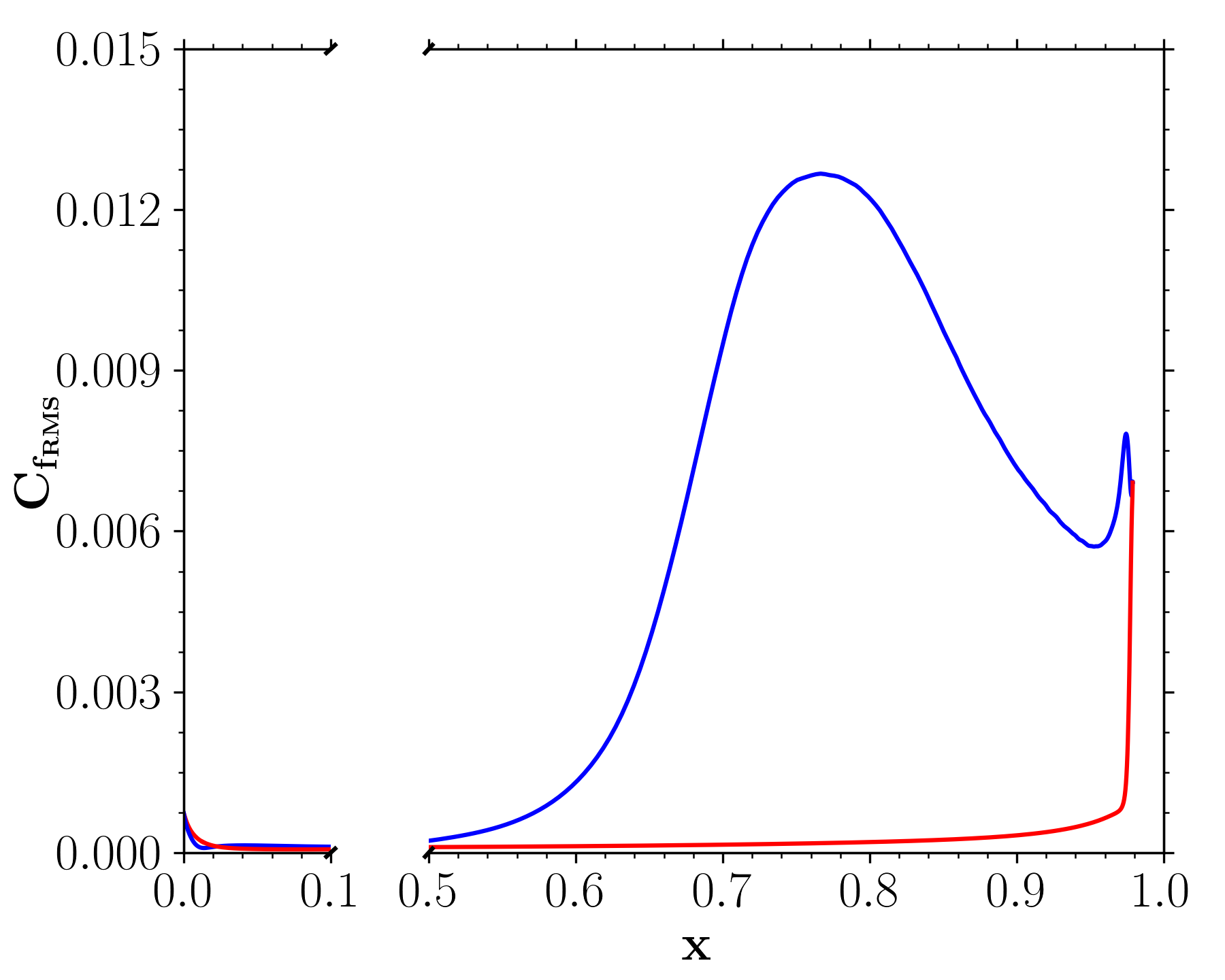}}
	\subfigure[$u_{t_{RMS}}$]{\includegraphics[width=0.32\textwidth]{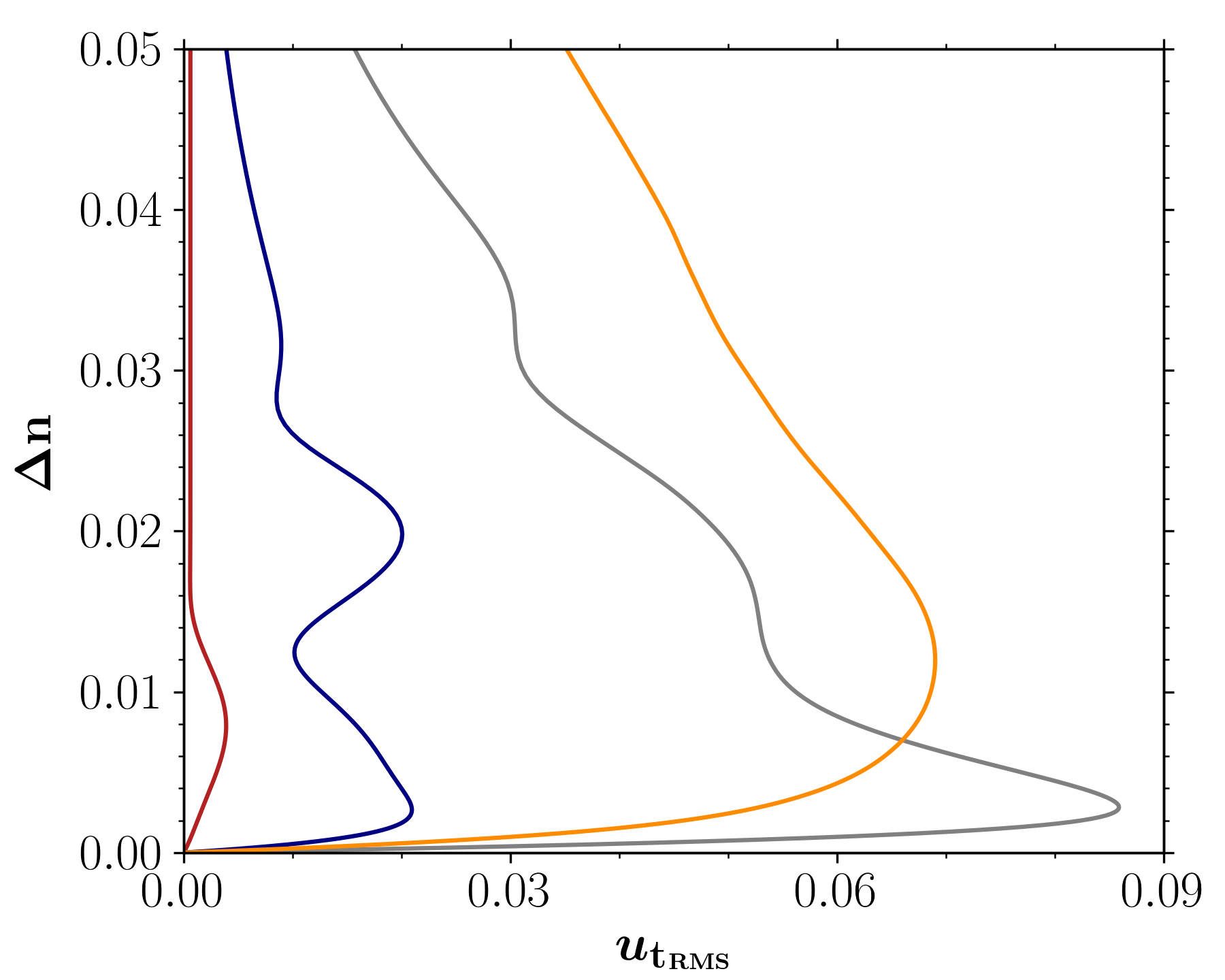}}
	\caption{Mean and RMS distributions of (a,d) pressure and (b,e) friction coefficients along the airfoil surface, and (c,f) tangential velocity profiles computed in the wall-normal direction $\Delta n$.}
	\label{fig:cp_and_cf}
\end{figure}

The negative mean pressure coefficient $-C_p$ is shown in figure \ref{fig:cp_and_cf}(a) with blue and red solid lines for suction and pressure sides, respectively. To highlight the regions with intense fluctuations, the pressure coefficient is presented as $-C_p \mp C_{p_{\mbox{\tiny RMS}}}$, where the blue dashed and red dotted lines correspond to the suction and pressure sides, respectively. The peak value of pressure coefficient on the suction side is observed near the leading edge at $x = 0.02$.
From this point onward, a pressure increase leads to a drop in $-C_p$ until reaching a plateau along the laminar separation bubble. The plateau extends up to $x = 0.7$, where the pressure further increases towards the trailing edge, causing another drop in $-C_p$. Pressure fluctuations are shown in figure \ref{fig:cp_and_cf}(d) for both suction and pressure sides. The RMS values of $C_p$ are low along the entire pressure side, and upstream of the bubble ($x \lesssim 0.5$) on the suction side. However, they increase towards the peak at $x \approx 0.78$ on the suction side.

The mean skin friction $C_f$ distribution over the airfoil is presented in figure \ref{fig:cp_and_cf}(b). The RMS values of skin friction $C_{f_{\mbox{\tiny RMS}}}$, shown in figure \ref{fig:cp_and_cf}(e), are also added to the mean values. A magnified view highlights the presence of the laminar separation bubble which starts at $x = 0.34$ and reattaches at $x = 0.67$. The reattachment is followed by a second detachment which reaches a minimum at $x = 0.75$ and reattaches at $x = 0.82$. \green{The nature of the double detachment profile is discussed in details by \cite{vassilios2000}.}
The RMS peaks at $x \approx 0.76$ due to the high value of $k$ shown in figure \ref{fig:k_rms}(a).

Mean tangential velocity profiles extracted along the wall normal direction $\Delta n$ are presented in figure \ref{fig:cp_and_cf}(c) at different locations along the chord. \blue{At the detachment point, $x = 0.34$, the red curve indicates the zero wall-normal derivative in the velocity profile.} At $x = 0.73$, the velocity profile shown as a grey line exhibits the maximum reversed flow of $-0.13 U_\infty$. The orange line shows a velocity profile at $x = 0.95$, where three-dimensional effects are important due to turbulent transition. In figure \ref{fig:cp_and_cf}(f), RMS values of the tangential velocity are presented and a single peak is observed for the profile computed at $x = 0.34$. Velocity fluctuations increase inside the bubble at $x = 0.64$ and become dominant at $x = 0.73$. In both these locations, the velocity fluctuations present triple peak profiles as also observed by \citet{Nash1999}, \citet{Desquesnes2007} and \cite{vassilios2000}. Near the trailing edge at $x = 0.95$, where a turbulent regime may occur, mixing results in a smoother velocity fluctuation profile.

\subsection{Vortex dynamics}
\label{ssec:dynamics}

\begin{figure}
	\centering
	\begin{overpic}[width=0.99\textwidth]{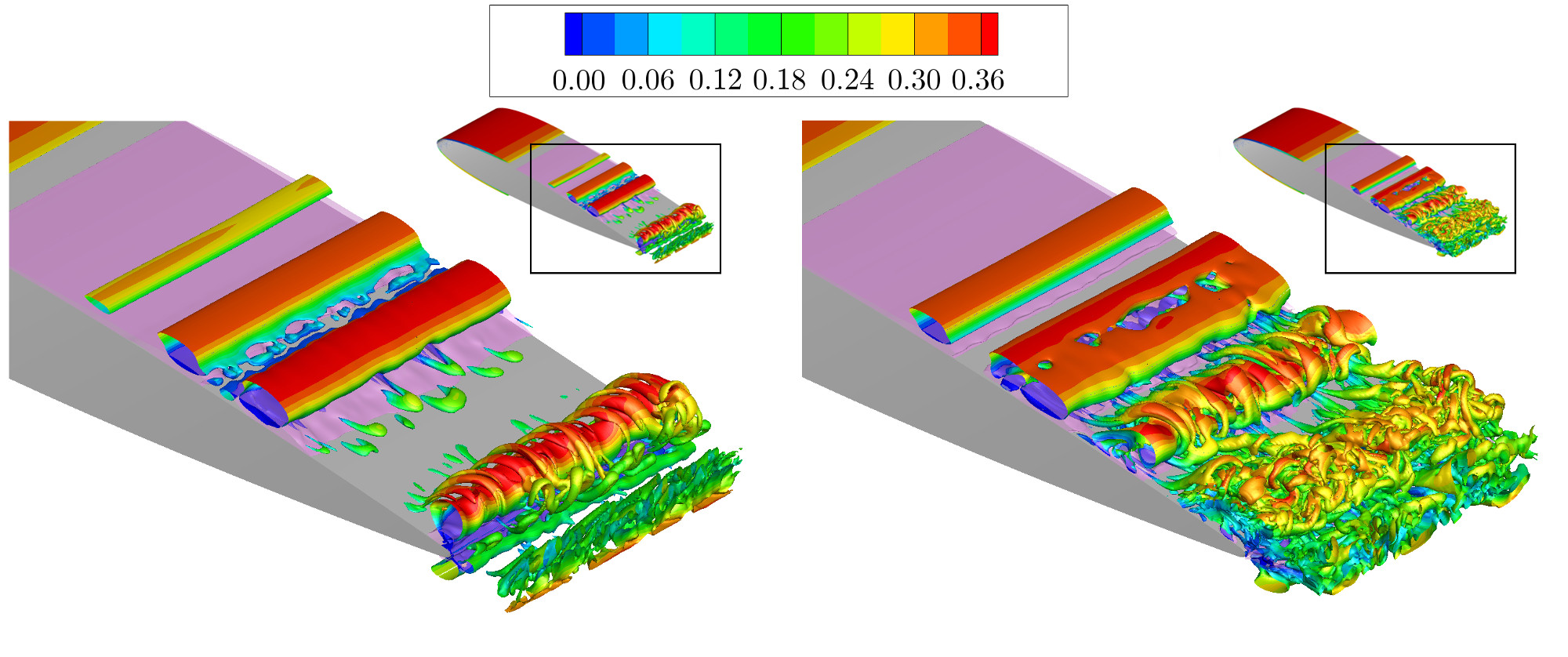}
		\put(4,0){(a) High spanwise coherence, $t = 20.82$}
		\put(54,0){(b) Low spanwise coherence, $t = 21.78$}
		\put(32.2,36.2){$u$:}
	\end{overpic}
	\caption{Instantaneous iso-surfaces of Q-criterion colored by $u$-velocity show regimes with (a) coherent structures at the trailing edge and (b) smaller-scale turbulent eddies.}
	\label{fig:vort_dynamics_3D_suction}
\end{figure}

Instantaneous three-dimensional flow fields over the airfoil are presented in figure \ref{fig:vort_dynamics_3D_suction} with iso-surfaces of Q-criterion colored by $u$-velocity. A magenta shade along the airfoil surface depicts the region of reversed flow along the separation bubble. This figure shows two-dimensional rolls that are visible along the separation bubble and amplify on the suction side, leading to vortex shedding.
Moreover, different flow regimes are observed at the trailing edge, where \blue{coherent structures} alternates with periods of turbulent packets. For example, figure \ref{fig:vort_dynamics_3D_suction}(a) shows a single \blue{laminar-like} roll reaching the trailing edge at $t = 20.82$ while figure \ref{fig:vort_dynamics_3D_suction}(b) shows that, at $t = 21.78$, vortex breakdown leads to a turbulent regime near the trailing edge.
Readers are referred to the movie of the flow field provided as supplemental material (movie 1) for detailed inspection of the present flow dynamics.

The analysis of flow snapshots is important to understand the vortex dynamics over the airfoil suction side. Based on the acoustic analogy of \citet{FWHall1970}, efficient sound generation is achieved when pressure fluctuations are aligned with an edge, such that it will be maximum for 2D-like perturbations. In this sense, figure \ref{fig:vort_dynamics_2D_suction} shows the spanwise averaged $\omega_z$ vorticity, where blue and red contours represent negative and positive values, respectively.
A dark blue color indicates strong spanwise coherence of a 2D vortex, while light blue contours are representative of uncorrelated turbulence. A thin black line is used to identify individual vorticity packets, based on the iso-contour of an entropy measure given by $\sfrac{p}{\rho^\gamma} - (\sfrac{p_{\infty}}{\rho_{\infty}^\gamma}) = 0.1\%$. This separates the region where vorticity is important from where the flow is irrotational, outside the boundary layer. A magenta dashed line represents the border of averaged negative $\bar{u}$ velocity, as discussed in figure \ref{fig:mean_flow}. The simulation time is shown in convective time units on the lower left corner of all plots. A movie of figure \ref{fig:vort_dynamics_2D_suction} is provided as supplemental material (movie 2) to aid the dynamic visualization of the flow features.


\tikzstyle{stuff_fill}=[rectangle,draw,fill=white]
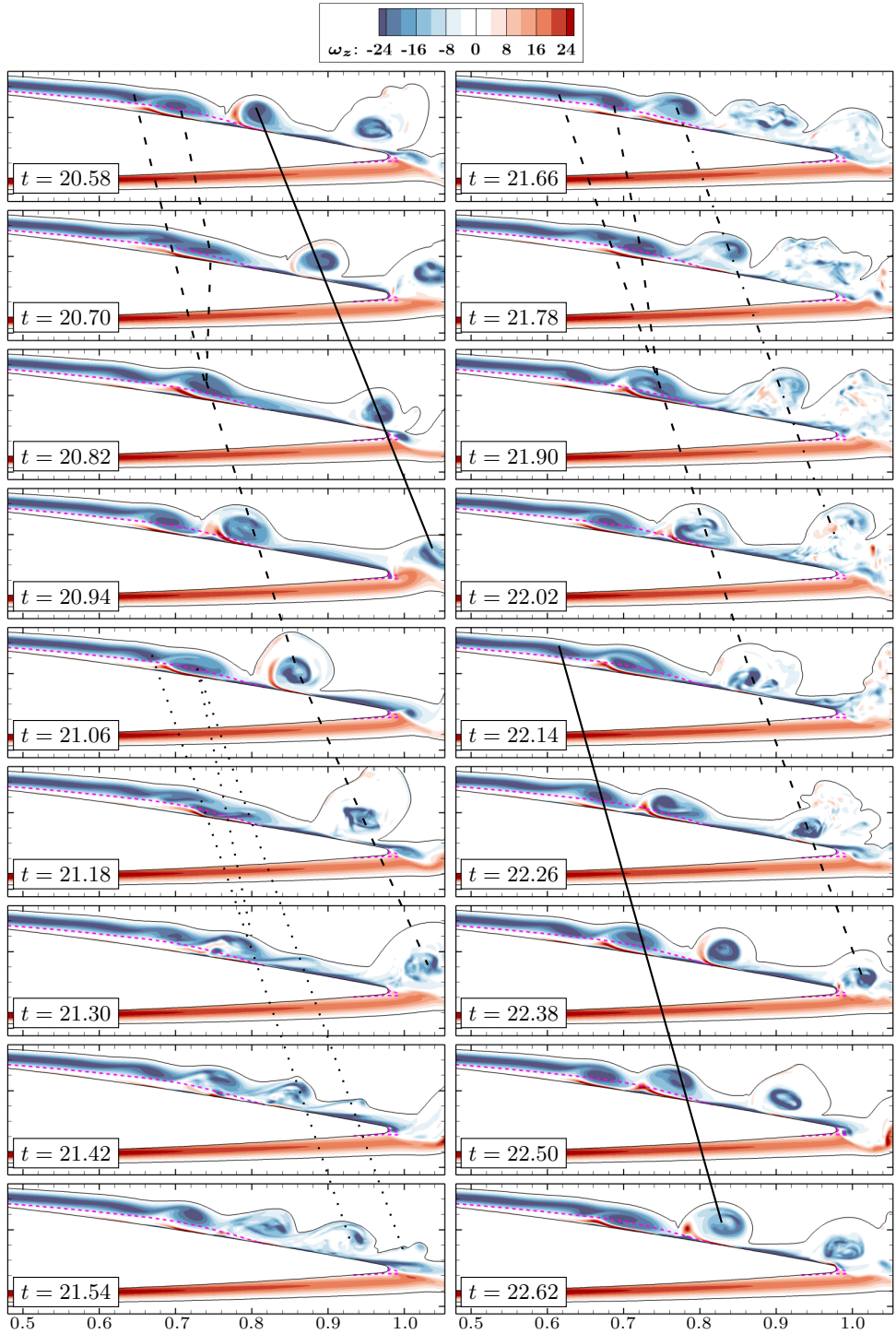
\begin{figure}
	\centering
	\begin{tikzpicture}
	\node[anchor=south west,inner sep=0] (image) at (0,0) {\includegraphics[width=0.96\textwidth]{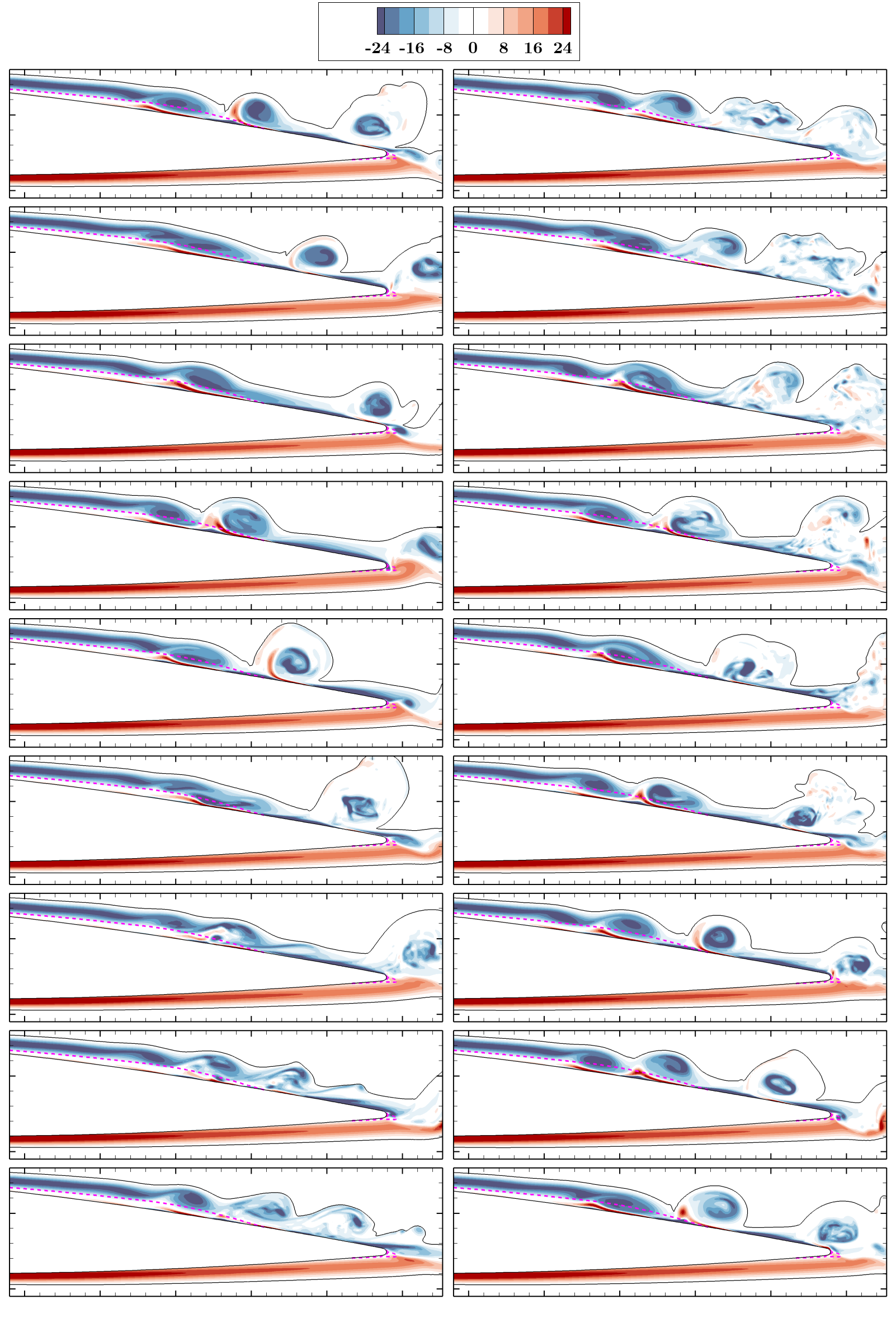}};
	\begin{scope}[x={(image.south east)},y={(image.north west)}]
	\draw [thick] (0.285,0.920) -- (0.48,0.59);
	\draw [thick,loosely dashed] (0.23,0.715) -- (0.33,0.5) -- (0.475,0.278);
	\draw [thick,loosely dashed] (0.15,0.930) -- (0.23,0.710) -- (0.235,0.810)  -- (0.20,0.925);
	\draw [thick,loosely dotted] (0.17,0.51) -- (0.39,0.07);
	\draw [thick,loosely dotted] (0.22,0.50) -- (0.28,0.30);
	\draw [thick,loosely dotted] (0.22,0.50) -- (0.45,0.06);
	\draw [thick,loosely dashed] (0.73,0.715) -- (0.83,0.5) -- (0.955,0.27);
	\draw [thick,loosely dashed] (0.62,0.930) -- (0.73,0.715) -- (0.71,0.815)  -- (0.68,0.925);
	\draw [thick,loosely dash dot] (0.75,0.920) -- (0.925,0.60);
	\draw [thick,solid] (0.62,0.517) -- (0.80,0.085);
	\node[] at (0.075,0.867) [stuff_fill] {\footnotesize $t=20.58$};
	\node[] at (0.075,0.763) [stuff_fill] {\footnotesize $t=20.70$};
	\node[] at (0.075,0.659) [stuff_fill] {\footnotesize $t=20.82$};
	\node[] at (0.075,0.554) [stuff_fill] {\footnotesize $t=20.94$};	
	\node[] at (0.075,0.450) [stuff_fill] {\footnotesize $t=21.06$};
	\node[] at (0.075,0.346) [stuff_fill] {\footnotesize $t=21.18$};
	\node[] at (0.075,0.242) [stuff_fill] {\footnotesize $t=21.30$};
	\node[] at (0.075,0.137) [stuff_fill] {\footnotesize $t=21.42$};
	\node[] at (0.075,0.033) [stuff_fill] {\footnotesize $t=21.54$};		
	\node[] at (0.570,0.867) [stuff_fill] {\footnotesize $t=21.66$};
	\node[] at (0.570,0.763) [stuff_fill] {\footnotesize $t=21.78$};
	\node[] at (0.570,0.659) [stuff_fill] {\footnotesize $t=21.90$};
	\node[] at (0.570,0.554) [stuff_fill] {\footnotesize $t=22.02$};	
	\node[] at (0.570,0.450) [stuff_fill] {\footnotesize $t=22.14$};
	\node[] at (0.570,0.346) [stuff_fill] {\footnotesize $t=22.26$};
	\node[] at (0.570,0.242) [stuff_fill] {\footnotesize $t=22.38$};
	\node[] at (0.570,0.137) [stuff_fill] {\footnotesize $t=22.50$};
	\node[] at (0.570,0.033) [stuff_fill] {\footnotesize $t=22.62$};
	\node[] at (0.382,0.962) {\scriptsize $\boldsymbol{\omega_z}$:};			
	\node[] at (0.027,0.01) {\scriptsize $0.5$};
	\node[] at (0.112,0.01) {\scriptsize $0.6$};
	\node[] at (0.198,0.01) {\scriptsize $0.7$};
	\node[] at (0.280,0.01) {\scriptsize $0.8$};
	\node[] at (0.365,0.01) {\scriptsize $0.9$};
	\node[] at (0.450,0.01) {\scriptsize $1.0$};			
	\node[] at (0.523,0.01) {\scriptsize $0.5$};
	\node[] at (0.607,0.01) {\scriptsize $0.6$};
	\node[] at (0.691,0.01) {\scriptsize $0.7$};
	\node[] at (0.775,0.01) {\scriptsize $0.8$};
	\node[] at (0.860,0.01) {\scriptsize $0.9$};
	\node[] at (0.945,0.01) {\scriptsize $1.0$};
	%
	\end{scope}
	\end{tikzpicture}
	\caption{Spanwise averaged vorticity $\omega_z$ shows vortex pairing over the airfoil. The magenta dashed line represents the boundary of reversed flow (negative $\bar{u}$). The black lines connecting the sub-figures mark the evolution of vortices with high ($\mathbf{-}$ and $- \; -$) or low ($\cdot \cdot \cdot$ and $- \cdot -$) spanwise coherence.}
	\label{fig:vort_dynamics_2D_suction}	
\end{figure}	

As shown in figure \ref{fig:vort_dynamics_2D_suction}, the flow dynamics is dominated by events on the suction side, \green{where vortices are shed from the laminar separation bubble}. During the shedding process, there is a possibility that the 2D laminar vortices undergo a process of vortex pairing. In case this process is not successful, vortex bursting leads to turbulent transition and low coherence along the span, represented by dotted lines. These turbulent packets may interact with other vortices and further reduce their coherence, as indicated by the dashed-dotted line. On the other hand, if the vortices merge or only a solitary vortex is shed from the bubble, the trend is for either structure to keep its high coherence up to the trailing edge. Both of these processes are represented by the dashed (merging) and solid (single vortex) lines in figure \ref{fig:vort_dynamics_2D_suction}. \green{A detailed discussion on the consequence and cause of this behavior is presented in sections \ref{ssec:intermittency} and \ref{ssec:merging}, respectively.} Similar dynamics of vortex shedding from LSBs are observed experimentally by \cite{yarusevich2016_coherent}.

\begin{figure}
	\centering
	\begin{tikzpicture}
	\node[anchor=south west,inner sep=0] (image) at (0,0) {\includegraphics[width=0.52\textwidth]{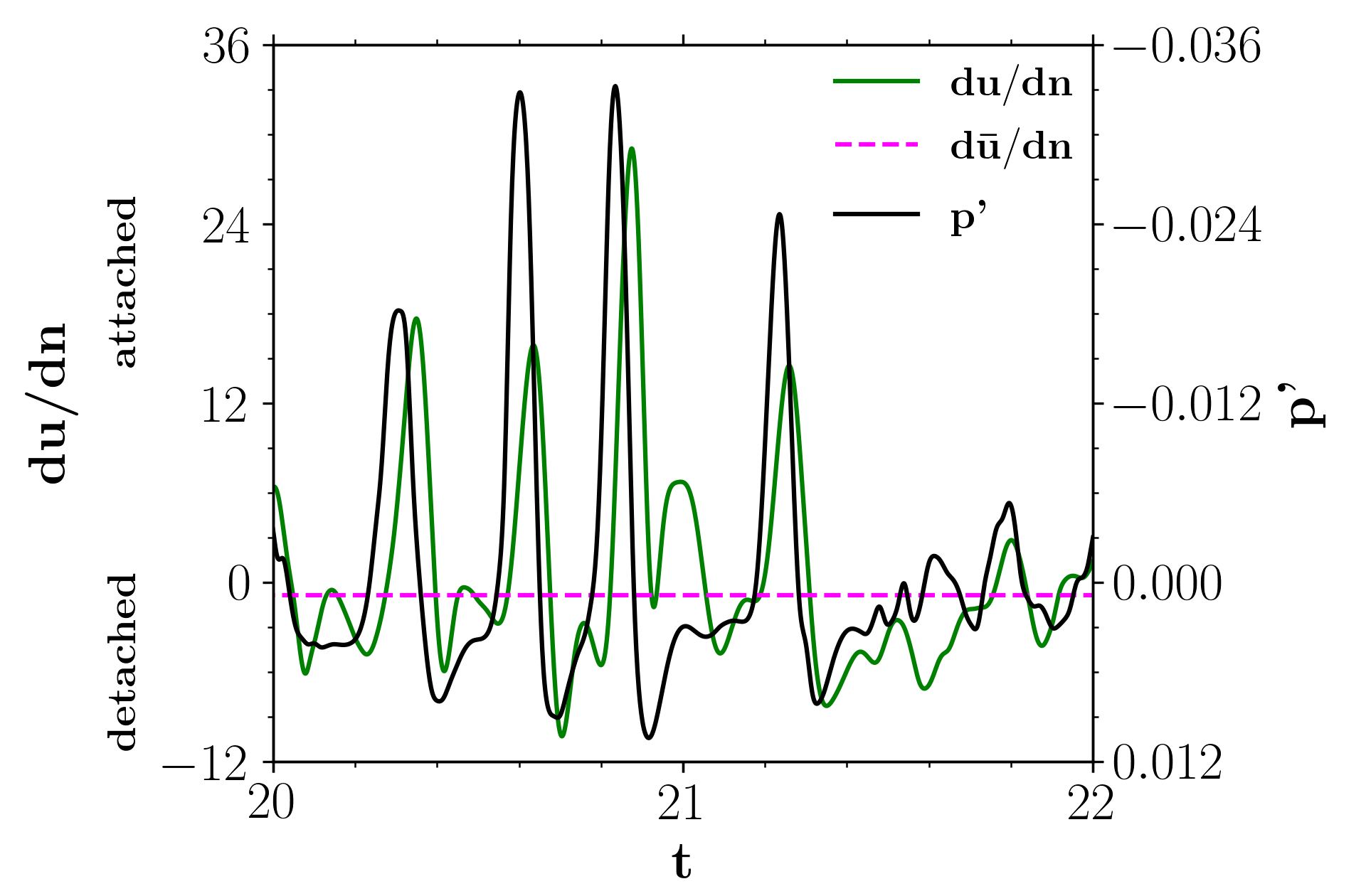}};
	\begin{scope}[x={(image.south east)},y={(image.north west)}]
	\node[] at (0.50,-0.06) {(a) Temporal signal};
	\end{scope}
	\end{tikzpicture}	
	\hspace{5mm}
	\begin{tikzpicture}
	\node[anchor=south west,inner sep=0] (image) at (0,0) {\includegraphics[width=0.375\textwidth]{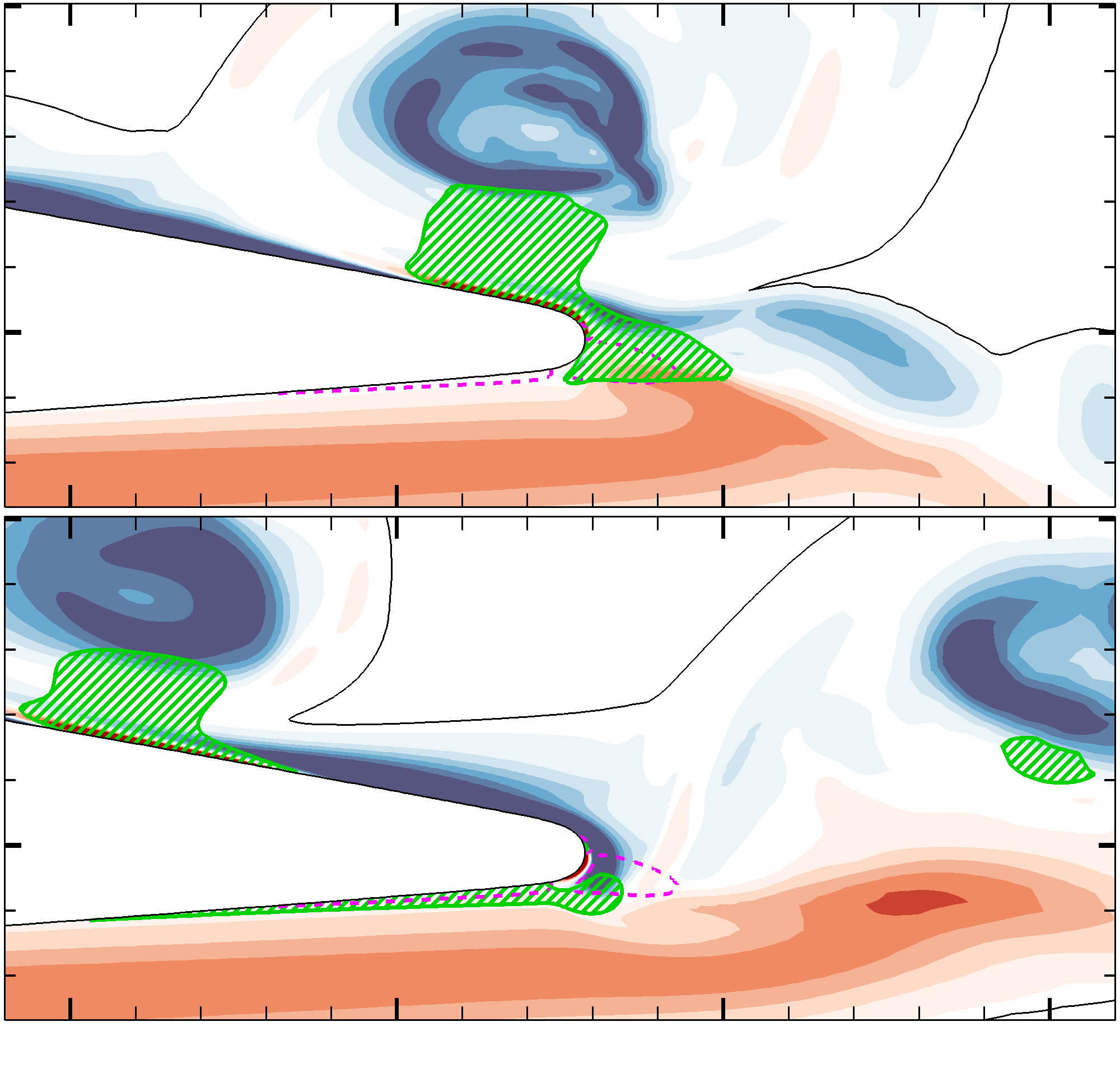}};
	\begin{scope}[x={(image.south east)},y={(image.north west)}]
	\node[] at (0.83,0.92) [stuff_fill] {\footnotesize $t=20.59$};
	\node[] at (0.83,0.44) [stuff_fill] {\footnotesize $t=20.73$};
	\node[] at (0.060,0.015) {\scriptsize $0.90$};
	\node[] at (0.355,0.015) {\scriptsize $0.95$};
	\node[] at (0.645,0.015) {\scriptsize $1.00$};
	\node[] at (0.935,0.015) {\scriptsize $1.05$};
	\node[] at (0.50,-0.06) {(b) Flow snapshots};
	\end{scope}
	\end{tikzpicture}	
	\caption{Intermittency of the pressure side bubble exhibits dependence on the suction side vortices as presented for (a) temporal signals and (b) flow snapshots. In sub-figure (a), the pressure signal is extracted on the suction side at $x = 0.98$ and the velocity derivative is computed on the pressure side at $x = 0.95$. In sub-figure (b), the hatched green region indicates instantaneous values of negative $u$-velocity. In both plots, the magenta dashed line indicates negative values in the time-averaged $u$-velocity.}
	\label{fig:intermittency_PS_bubble}
\end{figure}

\green{As shown in the experimental investigations of  \cite{Probsting2015_bubble} and \cite{Probsting2015_regimes}, for this Reynolds number and angle of attack, it is expected that the pressure side does not play an important role in terms of the flow dynamics. In order to confirm this observation, a temporal signal of pressure fluctuation is extracted at $x = 0.98$, marked by the blue dashed line in figure \ref{fig:k_rms}(b). A second signal is obtained for the velocity derivative in the wall-normal direction on the pressure side at $x = 0.95$. Both signals are presented in figure \ref{fig:intermittency_PS_bubble}(a) as well as the time averaged $d\bar{u}/dn$, indicating that the flow is, on average, recirculating ($d\bar{u}/dn<0$). In the velocity derivatives, the wall-normal direction is positive pointing outwards from the airfoil surface. Snapshots of the flow dynamics are presented in \ref{fig:intermittency_PS_bubble}(b), where the red-blue contours are the $\omega_z$ vorticity, similarly to the previous figures, the green hatched colors show instantaneous regions of negative $u$-velocity while the dashed magenta line indicates the boundaries of the time-averaged pressure side bubble. In this figure, it is possible to see that when a vortex arrive at the trailing edge from the suction side at $t = 20.59$ (negative $p'$ peak), the flow reattaches on the bottom side (positive $du/dn$ at the wall). As the vortex leaves the airfoil surface on the suction side at $t=20.73$, the pressure increases on the suction side (positive $p'$) and, as a consequence, the flow detaches on the pressure side (negative $du/dn$). Thus, the magenta dashed line in figure \ref{fig:intermittency_PS_bubble}(a) shows that the recirculation bubble exists on an time-averaged sense. Moreover, these figures show that the pressure side dynamics seems to be a reaction to the suction side events in the present flow.}

	\subsection{Intermittency and sound generation}
	\label{ssec:intermittency}
	
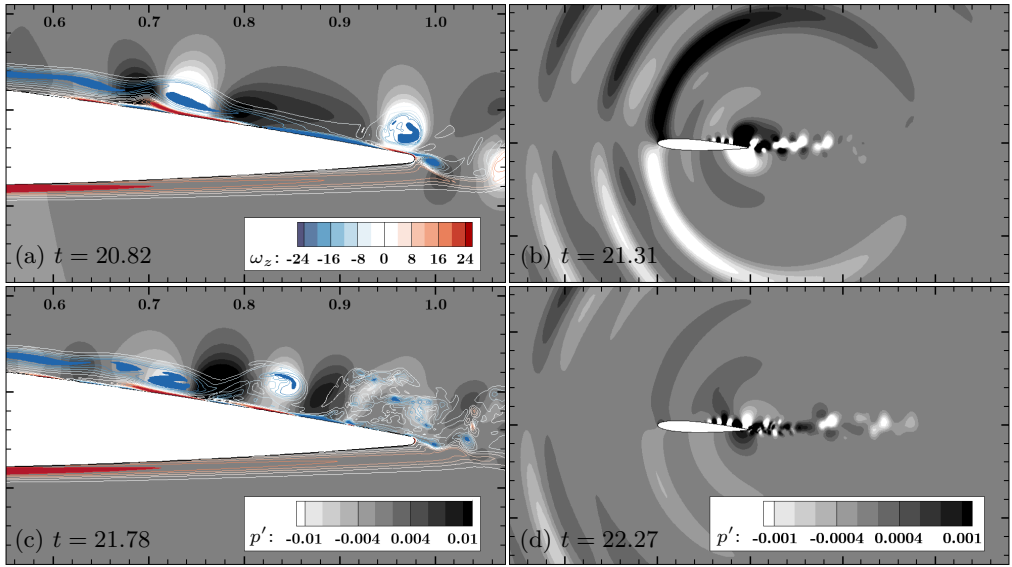
\begin{figure}
		\centering		
		\begin{overpic}[width=0.99\textwidth]{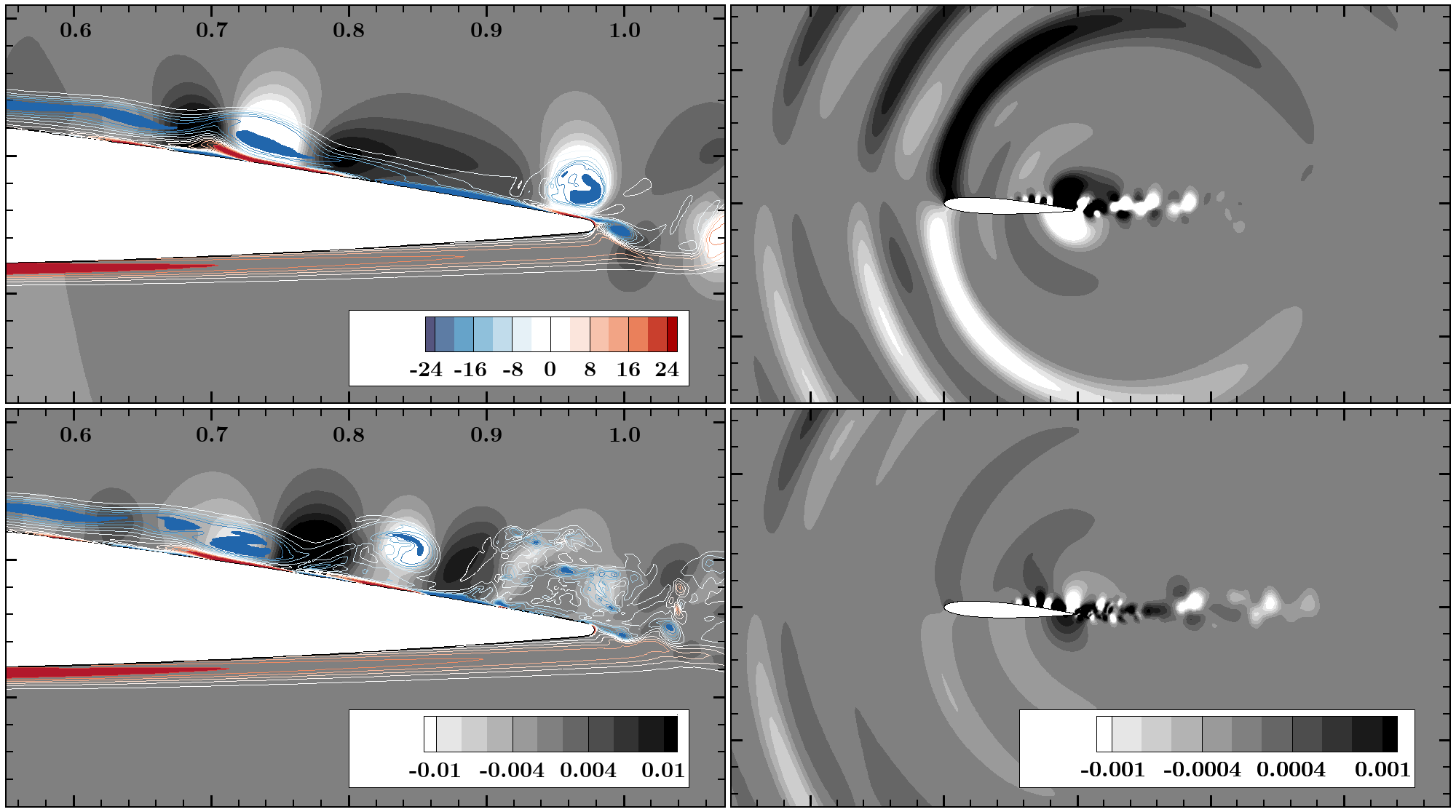}
			\put(1.2,30){(a) $t = 20.82$}
			\put(51.2,30){(b) $t = 21.31$}
			\put(1.2,2){(c) $t = 21.78$}
			\put(51.2,2){(d) $t = 22.27$}
			\put(24.5,30){\scriptsize $\omega_z$:}
			\put(24.5,2.5){\scriptsize $p'$:}
			\put(70.7,2.5){\scriptsize $p'$:}
		\end{overpic} 	
		\caption{Acoustic radiation from the airfoil. Sub-figure (a) shows a coherent structure near the trailing edge at $t = 20.82$ and its subsequent intense acoustic wave reaching the leading edge in sub-figure (b), at a retarded time $t = 21.31$. Sub-figure (c) shows an uncorrelated turbulent packet near the trailing edge at $t = 21.78$ and its subsequent weaker acoustic wave reaching the leading edge in sub-figure (d), at a retarded time $t = 22.27$. Note that the levels from figures (b,d) are 10 times lower than those from figures (a,c).}
		\label{fig:acoustic_waves}
	\end{figure}

		\green{The acoustic analogy of \citet{FWHall1970} states that efficient acoustic scattering occurs due to flow fluctuations near a trailing edge. In this sense, the hydrodynamic nearfield with pressure fluctuations is presented in figures \ref{fig:acoustic_waves}(a,c) with black-white contours, while the acoustic field is presented in figures \ref{fig:acoustic_waves}(b,d). In the latter, the levels are presented 10 times lower compared to the former.} The figures also display $\omega_z$ vorticity in red-blue colors. To highlight the spanwise coherence, regions of $|\omega_z| > 24$ are plotted as filled contours. The vortex cores are related to lower pressures (white background contours) while the gaps between the cores display higher pressures (black contours). A coherent vortex at the trailing edge is presented in figure \ref{fig:acoustic_waves}(a) at a time instant $t=20.82$. The highly coherent hydrodynamic structure leads to efficient acoustic scattering at the trailing edge which, in turn, induces the emission of an intense acoustic wave. The wave is scattered with a $\upi$ phase opposition relative to the incident fluctuations and it reaches the leading edge at a retarded time $t = 21.31$, as shown in figure \ref{fig:acoustic_waves}(b). When the flow is turbulent, at $t = 21.78$, a less coherent vorticity packet reaches the trailing edge as shown in figure \ref{fig:acoustic_waves}(c). In this case, a weaker acoustic wave is emitted from the trailing edge, reaching the leading edge at a retarded time of $t = 22.27$, as exhibited in figure \ref{fig:acoustic_waves}(d). A movie of figure \ref{fig:acoustic_waves} is submitted as supplemental material (movie 3) to show the relation between the acoustic field and vortex dynamics.	
	
\begin{figure}
	\centering
	\subfigure[Pressure fluctuation signal at the trailing edge, $x = 0.98$]{\includegraphics[width=0.675\textwidth]{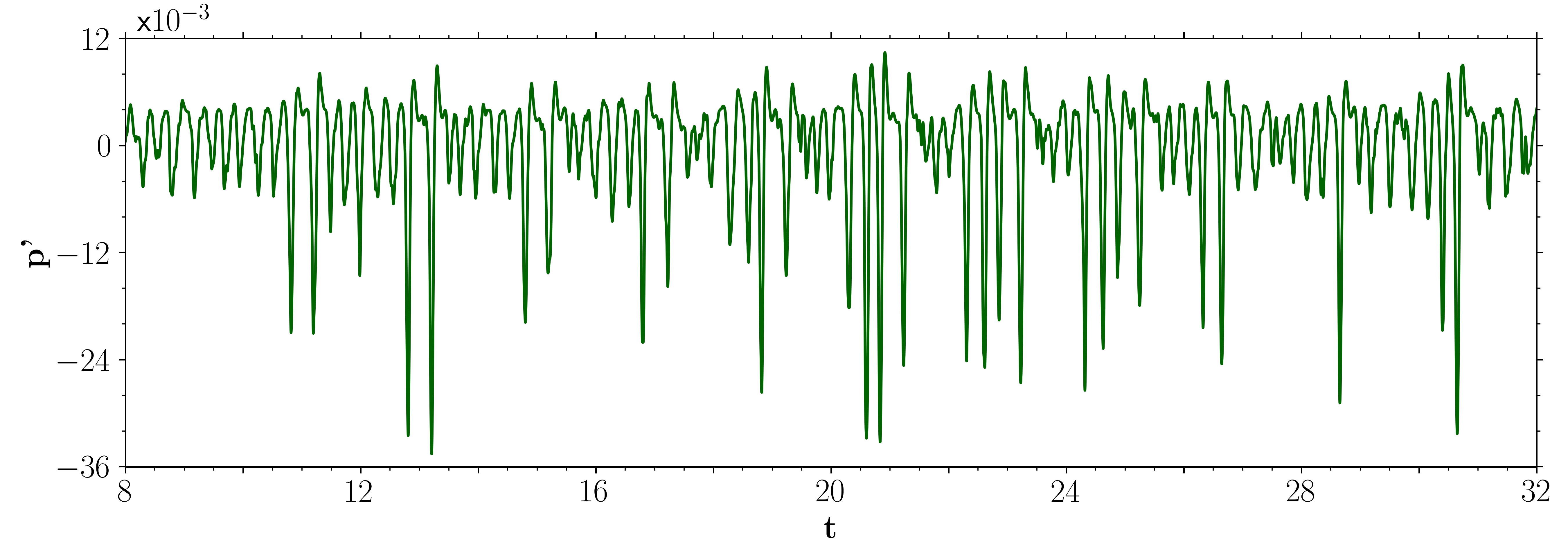}}
	\hfill
	\subfigure[Fourier spectra]{\includegraphics[width=0.315\textwidth]{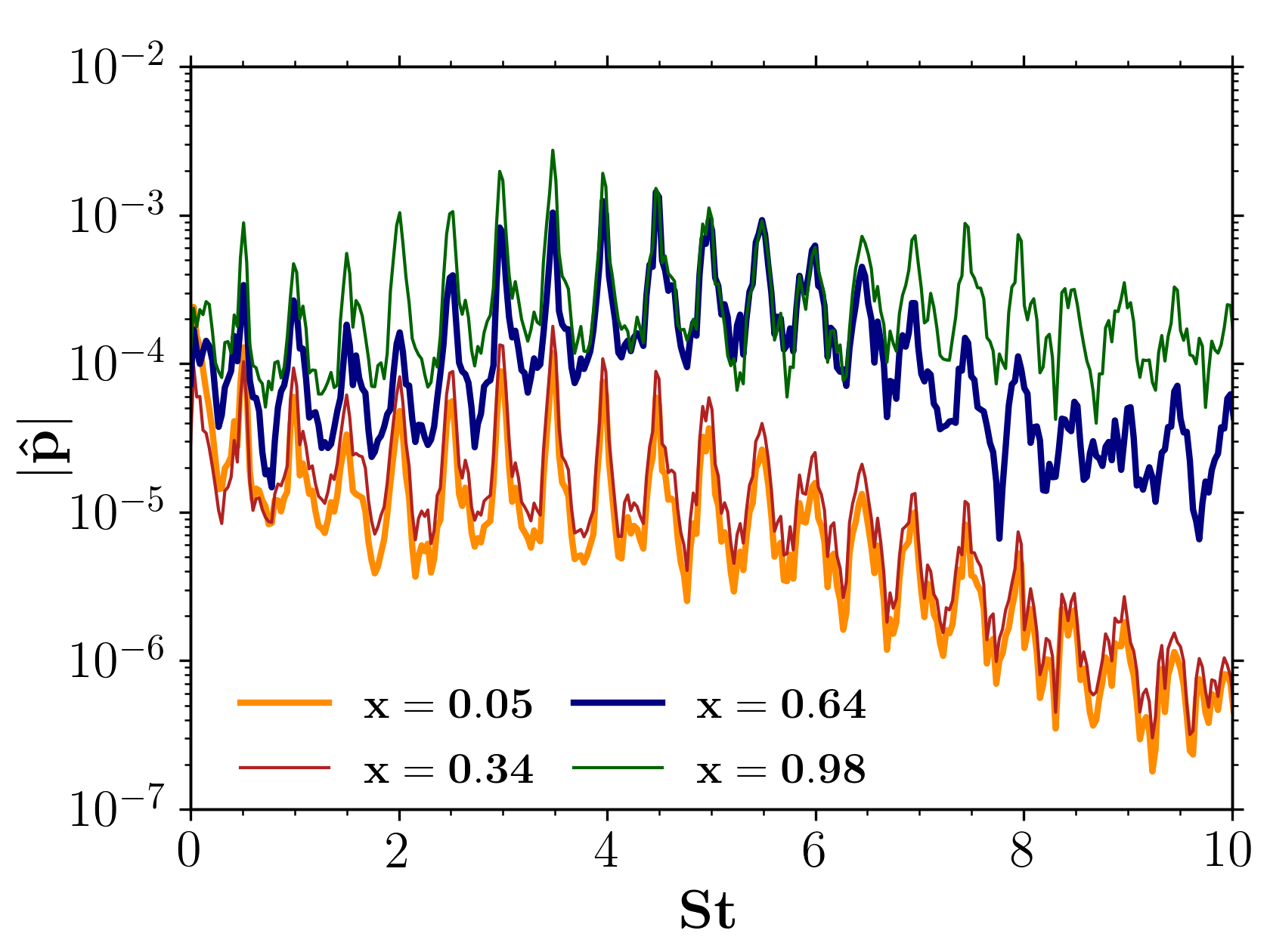}}
	\subfigure[Pressure fluctuation signal in the acoustic near-field at $(x,y) = (1c,1c)$]{\includegraphics[width=0.675\textwidth]{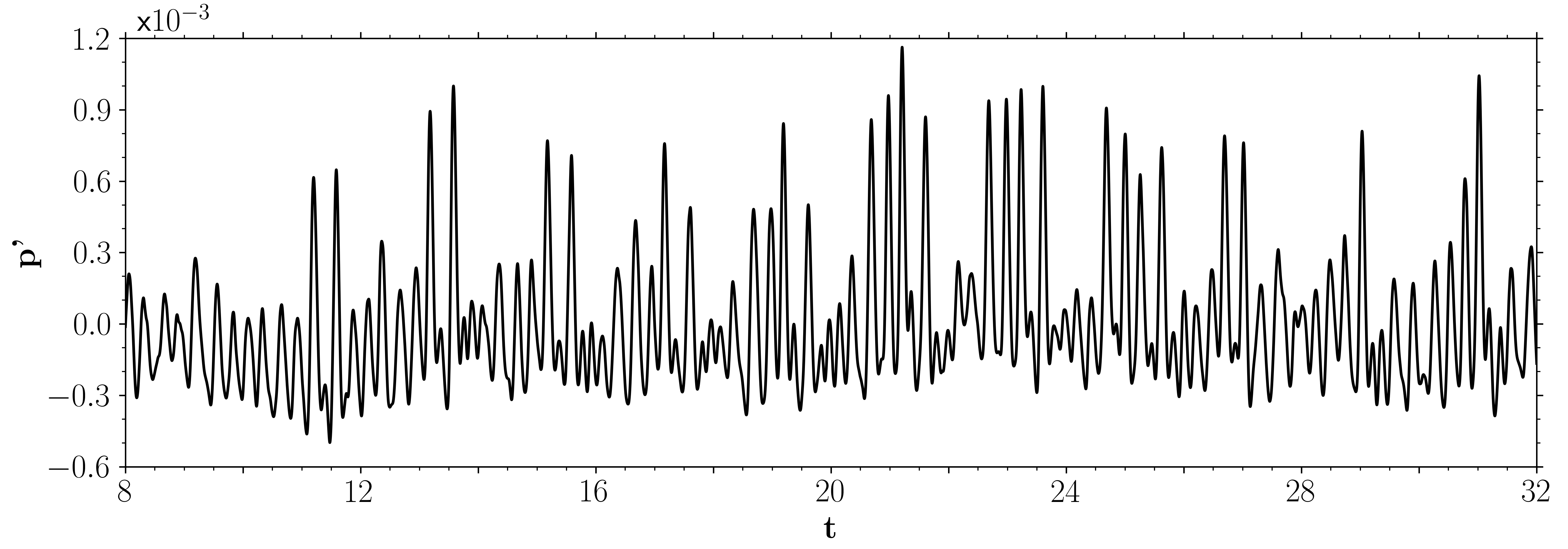}}	
	\hfill
	\subfigure[Fourier spectra]{\includegraphics[width=0.315\textwidth]{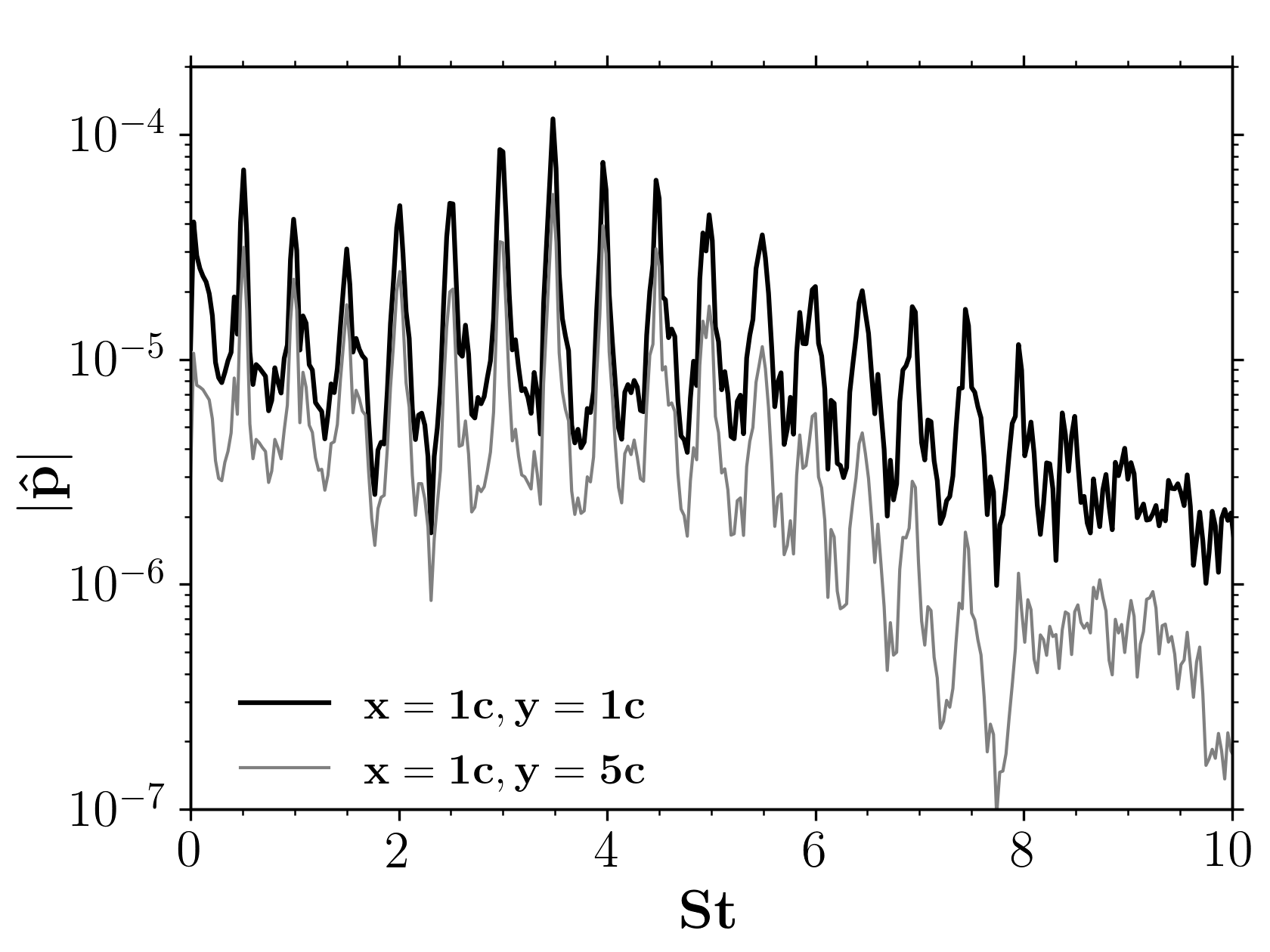}}
	\caption{Time-signal and Fourier transform of spanwise averaged pressure at different locations (a,b) inside the boundary layer along the chord and (c,d) in the acoustic field.}
	\label{fig:fft_probes}
	\end{figure}

	Temporal signals are acquired at different positions along the blue dashed line in figure \ref{fig:k_rms}(b) in terms of spanwise averaged \red{non-dimensional} pressure. The signal at the trailing edge is presented in figure \ref{fig:fft_probes}(a) and displays intermittent deep valleys related to the passage of coherent structures as shown in figure \ref{fig:acoustic_waves}(a). The signal also shows low amplitude oscillations related to the turbulent packets that carry a negative pressure envelope, as described in figure \ref{fig:acoustic_waves}(b). \blue{Such behavior can be characterized as an amplitude modulation of the signal.} The same trends are also observed in the acoustic pressure signal, which is extracted one chord above the trailing edge, at $(x,y) = (1c,1c)$, and it is presented in figure \ref{fig:fft_probes}(c). It is possible to see that the low pressure valleys from figure \ref{fig:fft_probes}(a) result in high pressure peaks that reach the observer position at a retarded time.
	
	The \red{magnitude of the Fourier coefficients is} presented in figures \ref{fig:fft_probes}(b,d) for the hydrodynamic and acoustic fields, respectively, displaying equidistant multiple tones.	\red{The Fourier transforms employ a frequency domain averaging with an overlap of $67\%$ amongst 4 bins covering the entire period. The frequency resolution is $\Delta f = 0.03$ and a Hanning window function with energy correction is applied.} \green{All tonal peaks are integer multiples of the lowest frequency tone, $St \approx 0.5$.} As it can be seen in figure \ref{fig:fft_probes}(b), the dominant peak at the trailing edge is at $St \approx 3.5$ for probes located close to the leading edge ($x = 0.05$), at the detachment point ($x = 0.34$) and the trailing edge ($x = 0.98$). This frequency is related to the passage of low pressure packets, either turbulent or coherent. At $x = 0.64$, in the shear-layer roll-up region, the dominant peak moves to a higher frequency, at $St \approx 4.5$, which may be related to the smaller scale instabilities yet to be merged into larger vortices. \cyan{In the acoustic field, figure \ref{fig:fft_probes}(d) presents results extracted at 1 and 5 chords above the trailing edge. The same frequencies are observed where $St = 3.5$ is still the most prominent peak.}
	
\begin{figure}
		\centering 	
		\includegraphics[width=0.99\textwidth]{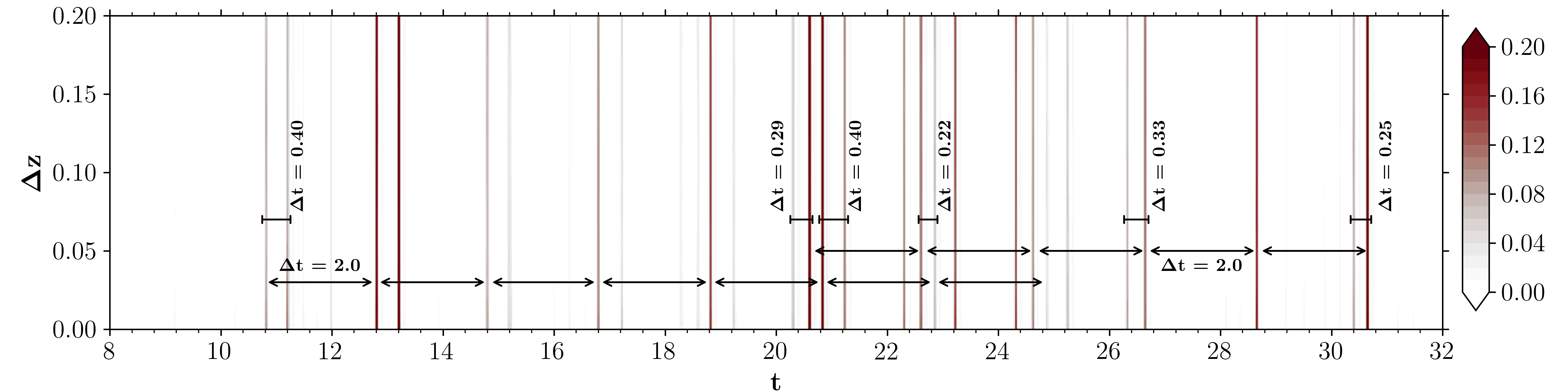}
		\caption{Two-point, \red{one-time pressure autocovariance} along the spanwise direction $(\Delta z)$ as a function of time ($t$) at the trailing edge. High correlation repeats about every 2 convective time units with intermittent events in between.}
		\label{fig:zoom_correlation}
	\end{figure}

	\red{The identification of time instants when coherent vortical structures pass by the trailing edge is performed using the two-point, one-time autocovariance of pressure fluctuations along the spanwise direction $\Delta z$ at each time $t$ according to
	\begin{equation}
	R(x,y,\Delta z,t) = \left\langle p'(x,y,z,t) \, p'(x,y,z + \Delta z,t) \right\rangle  \mbox{ .}
	\label{eq:correlation}
	\end{equation}
	Calculations are performed near the trailing edge at $x = 0.98$ and results are presented in figure \ref{fig:zoom_correlation}. The dark lines display higher spanwise coherence which are almost constant and independent of $\Delta z$. These corresponds to the two-dimensional vortex rolls illustrated in figure \ref{fig:vort_dynamics_3D_suction}(a). On the other hand, in moments where the flow transitions and only small scale eddies are observed at the trailing edge, such as shown in figure \ref{fig:vort_dynamics_3D_suction}(b), the spanwise coherence presents very low levels. Furthermore, in these time instants, the pressure fluctuations are considerably lower compared to the low pressure-core of the vortices, which are responsible for large pressure fluctuations. Hence, the products of small $p'$ in equation \ref{eq:correlation} results in even lower values of {autocovariance}.}

	As it can be seen from figure \ref{fig:zoom_correlation}, coherent structures often reach the trailing edge with a slow time-scale of $\Delta t \approx 2.0$ convective time units, which corresponds to a frequency $St = 0.5$. However, due to the chaotic motion, this process is not perfectly periodic and multiple vortices may reach the trailing edge within this period, as indicated by the time differences $\Delta t = 0.22$, 0.25, 0.29, 0.33 and 0.40 in the figure. In this faster time-scale, the different time intervals are related to frequencies $St =$ 4.5, 4.0, 3.5, 3.0 and 2.5, respectively, which are integer multiples of the low frequency tone at $St = 0.5$.
	
\begin{figure}
		\centering
		{\includegraphics[trim=0mm 1mm 0mm 10mm,clip,width=0.32\textwidth]{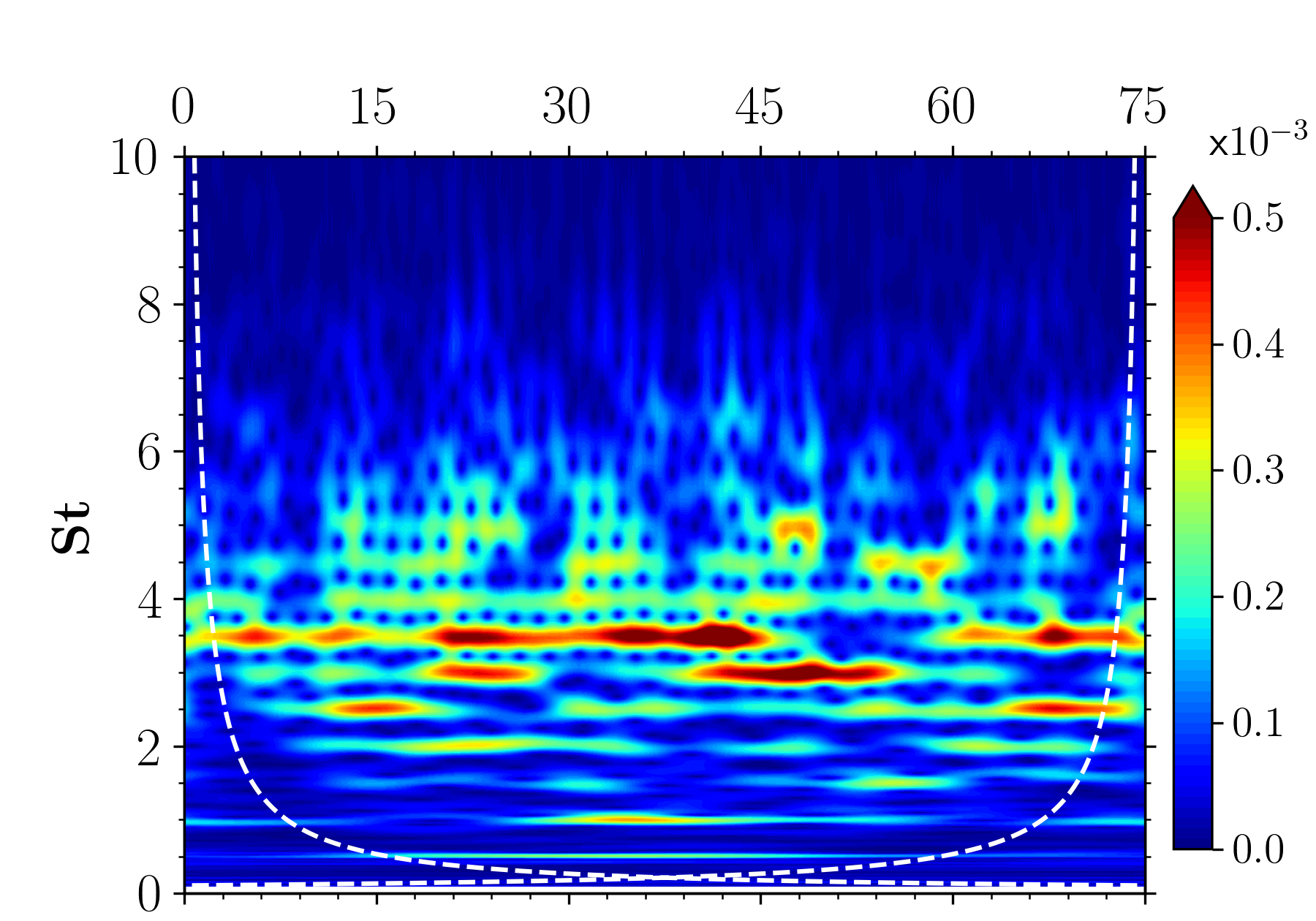}}
		{\includegraphics[trim=0mm 1mm 0mm 10mm,clip,width=0.32\textwidth]{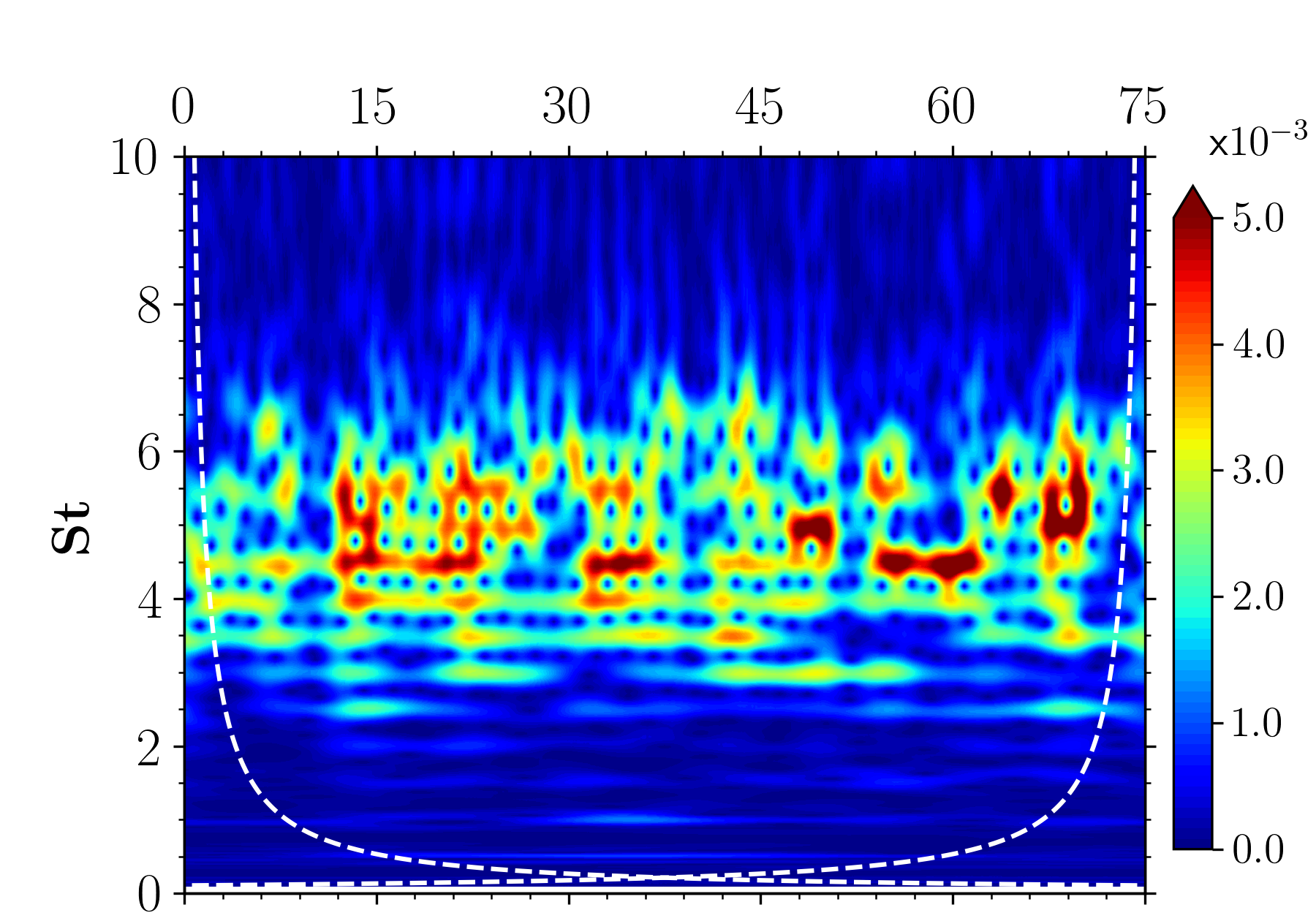}}
		{\includegraphics[trim=0mm 1mm 0mm 10mm,clip,width=0.32\textwidth]{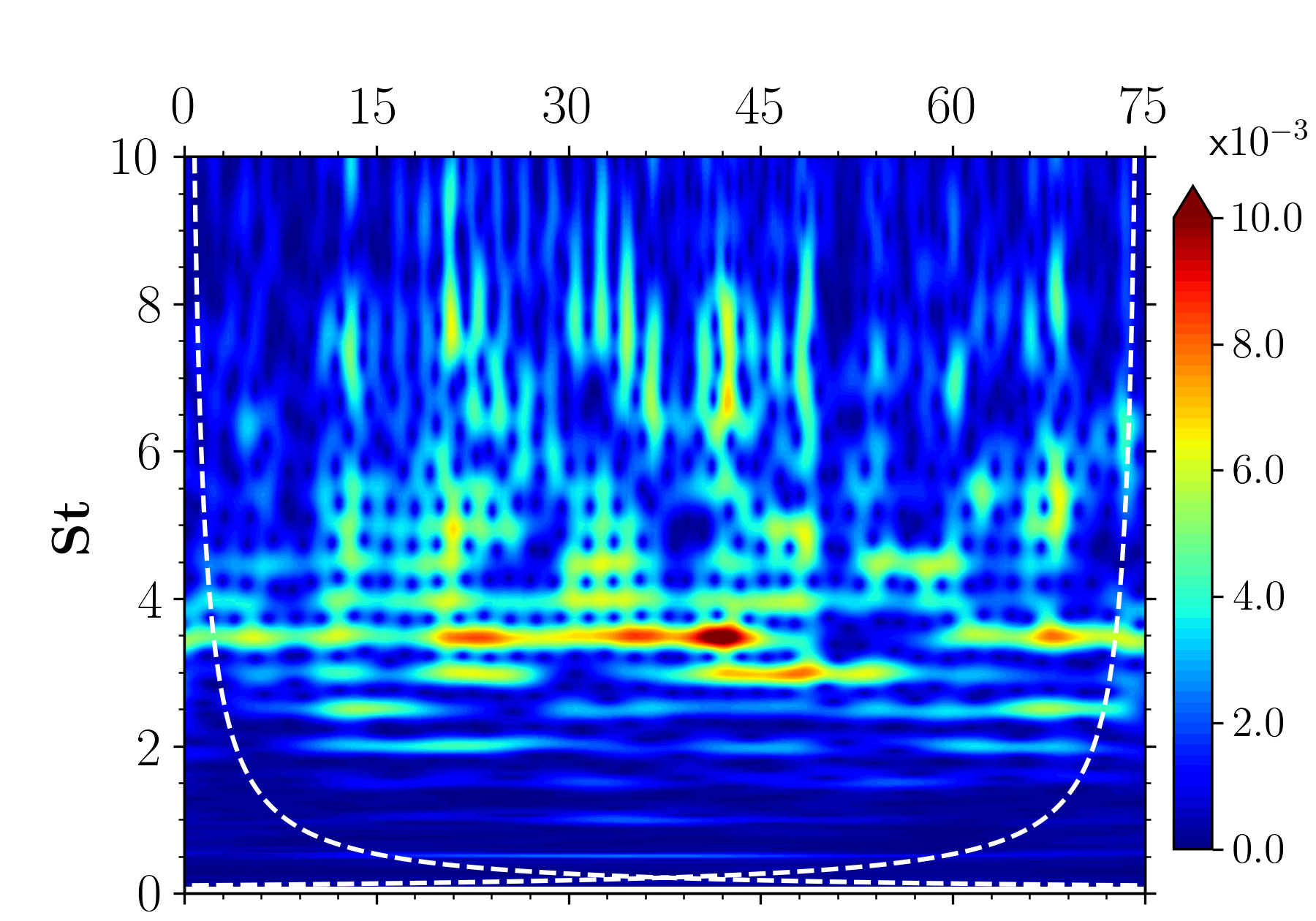}}
		\subfigure[$x = 0.34$]{\includegraphics[trim=0mm 0mm 0mm 2mm,clip,width=0.32\textwidth]{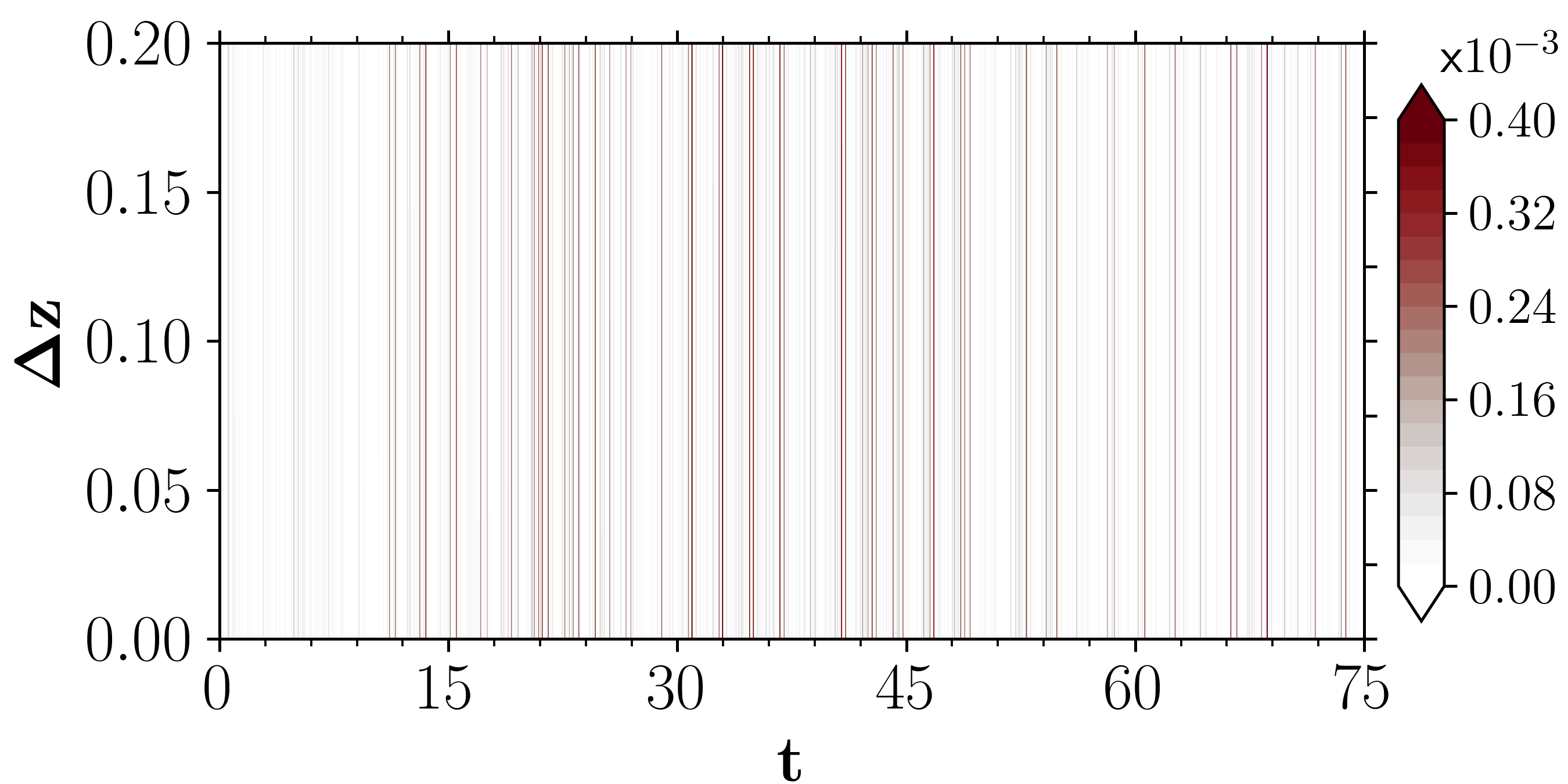}}
		\subfigure[$x = 0.64$]{\includegraphics[trim=0mm 0mm 0mm 2mm,clip,width=0.32\textwidth]{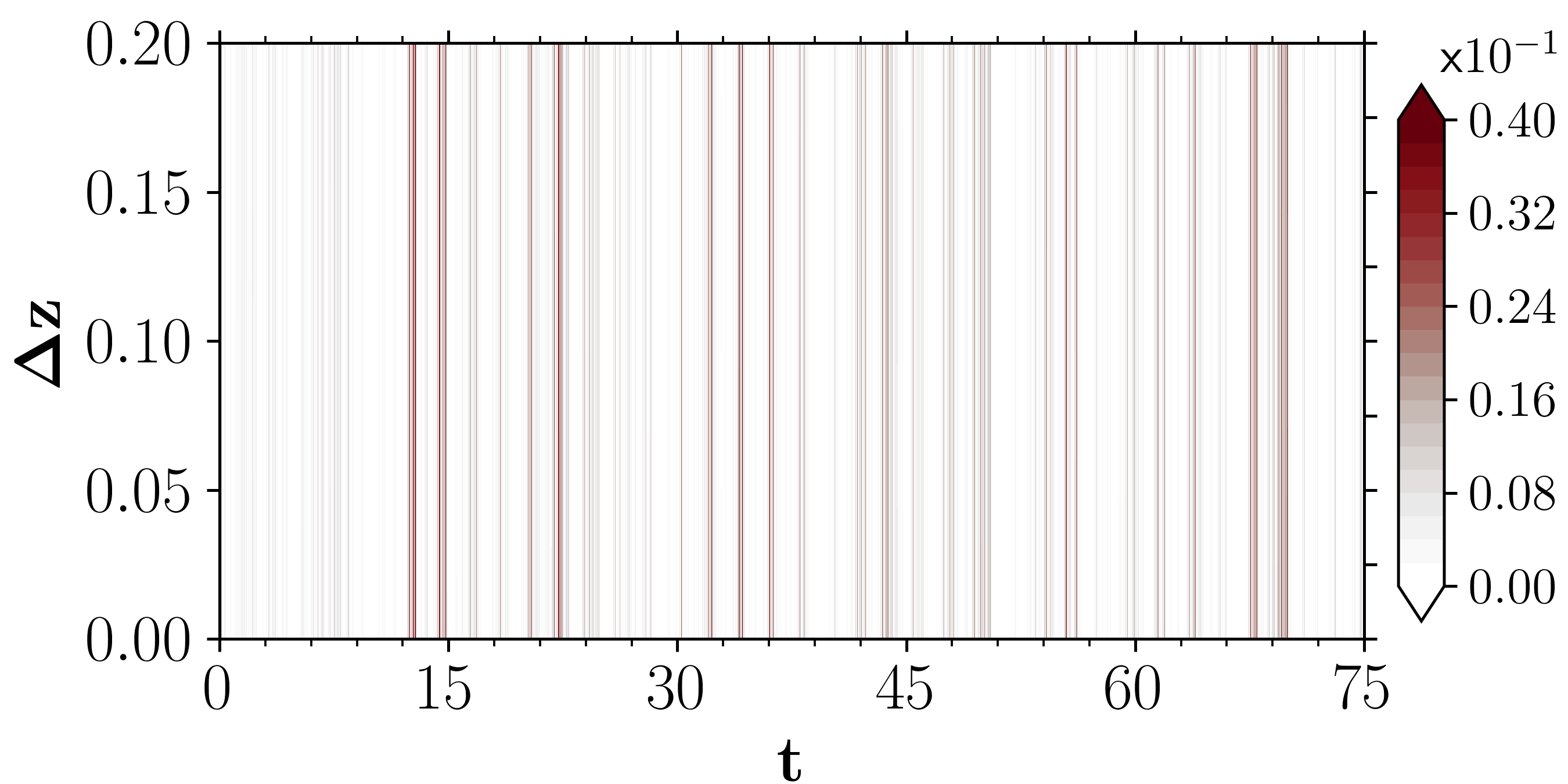}}
		\subfigure[$x = 0.98$]{\includegraphics[trim=0mm 0mm 0mm 2mm,clip,width=0.32\textwidth]{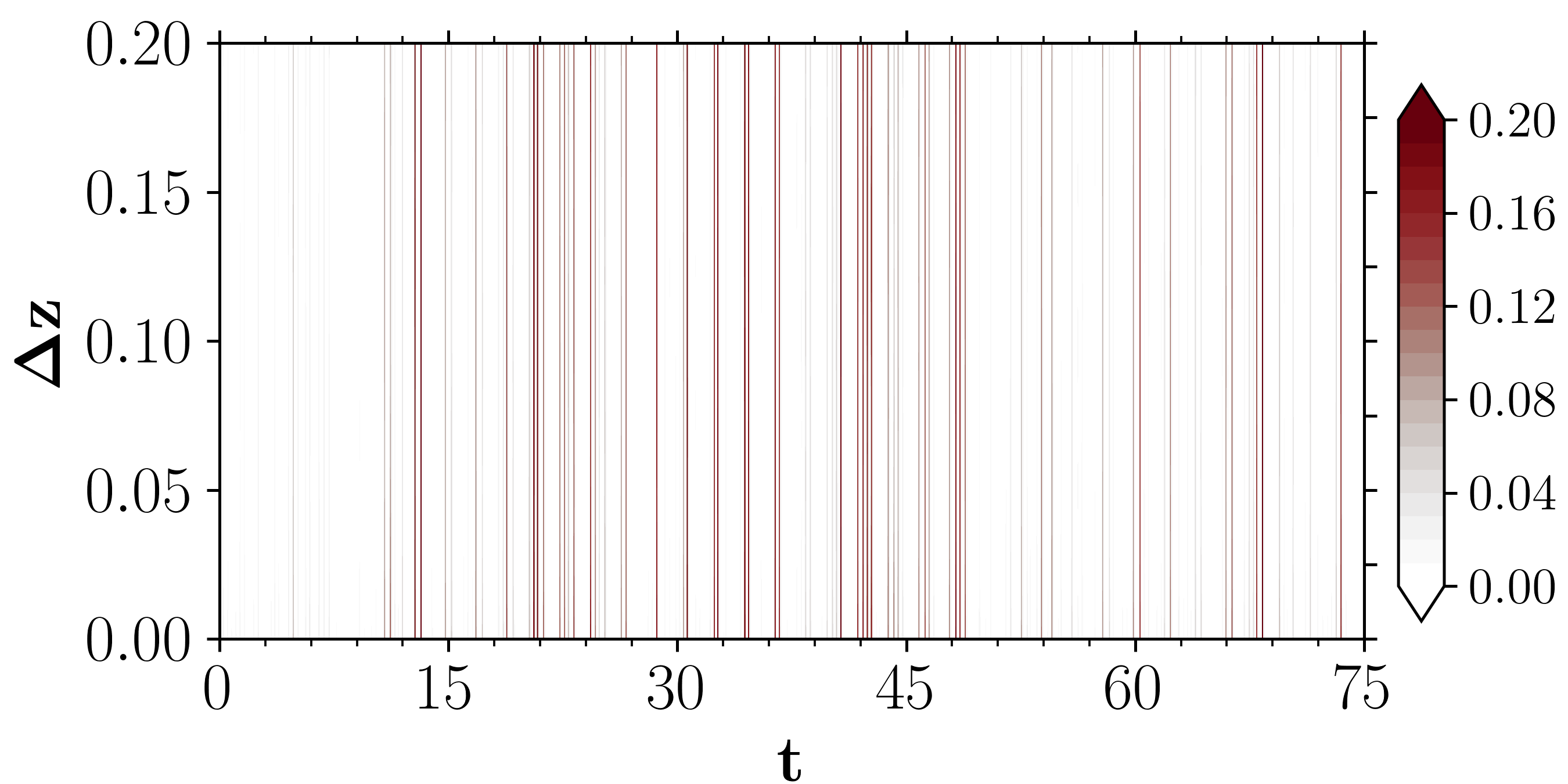}}
		\caption{Time-frequency analysis of pressure fluctuations using a continuous wavelet transform (top) and spanwise \red{two-point, one-time autocovariance} (bottom) at (a) the detachment point $x = 0.34$, (b) shear-layer roll-up $x = 0.64$ and (c) trailing edge $x = 0.98$.}
		\label{fig:wavelet_probes}
	\end{figure}

	To better highlight this behavior, a time-frequency analysis is also performed and results are presented in figure \ref{fig:wavelet_probes}. The top row shows the continuous wavelet transform, CWT, using the complex Morlet wavelet \citep{farge1992wavelet}. This function is the product of a monochromatic complex exponential with a Gaussian envelope. Its spectral response is an exponentially decaying \green{band-pass} filter centered around the frequency of the complex exponential. The filter width depends on the Gaussian standard deviation. The scalogram presents how the correlation of the wavelet with a reference signal changes in time based on the wavelet frequency. Hence, higher values indicate strong fluctuations of the signal at a particular frequency and time. Results are obtained for the same probe locations considered in figure \ref{fig:fft_probes}(b). The scalograms from figures \ref{fig:wavelet_probes}(a) and (c) exhibit longer periods of intense pressure fluctuations at $St = 3.5$, while in figure \ref{fig:wavelet_probes}(b), a broader range of frequencies is excited more intensely. As it can be seen from the probe at $x=0.64$, frequencies higher than $St = 3.5$ are excited, particularly $St = 4.5$, $5.0$ and $5.5$. One should note that the color levels are different for each location, with lower and higher amplitudes for probes at $x=0.34$ and $0.98$, respectively. These figures show that the secondary tones present intermittent behavior, with each frequency experiencing periods of silence (blue contours) and loud noise emission (green, yellow and red contours).
	\blue{Another important aspect captured by the scalograms is the amplitude modulation of the signals. Such behavior has an important role in the generation of secondary tones, as discussed by \citet{Desquesnes2007,Probsting2014} and \citet{Ricciardi2019_tones}.}
	
	\red{The bottom row shows the two-point, one-time pressure autocovariance along the span, computed for each time step over the entire simulation period.}	It is possible to see that the dark lines in the \red{covariance} plots are associated to the excitation of the secondary tones in the wavelet scalogram, indicating that specific frequencies are related to advection of quasi-2D flow structures. The coherence level and the time interval between successive vortices impacts not only the instantaneous magnitude of the wavelet transform but also may change its dominant frequency.
	
	\subsection{Linear analysis and amplification mechanisms}
	\label{ssec:linear}
	
\begin{figure}
		\centering
		\subfigure[$\hat{u}_n(St,x)$]{\includegraphics[width=0.300\textwidth]{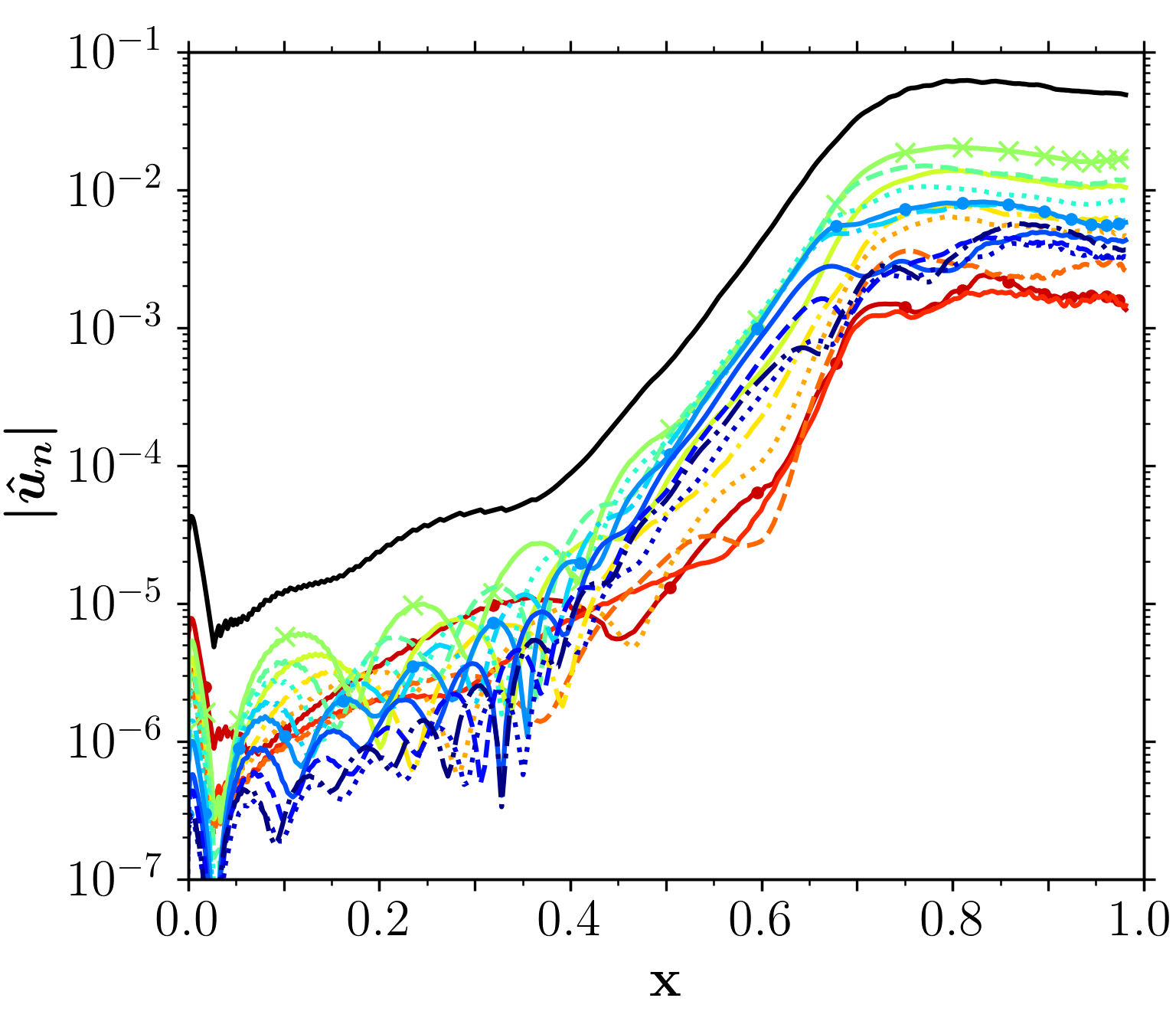}}
		\subfigure[$\hat{u}_t(St,x)$]{\includegraphics[width=0.300\textwidth]{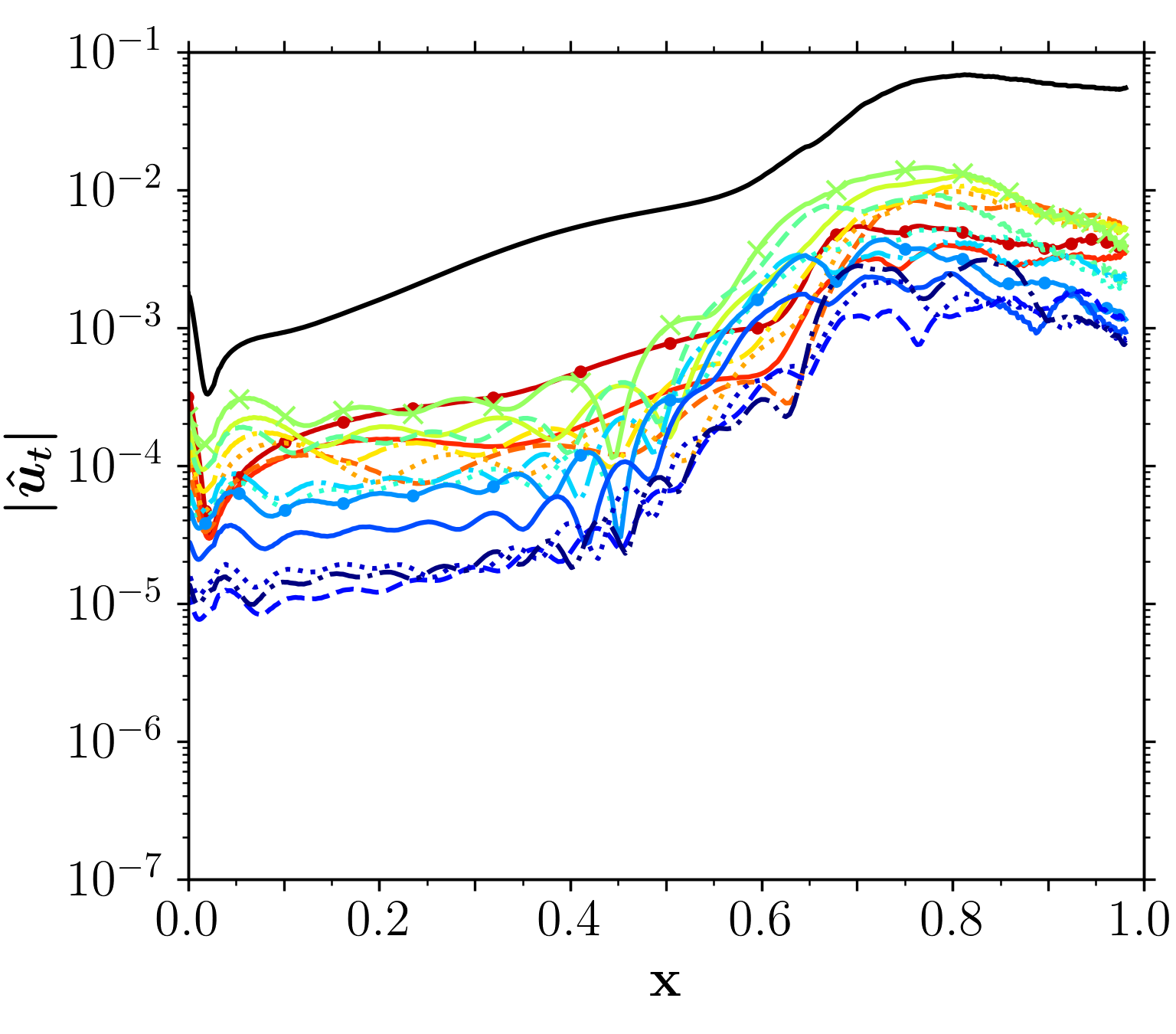}}
		\subfigure[$\hat{p}(St,x)$]{\includegraphics[width=0.375\textwidth]{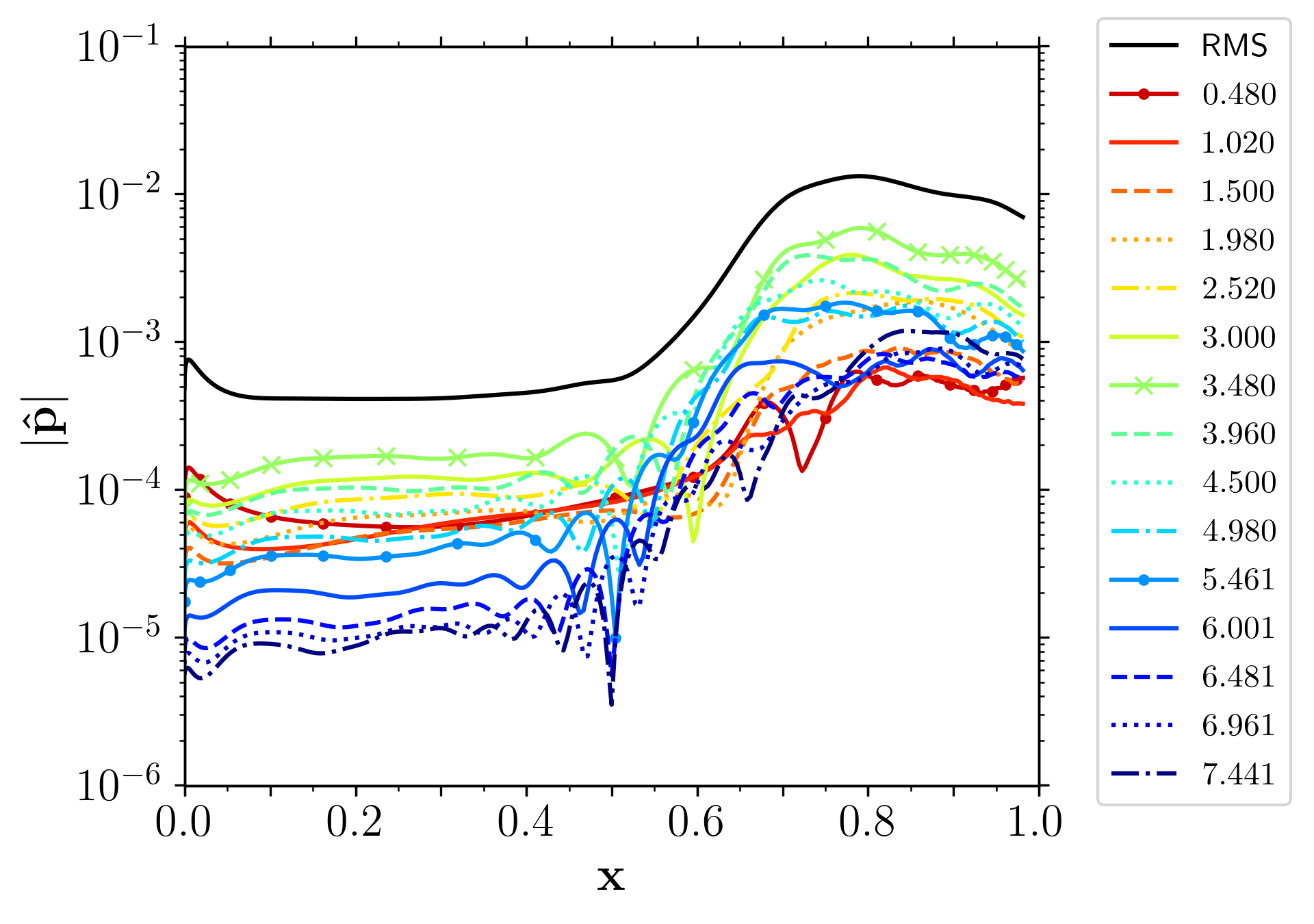}}
		\caption{Magnitude of Fourier transform coefficients for (a) normal velocity, (b) tangential velocity and (c) pressure computed along the green (velocities) and blue (pressure) lines from figures \ref{fig:k_rms}(a) and (b), respectively. \green{Each color represents a tonal frequency, identified in the colorbar, while the RMS is presented as the black line.}}
		\label{fig:FFT_along_lines}
	\end{figure}

	In order to investigate the growth of velocity and pressure fluctuations at the tonal frequencies, figure \ref{fig:FFT_along_lines} presents Fourier transforms of the temporal signals extracted along the airfoil suction side, following the green (velocities) and blue (pressure) dashed lines from figures \ref{fig:k_rms}(a) and (b), respectively. In this analysis, the temporal signals are computed for the spanwise averaged normal and tangential velocity components $u_n$ and $u_t$, respectively, besides pressure $p$. Results are plotted together with the RMS values (solid black line) extracted from figure \ref{fig:k_rms}. Both $\hat{u}_t$ and $\hat{p}$ show strong amplification downstream of $x = 0.5$, which coincides with the location of the recirculation bubble. Furthermore, a plateau is observed upstream of $x = 0.5$ for pressure fluctuations, which indicates that acoustic waves dominate over hydrodynamic disturbances along this region. Pressure fluctuations at $St \approx 3.5$ exhibit higher amplitudes along the entire chord and similar observations can be made for the velocity components. As discussed by \cite{yarusevich2019_merging}, it is possible that these fluctuations act as a sub-harmonic forcing that stimulates the vortex merging process. 
	
	The mechanisms of generation and amplification of flow instabilities is explained by linear stability theory. In this regard, bi-global linear stability and resolvent analyses are employed to understand the most sensitive frequencies in terms of optimal forcing and response characteristics for 2D perturbations, i.e., only for spanwise wavenumber $\beta = 0$. This is justified since 2D structures are expected to radiate noise more efficiently \citep{sanoprf2019}. In the linearized NS operator, the off-diagonal terms, related to velocity gradients and shear, lead to a non-normal operator. Hence, orthogonality of the stability modes is not expected and interaction among modes within the system may lead to transient response that affect its short-term dynamics \citep{Schmid2007annrev}. This behavior is properly captured by the resolvent analysis, which provides the information on eigenvalue sensitivity and transient energy amplification rate.
	
	\begin{figure}
		\centering
		\includegraphics[width=0.99\textwidth]{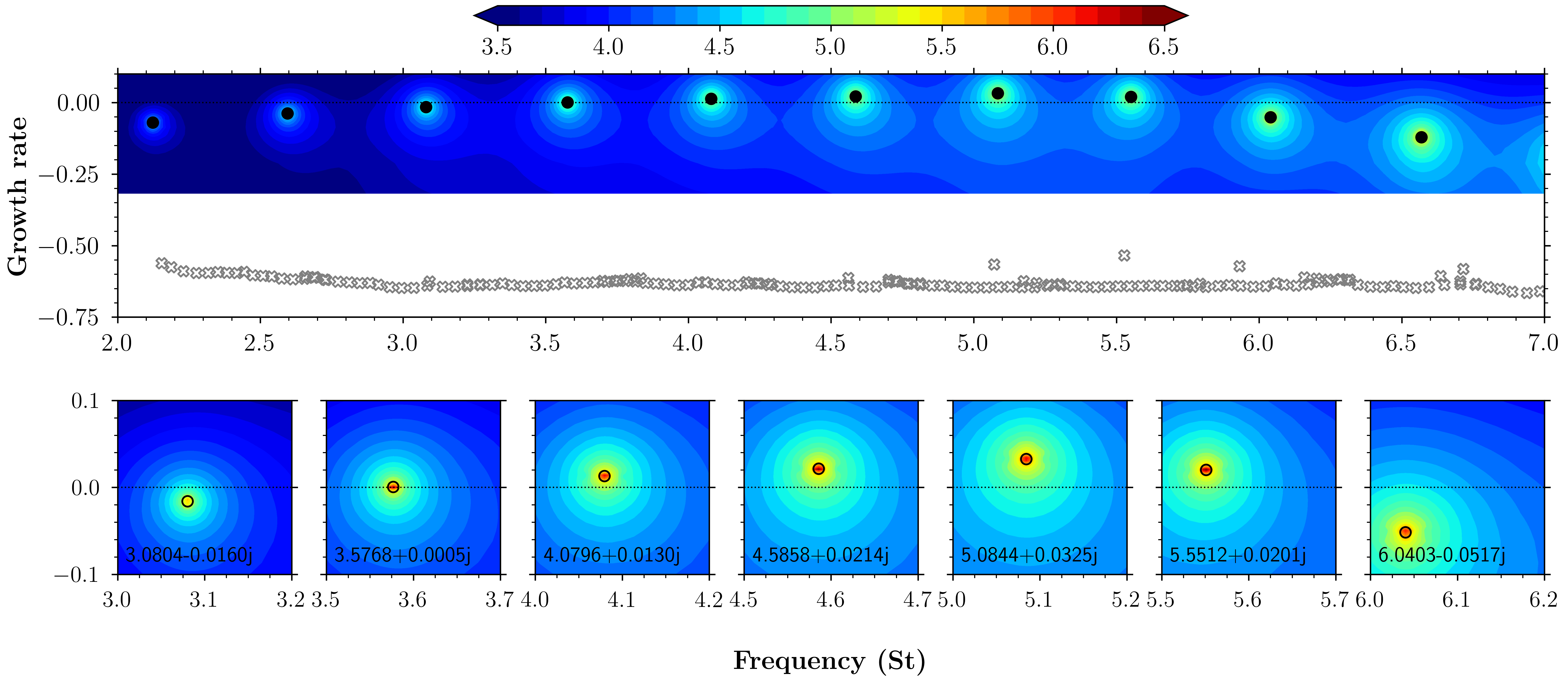}
		\caption{\green{Spectrum (symbols) and pseudospectrum (contours in logarithmic scale) of the linearized compressible Navier-Stokes equations for 2D spanwise perturbations $(\beta = 0)$. Higher frequencies present larger sensititivity as shown in light blue and green colors.}}
		\label{fig:pseudospectrum}
	\end{figure}
	
	Following the theoretical and numerical methodology described in section \ref{sec:LinearStabAnalysis}, the eigenvalue spectrum from a bi-global stability analysis is presented in figure \ref{fig:pseudospectrum}. The black circles represent the physically meaningful eigenvalues while the empty grey symbols are spurious eigenvalues. The latter could be identified due to their large displacement in the complex plane during the grid/sponge convergence study (not shown) as well as by visualization of the associated eigenvectors (modes).
	Sensitivity of the operator to external disturbances is computed by the resolvent analysis along both real (frequency) and imaginary (growth rate) axes and it is measured by the pseudospectrum, presented as the contour plot in figure \ref{fig:pseudospectrum}. The region of analysis in the resolvent operator is limited by means of matrices $\mathsfbi{B}$ and $\mathsfbi{C}$ in equation \ref{eq:final_resolvent}. Only the near-field grid points are considered, i.e., those inside the dashed red rectangle in figure \ref{fig:grid}. This is physically justified since the most intense quadrupole noise sources which are responsible for the incident field in the acoustic scattering process are close to the trailing edge \citep{Wolf2012}. As an additional positive side effect, this improved results by reducing spatial support of spurious eigenvectors.
	
	\blue{Results from the bi-global stability analysis in terms of the eigenspectrum are presented in figure \ref{fig:pseudospectrum}. It is possible to see multiple eigenvalues at frequencies close to those from the LES, previously discussed in figures \ref{fig:fft_probes}(b,d). For instance, the maximum relative deviation is less than 5\%, at $St \approx 2.0$, and around 2\% for the most unstable frequency $St \approx 5.0$. Thus, the vortex dynamics and the multiple tones seem to be triggered by linear mechanisms.
	The pseudospectrum, obtained from the resolvent analysis, is shown as a contour plot in logarithmic scale of the leading singular value from the SVD according to equation \ref{eq:svd_resolvent}.}
	The input-output amplification ratio depends on both resonance and pseudoresonance (see equation \ref{eq:ressonance}) such that peaks appear when approaching an eigenvalue. The pseudoresonance levels depend on the mutual excitation of eigenvectors with more sensitive eigenvalues presenting higher peaks.
	In this sense, it is possible to see in figure \ref{fig:pseudospectrum} that higher frequencies are more sensitive compared to the lower ones. Moreover, the stable eigenvalues may impact the flow dynamics in terms of transient energy amplification given their sensitivities to disturbances.
	
	\blue{The most unstable frequencies of $St \approx 4.5$, $5.0$ and $5.5$ are observed in the LES on the shear-layer roll-up region at $x=0.64$, as illustrated by figure \ref{fig:wavelet_probes}(b). The interaction among multiple frequencies of the flow instabilities and the vortex merging reduce the shedding frequency in the LES, where the dominant tonal peak becomes $St \approx 3.5$. This non-linear phenomenon is captured by the LES but not by the linear analysis. A more detailed discussion regarding the necessary conditions for vortex merging is presented in section \ref{ssec:merging}.}

\begin{figure}
		\centering
		\begin{tikzpicture}
		\node[anchor=south west,inner sep=0] (image) at (0,0) {\includegraphics[width=0.99\textwidth]{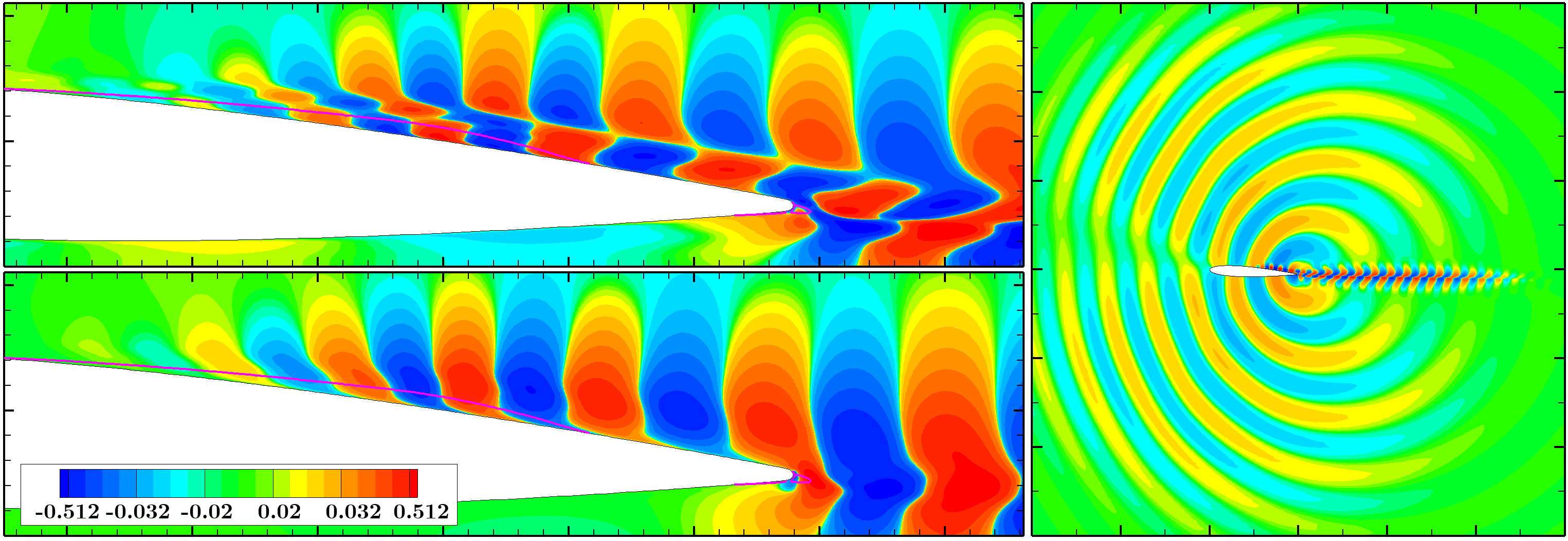}};
		\begin{scope}[x={(image.south east)},y={(image.north west)}]
		\node[] at (0.030,0.93) {(a)};
		\node[] at (0.030,0.43) {(b)};
		\node[] at (0.685,0.93) {(c)};		
		\node[] at (0.043,-0.03) {\scriptsize $0.4$};
		\node[] at (0.123,-0.03) {\scriptsize $0.5$};
		\node[] at (0.203,-0.03) {\scriptsize $0.6$};
		\node[] at (0.283,-0.03) {\scriptsize $0.7$};						
		\node[] at (0.363,-0.03) {\scriptsize $0.8$};
		\node[] at (0.443,-0.03) {\scriptsize $0.9$};
		\node[] at (0.523,-0.03) {\scriptsize $1.0$};
		\node[] at (0.603,-0.03) {\scriptsize $1.1$};	
		\node[] at (0.715,-0.03) {\scriptsize $-1.0$};
		\node[] at (0.772,-0.03) {\scriptsize $0.0$};
		\node[] at (0.828,-0.03) {\scriptsize $1.0$};
		\node[] at (0.885,-0.03) {\scriptsize $2.0$};	
		\node[] at (0.943,-0.03) {\scriptsize $3.0$};			
		\end{scope}
		\end{tikzpicture}
		\caption{Global stability modes of the most unstable frequency $St = 5.08$ for primitive variables (a) $u'$, (b) $v'$ and (c) $p'$.}
		\label{fig:global_modes}
	\end{figure}
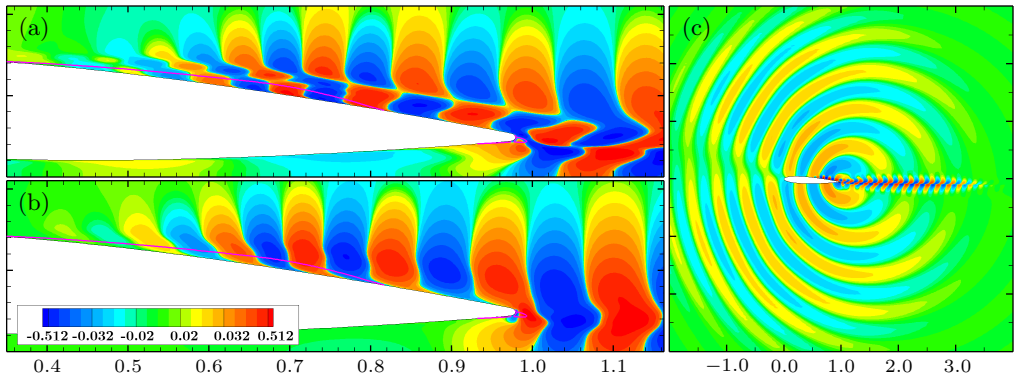

	The modal shape of each frequency is crucial to understand the system dynamics and to identify amplification regions within the flow. In this sense, figures \ref{fig:global_modes}(a), (b) and (c) present the real part of the global stability modes for the most unstable frequency $St = 5.08$ in terms of $u'$, $v'$ and $p'$, respectively. The modes are normalized with respect to the maximum value for each variable and the magnitude doubles every level to better represent lower values.
	The modal shapes computed for other relevant frequencies in the spectrum (resonances) show similar patterns compared to the one described here and differences are only related to their characteristic wavelengths.		
	Along the separation bubble, the modal structures of $u'$ and $v'$ are tilted against the mean shear, suggesting an Orr mechanism for transient growth of energy and amplification of disturbances \citep{schmidYellowBook}. After flow reattachment, the structures become aligned with the wall-normal direction.
	The pressure disturbances $p'$ display upstream traveling acoustic waves radiated from the trailing edge due to scattering mechanism.	
	The modes displayed in the figure originate as shear layer instabilities on the airfoil suction side and extends along the wake. Hence, wake-boundary layer coupling mechanisms cannot be discarded.	
	
	\tikzstyle{stuff_fill}=[rectangle,draw,fill=white]
	
\begin{figure}
		\centering
		\begin{tikzpicture}
		\node[anchor=south west,inner sep=0] (image) at (0,0) {\includegraphics[width=0.99\textwidth]{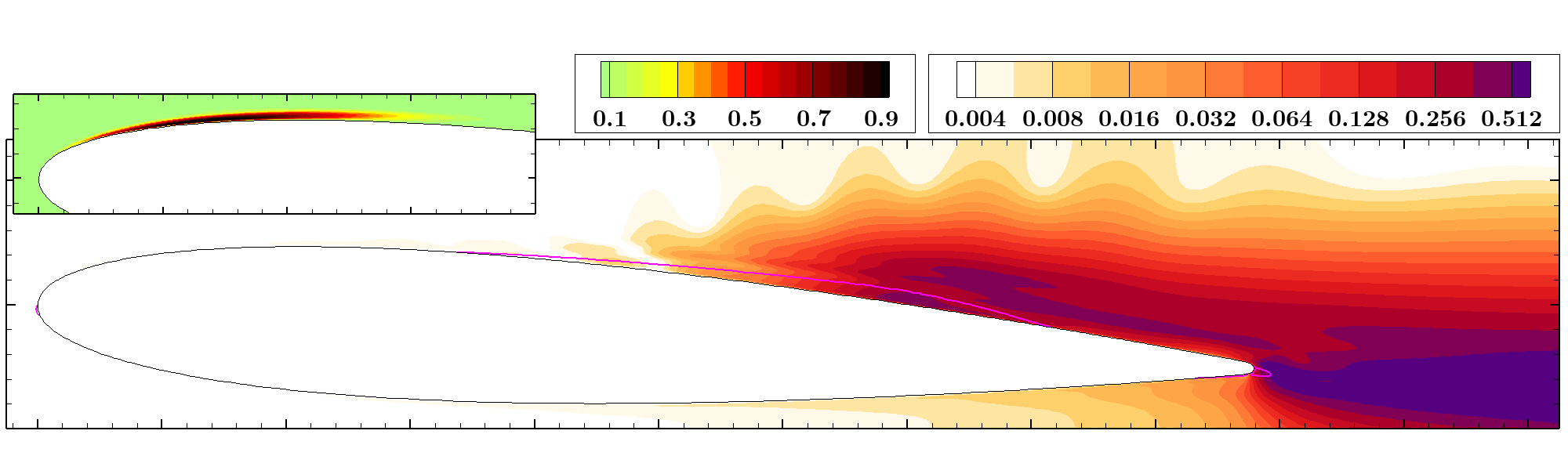}};
		\begin{scope}[x={(image.south east)},y={(image.north west)}]
		\node[] at (0.475,0.92) {\scriptsize \bf Forcing};
		\node[] at (0.793,0.92) {\scriptsize \bf Response};
		\node[] at (0.025,0.83) {\scriptsize $0.0$};
		\node[] at (0.104,0.83) {\scriptsize $0.1$};
		\node[] at (0.183,0.83) {\scriptsize $0.2$};
		\node[] at (0.262,0.83) {\scriptsize $0.3$};
		\node[] at (0.341,0.83) {\scriptsize $0.4$};
		\node[] at (0.024,0.02) {\scriptsize $0.0$};
		\node[] at (0.104,0.02) {\scriptsize $0.1$};
		\node[] at (0.183,0.02) {\scriptsize $0.2$};
		\node[] at (0.262,0.02) {\scriptsize $0.3$};						
		\node[] at (0.341,0.02) {\scriptsize $0.4$};
		\node[] at (0.420,0.02) {\scriptsize $0.5$};
		\node[] at (0.500,0.02) {\scriptsize $0.6$};
		\node[] at (0.579,0.02) {\scriptsize $0.7$};
		\node[] at (0.658,0.02) {\scriptsize $0.8$};
		\node[] at (0.737,0.02) {\scriptsize $0.9$};
		\node[] at (0.816,0.02) {\scriptsize $1.0$};
		\node[] at (0.895,0.02) {\scriptsize $1.1$};
		\node[] at (0.974,0.02) {\scriptsize $1.2$};						
		\end{scope}
		\end{tikzpicture}
		\caption{Response (white-purple contours) and forcing (upper-left plot with \green{white-blue contours}) modes of kinetic energy $k$ from the resolvent analysis along the neutral stability axis for $St = 5.08$.}
		\label{fig:resolvent_modes_abs}
	\end{figure}
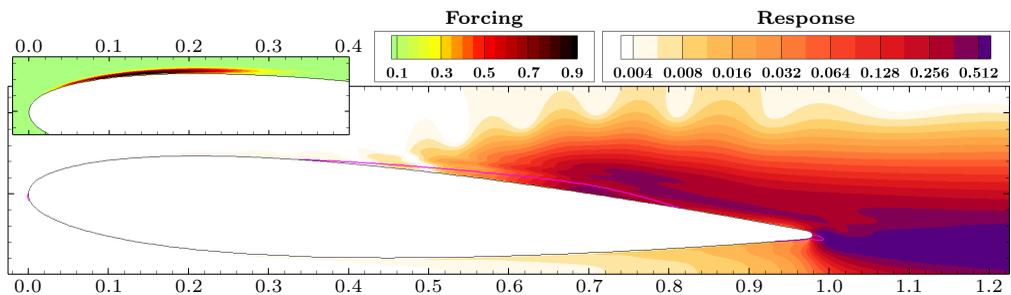

	The dominant resolvent modes are computed along the neutral stability axis $(\omega_i = 0.0)$ for $St = 5.08$ in terms of kinetic energy $k$ and presented in figure \ref{fig:resolvent_modes_abs} normalized with respect to their maximum value. The response mode, shown in white-purple contours, depicts the location where disturbances are more energetically relevant. Here, its modal shape magnitude is presented such that it doubles every 2 levels. It is possible to see negligible levels from the leading-edge until the recirculation bubble. Then, for $x > 0.4$, its magnitude increases quickly, indicating that the suction side bubble is an amplifier of disturbances.
	The forcing mode, \green{upper-left plot with white-blue contours}, highlights the most receptive location, where minimal disturbances result in larger transient growth by the system. This mode is related to the adjoint operator of the linearized NS equations and it is introduced in the context of airfoil secondary tones by \cite{FosasdePando2017}. In the present results, the leading edge region on the suction side shows the higher values of forcing modes. Furthermore, these modes are absent in the pressure side which corroborates to the absence of pressure side-driven events for this flow configuration.
	
	\tikzstyle{stuff_fill}=[rectangle,draw,fill=white]
	
	
	\subsection{Acoustic feedback loop mechanism}
	\label{ssec:feedback}
	
	As discussed by several authors \citep{Tam1974, Lowson1994,  Desquesnes2007, Jones2011, FosasdePando2014}, airfoil flows at transitional Reynolds numbers are subject to an acoustic feedback loop mechanism that is related to the multiple tones. In this feedback process, hydrodynamic perturbations are excited in the vicinity of the leading edge and convected downstream. The disturbances amplify along the recirculation bubble on the airfoil surface and, upon reaching the trailing edge, scatter acoustic waves that travel upstream and trigger new disturbances.
	%
%
	In this context, figures \ref{fig:resolvent_modes_real}(a--h) present results from the resolvent analysis for different frequencies. The real part of the pressure forcing modes is shown in red-blue contours while black-white contours display the real component of the leading pressure response modes. The modes have been normalized with respect to their maximum values. To highlight the most receptive locations for each individual frequency, the yellow spots in the forcing modes are shown only for regions where the magnitude reaches higher values than 98\% of the maximum. As presented in figure \ref{fig:resolvent_modes_real}, the locations of maximum receptivity change with frequency and vary along the interval $0.10 < x < 0.18$.
	
	\tikzstyle{stuff_fill}=[rectangle,draw,fill=white]
	\begin{figure}
		\centering
		\begin{tikzpicture}
		\node[anchor=south west,inner sep=0] (image) at (0,0) {\includegraphics[width=0.99\textwidth]{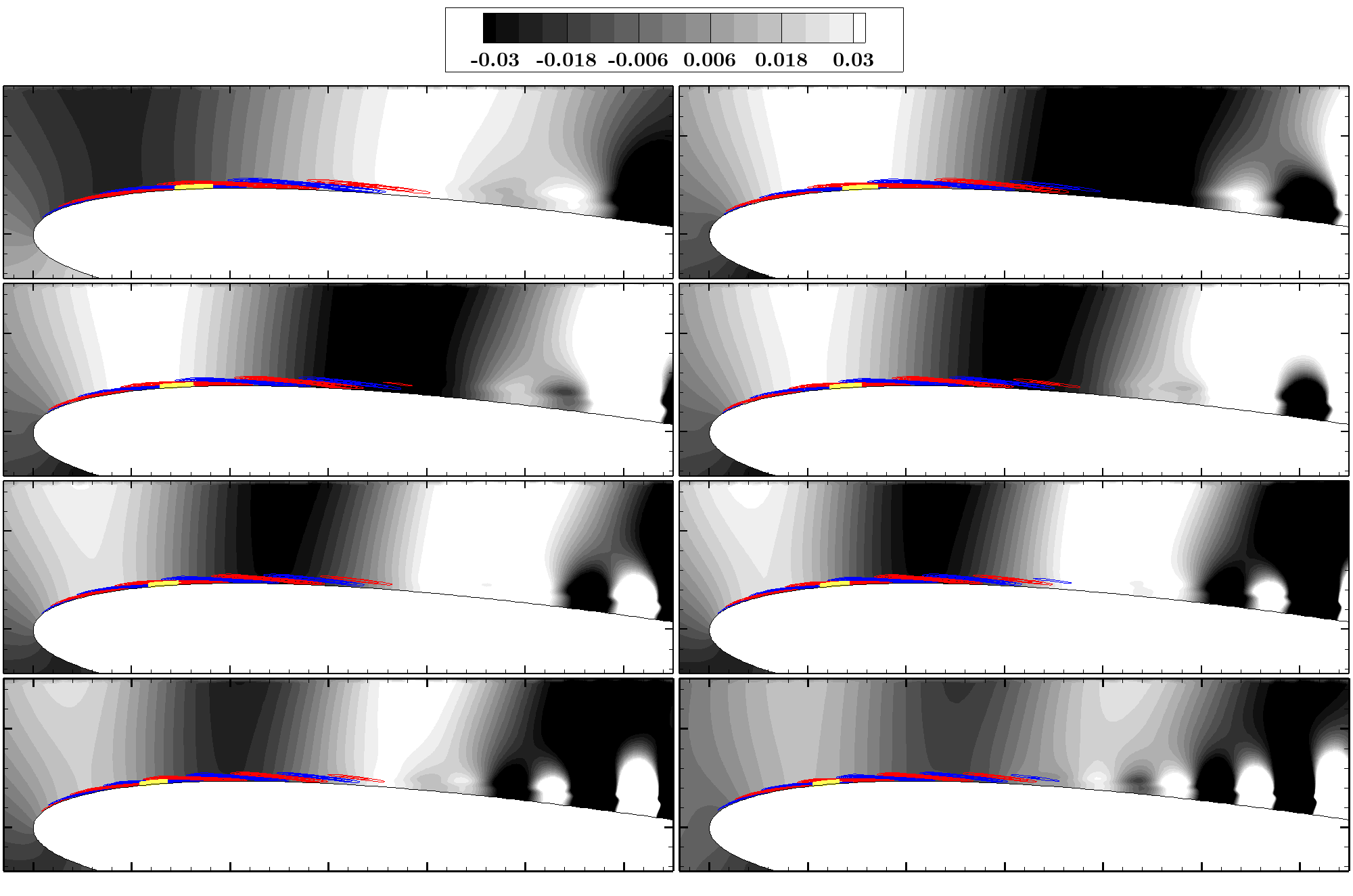}};
		\begin{scope}[x={(image.south east)},y={(image.north west)}]
		\node[] at (0.088,0.730) [stuff_fill] {(a) St = 3.08};
		\node[] at (0.088,0.510) [stuff_fill] {(c) St = 4.08};
		\node[] at (0.088,0.290) [stuff_fill] {(e) St = 5.08};
		\node[] at (0.088,0.070) [stuff_fill] {(g) St = 6.04};		
		\node[] at (0.588,0.730) [stuff_fill] {(b) St = 3.57};
		\node[] at (0.588,0.510) [stuff_fill] {(d) St = 4.58};
		\node[] at (0.588,0.290) [stuff_fill] {(f) St = 5.55};
		\node[] at (0.588,0.070) [stuff_fill] {(h) St = 6.56};
		\node[] at (0.025,0.01) {\scriptsize $0.0$};
		\node[] at (0.098,0.01) {\scriptsize $0.1$};
		\node[] at (0.170,0.01) {\scriptsize $0.2$};
		\node[] at (0.243,0.01) {\scriptsize $0.3$};
		\node[] at (0.316,0.01) {\scriptsize $0.4$};
		\node[] at (0.388,0.01) {\scriptsize $0.5$};
		\node[] at (0.461,0.01) {\scriptsize $0.6$};
		\node[] at (0.525,0.01) {\scriptsize $0.0$};
		\node[] at (0.598,0.01) {\scriptsize $0.1$};
		\node[] at (0.670,0.01) {\scriptsize $0.2$};
		\node[] at (0.744,0.01) {\scriptsize $0.3$};
		\node[] at (0.815,0.01) {\scriptsize $0.4$};
		\node[] at (0.888,0.01) {\scriptsize $0.5$};
		\node[] at (0.959,0.01) {\scriptsize $0.6$};
		\end{scope}
		\end{tikzpicture}
		\caption{Real component of pressure response (black-white contours) and forcing (blue-red contours) modes from resolvent analysis along the neutral stability axis. The yellow spots represent the most sensitive regions where the forcing modes achieve at least 98\% of the respective maximum values for each frequency.}
		\label{fig:resolvent_modes_real}
	\end{figure}
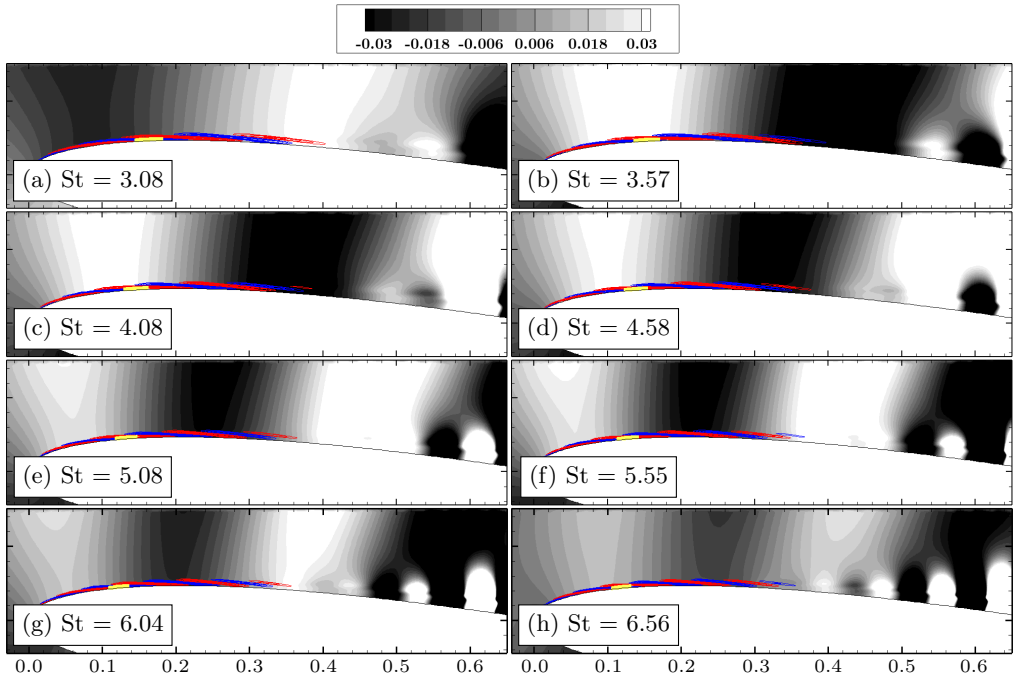

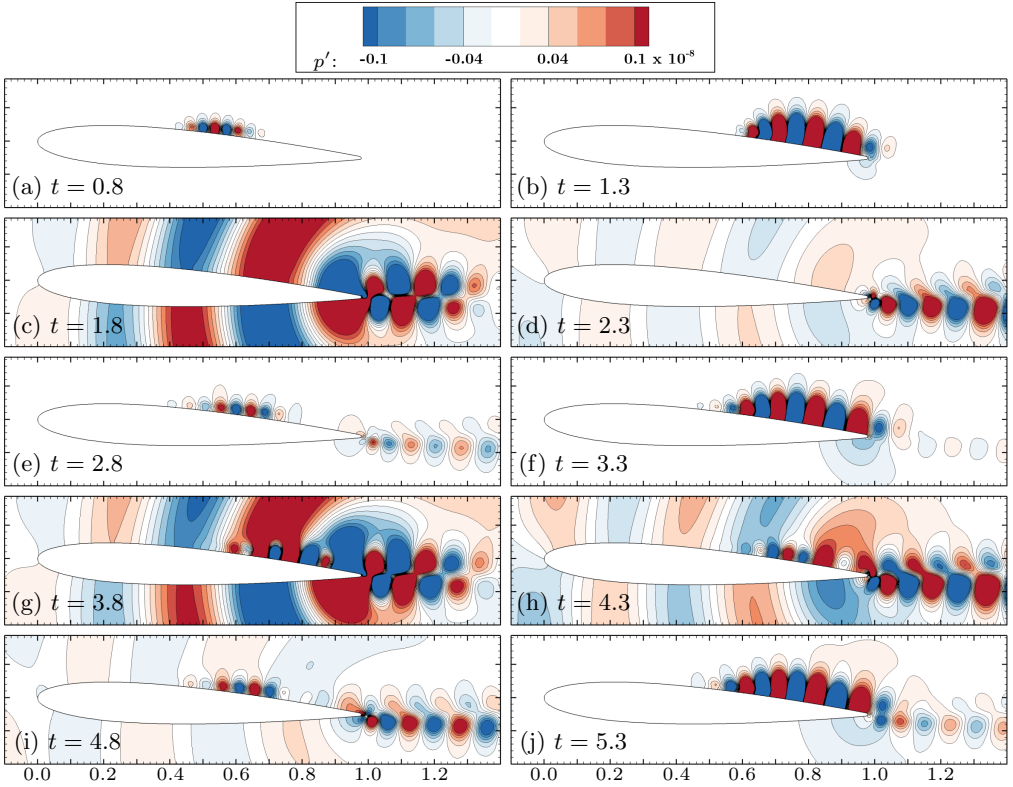
\begin{figure}
		\centering
		\begin{tikzpicture}
		\node[anchor=south west,inner sep=0] (image) at (0,0) {\includegraphics[width=0.99\textwidth]{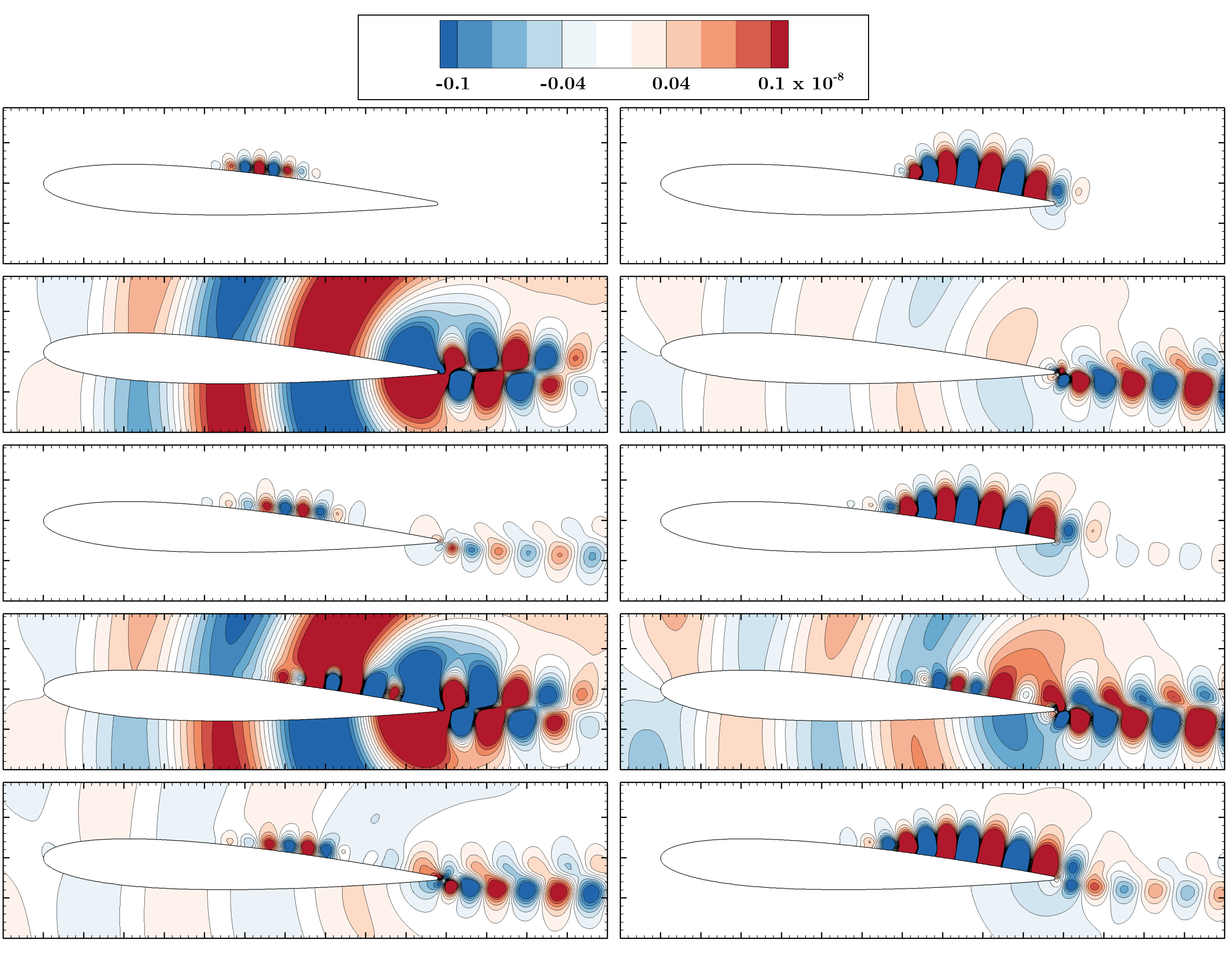}};
		\begin{scope}[x={(image.south east)},y={(image.north west)}]
		\node[] at (0.32,0.91) {\scriptsize $p'$:};
		\node[] at (0.065,0.75) {(a) $t = 0.8$};
		\node[] at (0.565,0.75) {(b) $t = 1.3$};
		\node[] at (0.065,0.575) {(c) $t = 1.8$};
		\node[] at (0.565,0.575) {(d) $t = 2.3$};
		\node[] at (0.065,0.400) {(e) $t = 2.8$};
		\node[] at (0.565,0.400) {(f) $t = 3.3$};
		\node[] at (0.065,0.225) {(g) $t = 3.8$};		
		\node[] at (0.565,0.225) {(h) $t = 4.3$};
		\node[] at (0.065,0.05) {(i) $t = 4.8$};		
		\node[] at (0.565,0.05) {(j) $t = 5.3$};				
		\node[] at (0.035,0.01) {\scriptsize $0.0$};
		\node[] at (0.100,0.01) {\scriptsize $0.2$};
		\node[] at (0.167,0.01) {\scriptsize $0.4$};
		\node[] at (0.232,0.01) {\scriptsize $0.6$};
		\node[] at (0.297,0.01) {\scriptsize $0.8$};
		\node[] at (0.362,0.01) {\scriptsize $1.0$};
		\node[] at (0.428,0.01) {\scriptsize $1.2$};
		\node[] at (0.536,0.01) {\scriptsize $0.0$};
		\node[] at (0.601,0.01) {\scriptsize $0.2$};
		\node[] at (0.667,0.01) {\scriptsize $0.4$};
		\node[] at (0.732,0.01) {\scriptsize $0.6$};
		\node[] at (0.797,0.01) {\scriptsize $0.8$};
		\node[] at (0.863,0.01) {\scriptsize $1.0$};
		\node[] at (0.928,0.01) {\scriptsize $1.2$};
		\end{scope}
		\end{tikzpicture}
		\caption{Snapshots of the linearized Navier-Stokes equations presented in terms of pressure fluctuations show periodic wavepackets composed by the multiple frequencies from the eigenspectrum of figure \ref{fig:pseudospectrum}.}
		\label{fig:snaps_MFP}
	\end{figure}

	Aspects such as the amplitude and phase between the acoustic waves and flow receptivity at different frequencies impact the onset of hydrodynamic disturbances. To understand such complex interrelation, the acoustic feedback is analyzed using the mean flow perturbation technique to solve the linearized NS equations around the 3D mean flow in time-domain \citep{Jones2011,FosasdePando2014}. \green{The method assumes that the base flow is steady and that it is not modified by the linear perturbations.}
	The dynamics of the pressure fluctuations is presented in figure \ref{fig:snaps_MFP}. For the initial condition, a pressure disturbance is applied at the leading edge, from $x = 0.0$ to $0.005$ over a short duration of $\Delta t = 1 \times 10^{-4}$ to excite all frequencies (impulse response). This triggers a wavepacket that is advected by the mean flow along the suction side. Figures \ref{fig:snaps_MFP}(a,b) present the growth of disturbances for $x > 0.4 $, where spatial amplification described by the response modes \green{becomes} important, as shown in figure \ref{fig:resolvent_modes_abs}. When the wavepacket reaches the trailing edge, acoustic waves are generated and travel upstream as shown in figure \ref{fig:snaps_MFP}(c). After the wavepacket reaches the wake, hydrodynamic disturbances are no longer observed on the suction side, as displayed by figure \ref{fig:snaps_MFP}(d). Afterwards, disturbances in the wake slowly decay while a new wavepacket amplifies on the suction side (figures \ref{fig:snaps_MFP}(e,f)). Similarly to the low frequency observed in the LES calculation, the wavepackets have a period of $\Delta t = 2.0$ and the next cycle is presented in figures \ref{fig:snaps_MFP}(g--j). The combined response from the suction side wavepacket, wake instability and upstream traveling acoustic waves keeps the fluctuation field active, amplifying the globally unstable frequencies in the eigenspectrum of figure \ref{fig:pseudospectrum}, which ultimately become dominant.
	Readers are referred to the movie provided as supplemental material (movie 4) presenting the evolution of the linearized flow dynamics.
	
	\tikzstyle{stuff_fill}=[rectangle,draw,fill=white]
	
\begin{figure}
	\centering
	\begin{tikzpicture}
	\node[anchor=south west,inner sep=0] (image) at (0,0) {\includegraphics[width=0.75\textwidth]{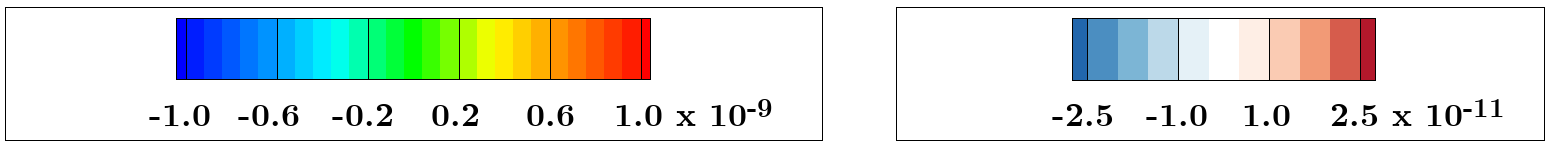}};
	\begin{scope}[x={(image.south east)},y={(image.north west)}]
	\node[] at (0.050,0.21) {\scriptsize $p'$:};
	\node[] at (0.626,0.21) {\scriptsize $(\sfrac{p}{\rho^{\gamma}})'$:};		
	\end{scope}1
	\end{tikzpicture}
	\begin{tikzpicture}
	\node[anchor=south west,inner sep=0] (image) at (0,0) {\includegraphics[width=0.99\textwidth]{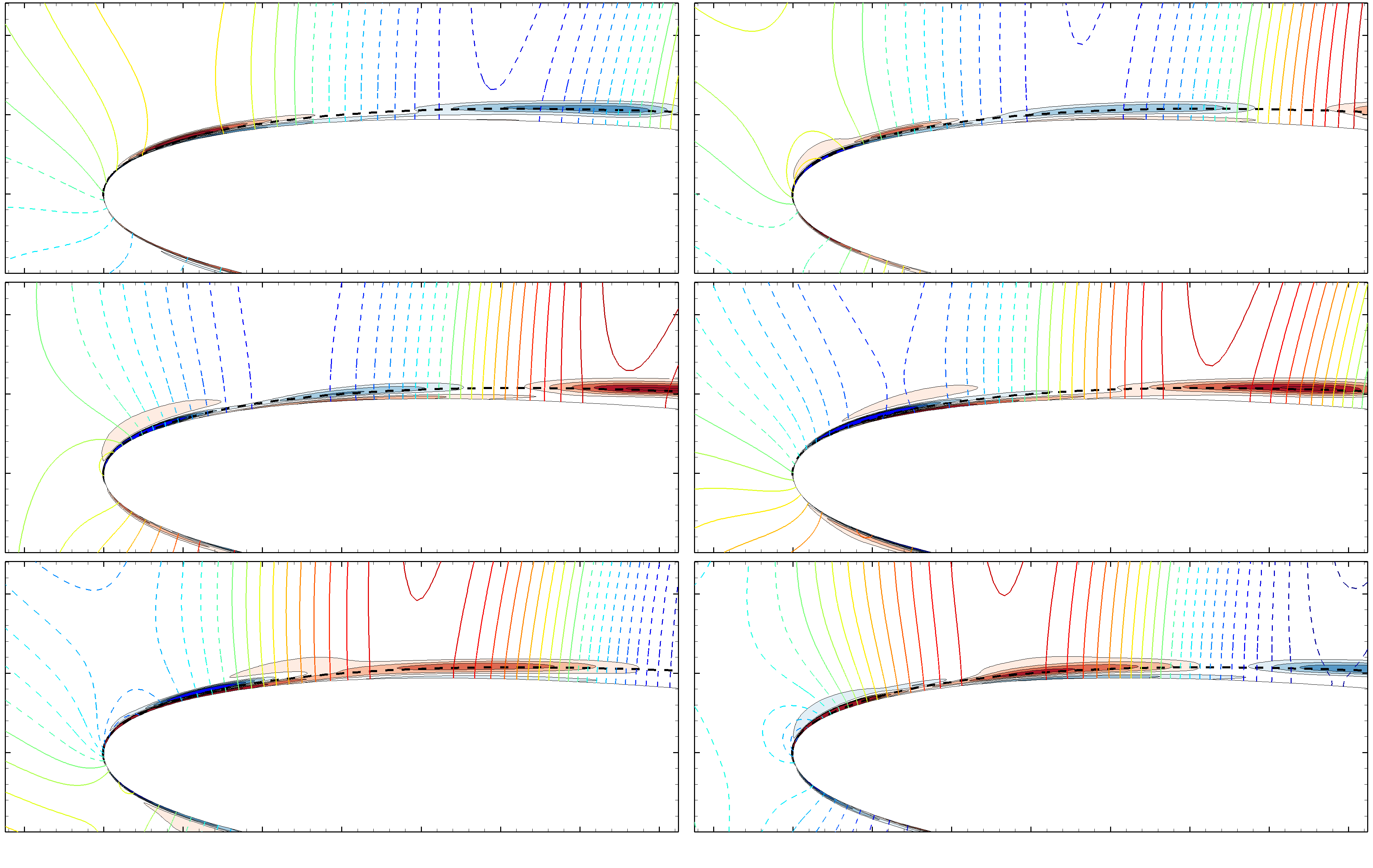}};
	\begin{scope}[x={(image.south east)},y={(image.north west)}]
	\draw[thick,dotted] (0.56,0.870)--(0.63,0.870)--(0.63,0.770)--(0.56,0.770)--(0.56,0.870);
	\draw[thick,dotted] (0.06,0.550)--(0.18,0.550)--(0.18,0.450)--(0.06,0.450)--(0.06,0.550);
	\draw[thick,dotted] (0.60,0.580)--(0.72,0.580)--(0.72,0.500)--(0.60,0.500)--(0.60,0.580);
	\draw[thick,dotted] (0.16,0.260)--(0.28,0.260)--(0.28,0.190)--(0.16,0.190)--(0.16,0.260);
	\draw[thick,dotted] (0.692,0.260)--(0.785,0.260)--(0.785,0.200)--(0.692,0.200)--(0.692,0.260);
	\node[] at (0.08,0.960) {(a) $t = 5.888$};
	\node[] at (0.58,0.960) {(b) $t = 5.920$};
	\node[] at (0.08,0.636) {(c) $t = 5.952$};
	\node[] at (0.58,0.636) {(d) $t = 5.984$};
	\node[] at (0.08,0.315) {(e) $t = 6.016$};			
	\node[] at (0.58,0.315) {(f) $t = 6.048$};
	\node[] at (0.33,0.78) {\scriptsize Incoming positive acoustic};
	\node[] at (0.33,0.74) {\scriptsize pressure wave (green contours)};	
	\node[] at (0.83,0.78) {\scriptsize Secondary diffraction at the leading edge};
	\node[] at (0.83,0.74) {\scriptsize and generation of entropy wave};	
	\node[] at (0.33,0.46) {\scriptsize Propagation of leading edge};
	\node[] at (0.33,0.42) {\scriptsize entropy wave downstream};
	\node[] at (0.83,0.48) {\scriptsize Leading edge entropy wave};
	\node[] at (0.83,0.44) {\scriptsize propagates downstream, while trailing edge};	
	\node[] at (0.83,0.40) {\scriptsize entropy wave moves upstream};	
	\node[] at (0.33,0.14) {\scriptsize Region of interaction of};
	\node[] at (0.33,0.10) {\scriptsize positive entropy waves};	
	\node[] at (0.83,0.16) {\scriptsize Interaction of positive};
	\node[] at (0.83,0.12) {\scriptsize upstream and downstream propagating};
	\node[] at (0.83,0.08) {\scriptsize entropy waves at most sensitive region};	
	\node[] at (0.015,0.02) {\scriptsize $-0.05$};
	\node[] at (0.080,0.02) {\scriptsize $0.00$};
	\node[] at (0.135,0.02) {\scriptsize $0.05$};		
	\node[] at (0.195,0.02) {\scriptsize $0.10$};
	\node[] at (0.250,0.02) {\scriptsize $0.15$};	
	\node[] at (0.310,0.02) {\scriptsize $0.20$};
	\node[] at (0.365,0.02) {\scriptsize $0.25$};
	\node[] at (0.425,0.02) {\scriptsize $0.30$};
	\node[] at (0.475,0.02) {\scriptsize $0.35$};
	\node[] at (0.525,0.02) {\scriptsize $-0.05$};
	\node[] at (0.580,0.02) {\scriptsize $0.00$};	
	\node[] at (0.640,0.02) {\scriptsize $0.05$};
	\node[] at (0.695,0.02) {\scriptsize $0.10$};	
	\node[] at (0.755,0.02) {\scriptsize $0.15$};
	\node[] at (0.810,0.02) {\scriptsize $0.20$};
	\node[] at (0.870,0.02) {\scriptsize $0.25$};
	\node[] at (0.930,0.02) {\scriptsize $0.30$};
	\node[] at (0.985,0.02) {\scriptsize $0.35$};
	\end{scope}
	\end{tikzpicture}	
	\caption{\green{Snapshots of the linearized Navier-Stokes equations presented in terms of entropy (contour) and pressure (lines) fluctuations. A downstream propagating entropy wave (highlighted by dotted line box) is generated due to leading edge secondary diffraction. This wave impinges an upstream propagating entropy fluctuation carried by an acoustic wave at the region of maximum receptivity.}}
	\label{fig:snaps_MFP_entropy}
\end{figure}
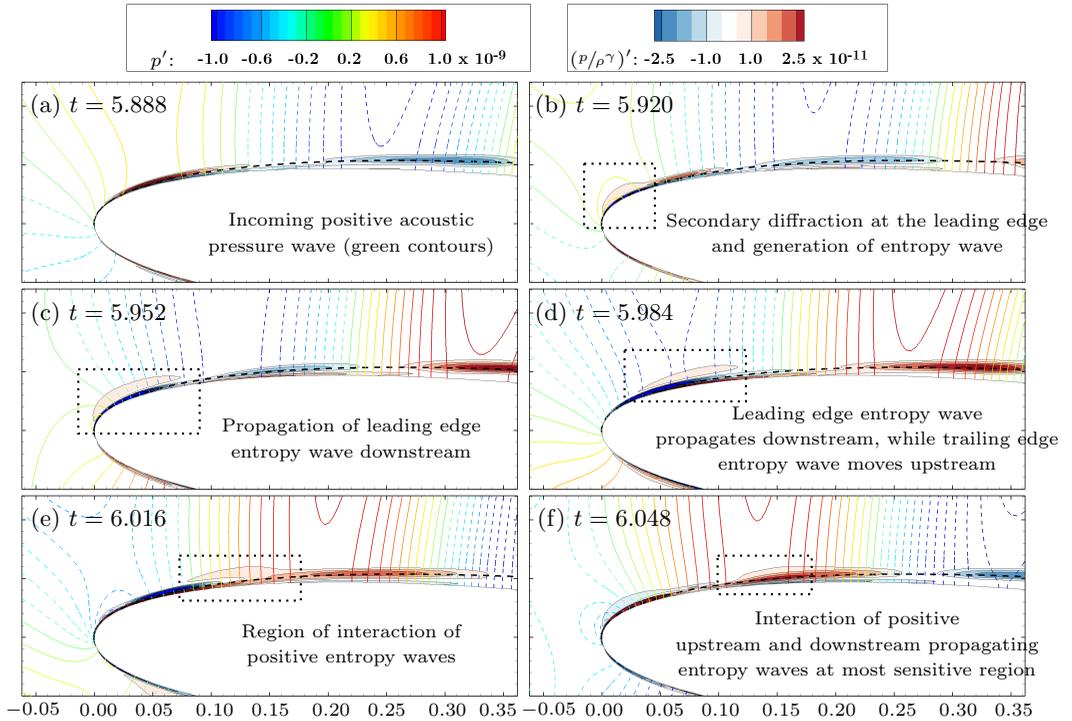

\green{The feedback loop mechanism is investigated here by analyzing the generation of flow instabilities by upstream propagating acoustic waves, as presented in figure \ref{fig:snaps_MFP_entropy}. In this figure, the entropy measure $\sfrac{p}{\rho^\gamma}$ is given by red-blue contours while pressure fluctuations are given by blue-green-red lines, where solid and dashed lines are positive and negative values, respectively. The entropy measure is used to partially filter out fluctuations associated to acoustic waves while pressure is used to highlight acoustic waves. In figure \ref{fig:snaps_MFP_entropy}(a), it is possible to see an incoming acoustic wave at $t = 5.888$ that, upon reaching the leading edge at $t = 5.920$, triggers an entropy fluctuation due to the secondary diffraction. This latter phenomenon is typical of airfoil noise as discussed by \cite{miottojsv2017}. The entropy fluctuation at the leading edge, presented in red color and highlighted by a dotted line box, is propagated downstream in figures \ref{fig:snaps_MFP_entropy}(b,c,d) at $t = 5.920$, $5.952$ and $ 5.984$, respectively. This disturbance overlaps with an incoming entropy wave moving upstream, carried by an acoustic wave depicted with blue dashed lines. Then, in figures \ref{fig:snaps_MFP_entropy}(e,f), respectively at $t = 6.016$ and $6.048$, the downstream propagating entropy fluctuation encounters a second positive acoustic wave in the region where the forcing modes are maximum, i.e., for $0.10 < x < 0.18$. This upstream propagating acoustic wave carries a positive entropy fluctuation which impinges on that generated at the leading edge. Thus, although the onset of instabilities occurs very close to the leading edge, the interaction between upstream traveling acoustic waves and downstream propagating boundary layer instabilities cannot be discarded since it occurs along the entire sensitive region of the flow. 
\blue{Albeit the analysis is conducted for pressure and entropy, local velocity fluctuations also occur and contribute to the incoming perturbations seen by the bubble.}
This overall process of the feedback loop mechanism is better visualized in movie 5 submitted as supplemental material.}
	
\begin{figure}
		\centering		
		\begin{overpic}[width=0.99\textwidth]{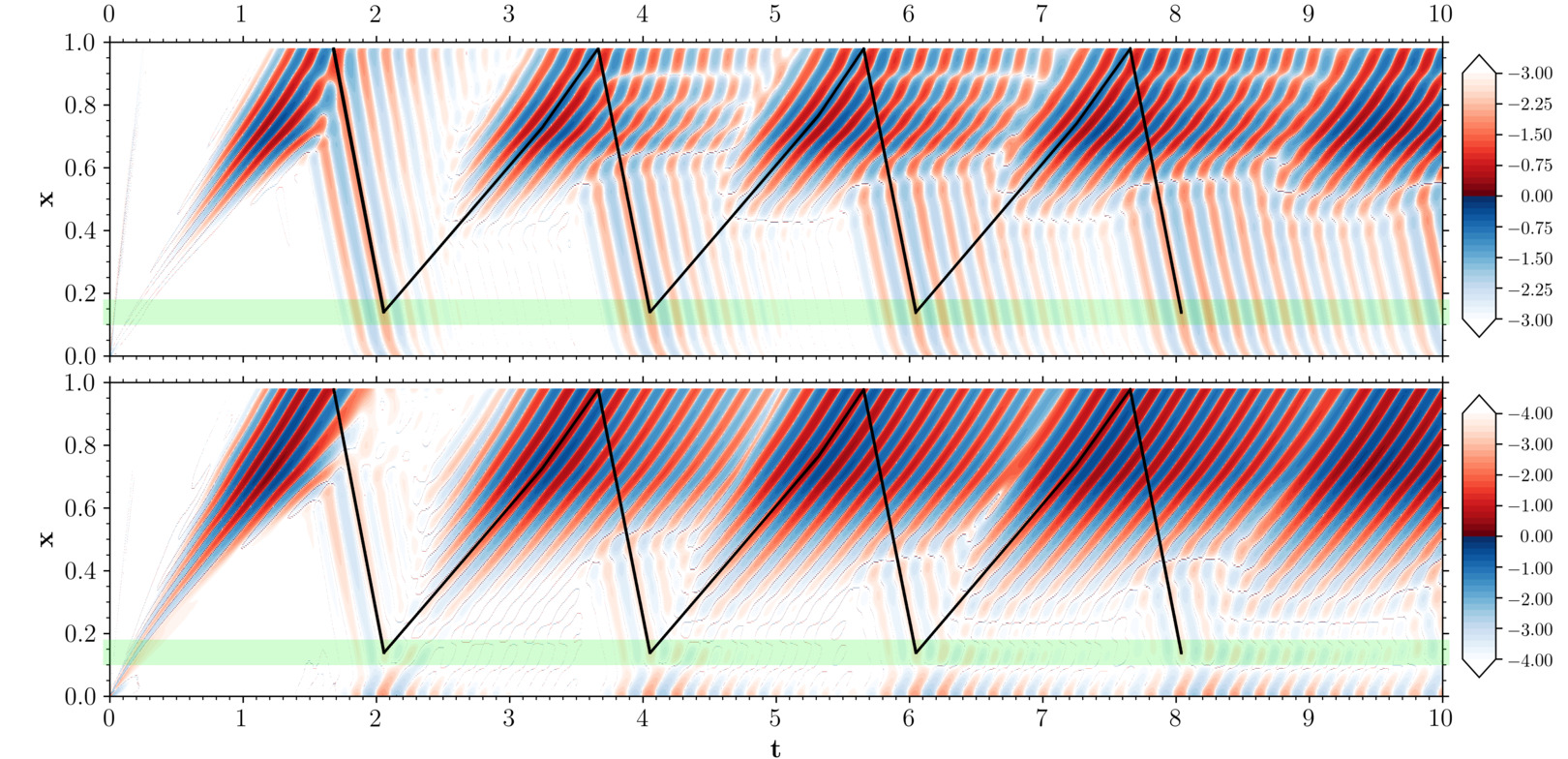}
			\put(8,26.7){(a) $p'$}
			\put(8,5.2){(b) $(\sfrac{p}{\rho^\gamma})'$}	
		\end{overpic} 	
		\caption{Space-time map of linear solution obtained by the mean flow perturbation for (a) pressure and (b) entropy. The map is normalized with respect to the maximum value and presented in a logarithmic scale. The positive and negative fluctuations are indicated by red and blue colors, respectively.}
		\label{fig:correlation_MFP}
	\end{figure}

	Flow disturbances are tracked along the black dashed line in figure \ref{fig:snaps_MFP_entropy} (also shown as blue dashed line in figure \ref{fig:k_rms}(b)) and presented as a function of time and space in figures \ref{fig:correlation_MFP}(a) and (b). The former shows pressure fluctuations $p'$ while the latter displays disturbances of an entropy measure $(\sfrac{p}{\rho^\gamma})'$. In these plots, the values are normalized with respect to the maximum and presented in a logarithmic scale. The positive and negative values are presented as red and blue colors, respectively.
	The first wavepacket shown in figures \ref{fig:snaps_MFP}(a,b) is observed traveling downstream with a local mean velocity $\bar{u}$ as bottom-up contours starting at $t = 0.0$.
	Upon reaching the trailing edge, at $t > 1.2$, the wavepacket generates acoustic waves propagating upstream with $\bar{u} - a$ speed. These waves are observed as top-down contours with lower magnitudes. The acoustic pressure has a phase opposition of $\pi$ compared to the hydrodynamic fluctuations on the trailing edge and, hence, an incident negative hydrodynamic pressure fluctuation (blue) creates a scattered positive acoustic wave (red). 
	
	Upstream of $x \approx 0.4$, the acoustic waves dominate and it is not possible to see the onset of instabilities. Hence, a measure of entropy $(\sfrac{p}{\rho^\gamma})'$ is shown in figure \ref{fig:correlation_MFP}(b) to partially filter the upstream acoustic disturbances and highlight the hydrodynamic content. In this figure, the contours of hydrodynamic fluctuations from bottom-up are better visualized.
	In both plots, a green region covering the entire period is placed at $0.1 \le x \le 0.18$, illustrating the maximum forcing region from figure \ref{fig:resolvent_modes_real}.
	\green{Along this region, disturbances generated in the vicinity of the leading-edge ($x \approx 0.0$) travel downstream and appear to interact with the incoming acoustic waves, as depicted in figures \ref{fig:snaps_MFP_entropy}(e,f). Such interaction results in a chess board-like interference pattern where the entropy levels carried by upstream propagating acoustic waves (inside the boundary layer) are of the same order of the entropy disturbances generated at the leading edge. We conjecture that the simultaneous combination of acoustic and hydrodynamic interactions at this location results in the maximum forcing.}
	Finally, black lines are drawn on the plots using the convective velocity information for both pressure and entropy disturbances. These lines are used to track the feedback dynamics and show that, if the feedback indeed closes along the most receptive region, the period of $\Delta t \approx 2.0$ is satisfied.

\begin{figure}
		\centering		
		\begin{overpic}[width=0.99\textwidth]{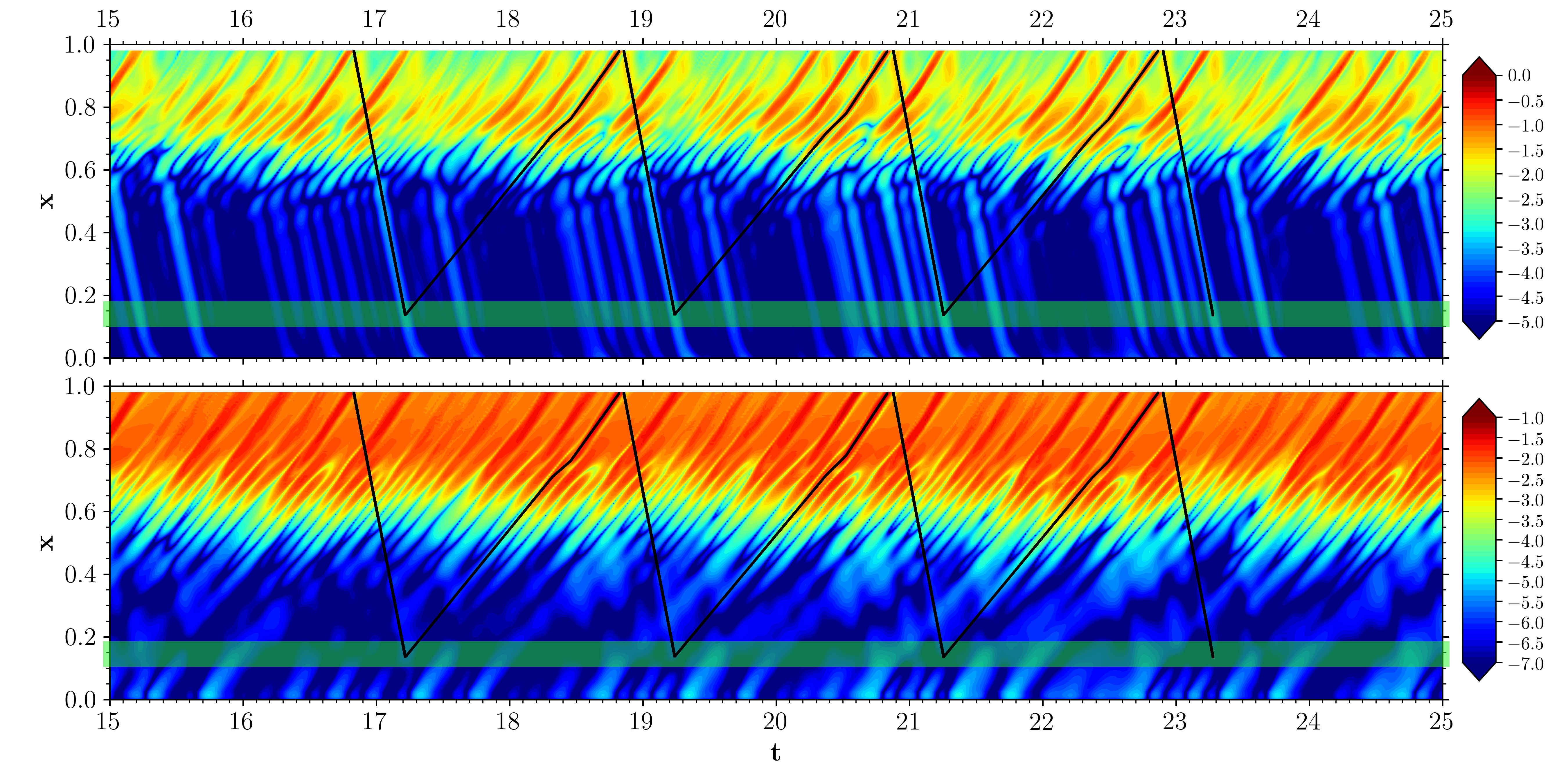}
			\put(8,28){\white{(a) $p'$}}
			\put(8,6){\white{(b) $(\sfrac{p}{\rho^\gamma})'$}}
		\end{overpic} 	
		\caption{Root mean square values along the spanwise direction computed as a function of chordwise position $x$ and time $t$ for (a) pressure and (b) entropy.}
		\label{fig:correlation_along_chord}
	\end{figure}

		The analysis of the feedback mechanism for the LES results is presented based on the RMS values of either the pressure $p$ or the entropy measure $\sfrac{p}{\rho^\gamma}$ averaged along the $z$-direction (spanwise direction).
		Similarly to the linear analysis, time signals are extracted along the blue line in figure \ref{fig:k_rms}(b) and the values of $R_0$ are presented in a logarithmic scale in figure \ref{fig:correlation_along_chord}(a) for pressure and (b) for the entropy measure. In both cases, it is possible to see high fluctuation values for $x > 0.4$ traveling downstream towards the trailing edge. For $x < 0.4$, smaller values of $p'$ propagate upstream towards the leading edge. Similarly to the linear analysis, the former and latter disturbances represent hydrodynamic and acoustic waves, respectively. The entropy measure partially filters the acoustic waves moving upstream. 
		\green{Based on linear analysis results, the onset of disturbances occurs very close to the leading edge in a position that coincides with the small hump observed in figure \ref{fig:cp_and_cf} for pressure and friction coefficients in terms of RMS values, respectively $C_{p_{\mbox{\tiny RMS}}}$ and $C_{f_{\mbox{\tiny RMS}}}$. Thus, the wave secondary diffraction at the leading edge excites boundary layer disturbances which are also observed in the LES. The entropy fluctuations generated at the leading edge appear to interact with the upstream-propagating acoustic waves in the region of maximum receptivity, similarly to the phenomenon observed in the linear analysis. In figure \ref{fig:correlation_along_chord}, black lines are used to track disturbances assuming that the feedback mechanism takes place at $0.1 < x < 0.18$, indicated by the green shaded rectangle which represents the region of maximum receptivity from figure \ref{fig:resolvent_modes_real}.}
	
	As a final remark, the repeating patterns observed in the simulations demonstrate that acoustic feedback occurs in the present flow. Results indicate that the loop depends on the leading edge secondary diffraction and the region of maximum sensitivity, located downstream of the leading edge and upstream of the separation bubble. However, the intermittent interaction of vortex merging and bursting on the suction side may eventually disrupt the cycle \blue{by altering the phase and magnitude of the tonal frequencies with a temporary impact on} the feedback loop.
	
	\subsection{On vortex merging and phase interference}
	\label{ssec:merging}
	
	From the previous sections, it is observed that the acoustic feedback loop is self-sustained by strong acoustic waves scattered from coherent flow structures advected past the trailing edge. These upstream-propagating acoustic waves create new flow instabilities which are amplified by the separation bubble and undergo a process of vortex pairing. Thus, it is important to understand which circumstances lead to generation of coherent structures or small-scale eddies, as discussed in section \ref{ssec:intermittency}. For this task, the dynamics of wavepackets in the linear simulation is first investigated, followed by a similar analysis in terms of the LES data.
	
\begin{figure}
	\centering		
	\begin{overpic}[width=0.45\textwidth]{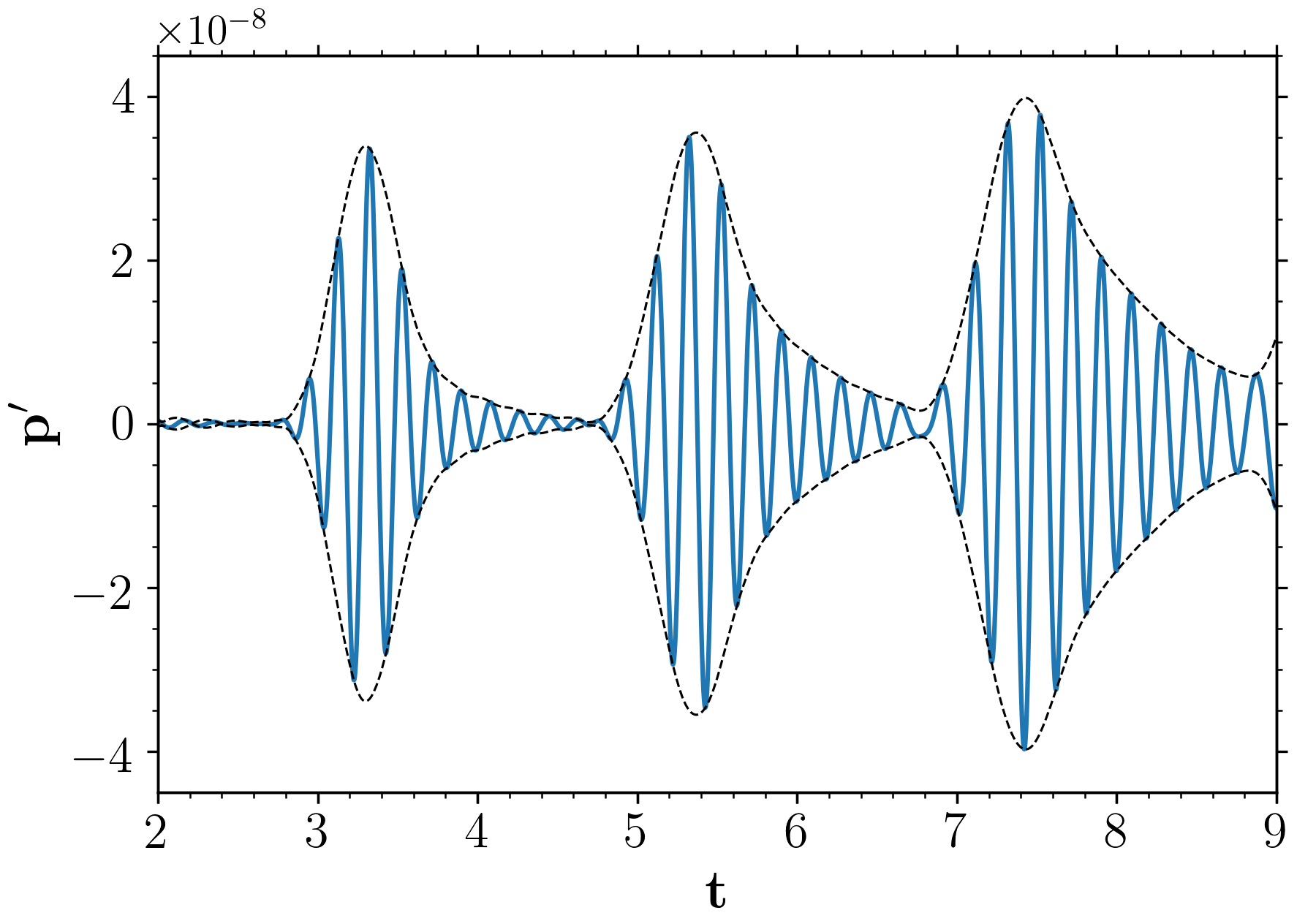}
		\put(14,12){(a)}	
	\end{overpic}
	\begin{overpic}[width=0.45\textwidth]{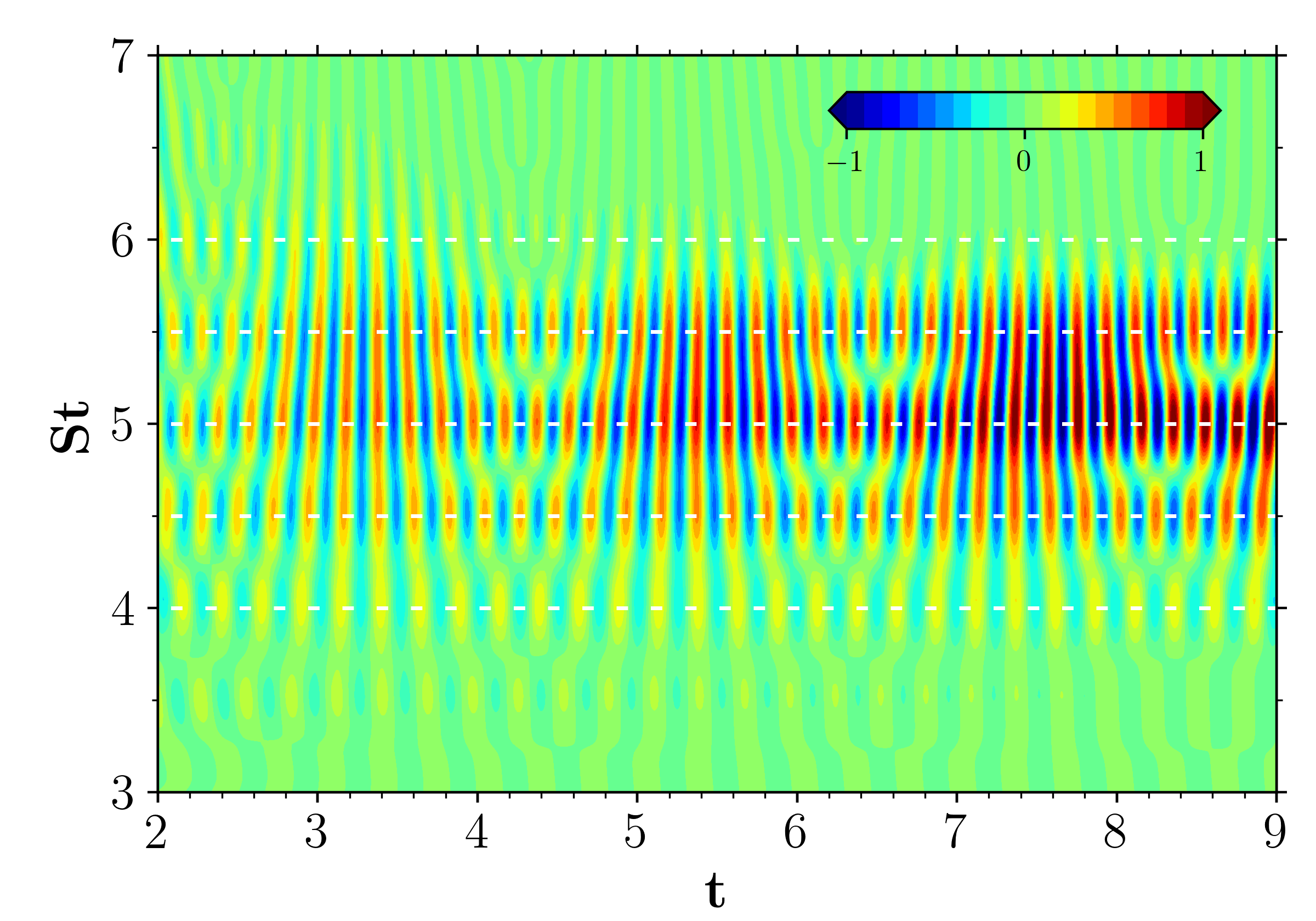}
		\put(14,12){(b)}
	\end{overpic}
		\caption{Wavepacket dynamics from linear solution including (a) temporal signal and (b) its phase scalogram at point of maximum amplitude (x = 0.72).}
	\label{fig:MFP_phase}
	\end{figure}

	The amplification of the wavepacket along the recirculation bubble results in a maximum value of the perturbations evaluated by the MFP at $x = 0.72$. The temporal signal of $p'$ at this location is presented in figure \ref{fig:MFP_phase}(a) depicting three envelopes. These are evaluated using the Hilbert transform to obtain the imaginary part and thus the magnitude of a real signal. Then, using the continuous wavelet transform, it is possible to extract the phase information of the previous signal. To highlight the relevant periods of the dynamics, the phase is weighted by the magnitude of the CWT coefficients as presented in figure \ref{fig:MFP_phase}(b). Comparing the temporal signal with the weighted phase, it is possible to see that the wavepacket envelopes peak when there is phase alignment across the multiple frequencies of the eigenspectrum shown in figure \ref{fig:pseudospectrum}. Hence, constructive interference of multiple modes is crucial for maximum energy amplification. When the waves are out of phase, the wavepacket magnitude drops.

\begin{figure}
		\centering		
		\begin{overpic}[width=0.99\textwidth]{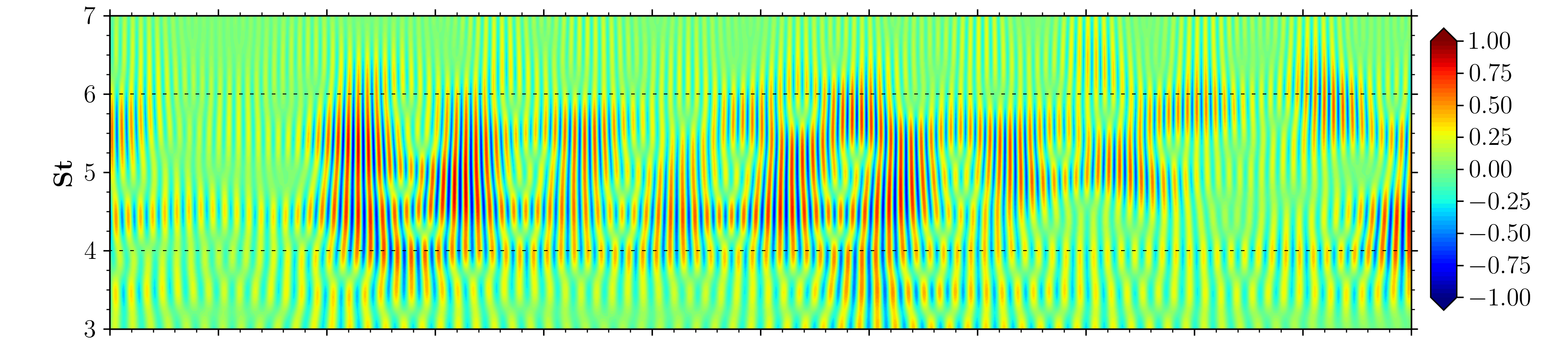}
			\put(8,3){(a)}	
		\end{overpic}
		\begin{overpic}[width=0.99\textwidth]{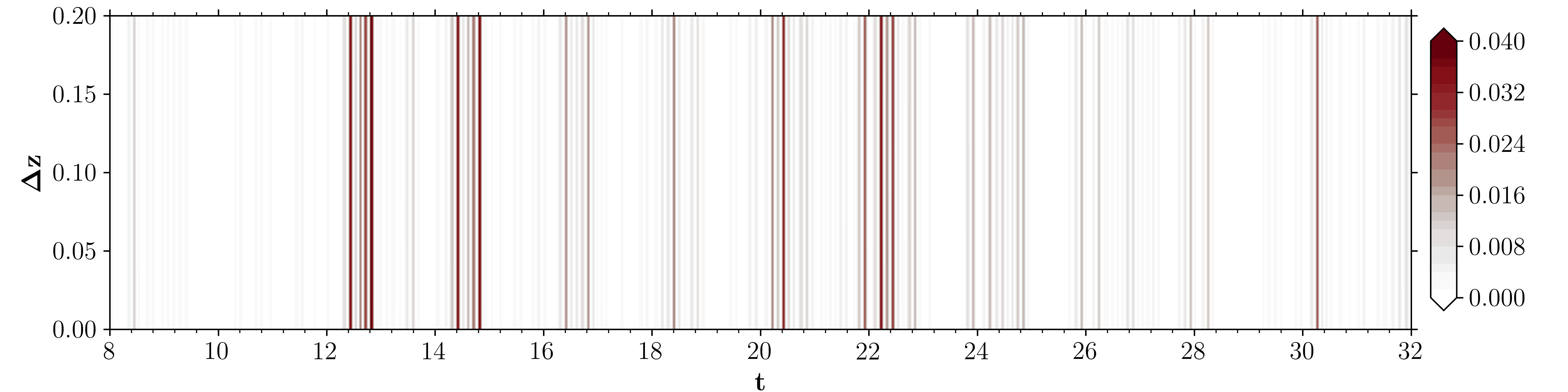}
			\put(8,5){(b)}
		\end{overpic}
		\caption{\green{Effects of constructive/destructive interference across multiple frequencies presented as the (a) phase weighted-scalogram and its impact in the (b) spanwise \red{two-point, one time autocovariance} levels.}}
		\label{fig:wavelet_phase}
	\end{figure}

	The phase information from the LES data is presented together with the spanwise pressure \red{autocovariance} in figures \ref{fig:wavelet_phase}(a) and (b) at $x = 0.64$. This location is representative of the vortex \green{pairing} region. These figures show that the high coherence depends on the phase alignment of frequencies $4 \lesssim St \lesssim 6$. This frequency range corresponds to that where the most amplified modes from the linear analysis are observed. If more (fewer) frequencies are aligned, a stronger (weaker) signal coherence is achieved. If the disturbances are in phase, the result is vortex merging and strong coherence. On the other hand, if they are out of phase, the vortex pairing fails and coherence is reduced. The repeating patterns of \red{high coherence} show that the feedback mechanism has a period of $\Delta t \approx 2.0$, related to the lower tone in the spectrum at $St = 0.5$.
	
	Destructive and constructive interference of vortical structures during the \green{pairing} process can be interpreted as an amplitude modulation of the signal \citep{Desquesnes2007,Ricciardi2019_tones}. The modulation mechanism from the 2D simulations performed in these references differs from the current 3D case, where turbulent transition occurs. Here, the destructive interference leads to vortex breakdown and smaller pressure fluctuations. On the other hand, constructive interference of multiple frequencies creates spikes in the pressure signal as shown for certain time instants in figures \ref{fig:fft_probes}(a) and (c).
	
	
	\section{Conclusions}
	\label{ssec:conclusions}
	
	A study of trailing edge tonal noise is conducted through a combination of large eddy simulation and linear \blue{stability theory}. The focus of this investigation is on the secondary tones and acoustic feedback loop which arise in airfoil flows at moderate Reynolds numbers. A compressible LES is performed for the flow over a NACA 0012 airfoil at $\alpha = 3^{\circ}$ immersed in a freestream flow with Mach number $M_{\infty} = 0.3$ and Reynolds number $Re = 5 \times 10^4$. Despite intermittent flow transition to turbulent regime, the noise spectrum depicts a main tone with multiple equidistant secondary tones, and all tonal peaks observed are integer multiples of the lowest frequency peak.
	
	The mean flow displays two recirculation regions, one on each side of the airfoil, where the separation bubble on the suction side is longer than that on the pressure side. The flow dynamics on the airfoil is dominated by events on the suction side, aft of the separation bubble, as indicated by the RMS values of kinetic energy and pressure. The mean friction coefficient shows a double detachment pattern, typical of large fluctuations downstream the recirculation bubble. Mean velocity profiles show that the maximum reversed velocity inside the separation bubble is 13\% of the freestream value.
	
	Linear stability analysis of the mean flow using a bi-global operator shows marginally stable and unstable eigenvalues at frequencies close to those of the tonal peaks observed in the non-linear simulation, indicating that the overall flow dynamics originates from linear mechanisms. \green{Results from resolvent analysis of the linear operator are displayed as the pseudospectrum, showing that the higher frequencies are more susceptible to be excited by pseudoresonances. The response modes show that the suction side recirculation bubble acts as an amplifier of disturbances. On the other hand, the forcing modes show a high sensitivity region within the flow, starting downstream of the leading edge and extending until upstream of the bubble.}

	In the large eddy simulation, the amplification of fluctuations along the laminar separation bubble results ultimately in vortex shedding on the suction side.
	The vortices undergo a pairing process that may lead to either merging or bursting. This process moves the dominant tonal peak in the LES to a lower frequency than that observed in the unstable eigenvalues of the linear spectrum. Another result of the vortex pairing is that either coherent vortices or turbulent spots are advected towards the trailing edge.
	The success of vortex \green{merging} aft of the bubble depends on the phase alignment between the unstable frequencies observed in the linear eigenspectrum. When such frequencies are in phase, it is likely that the vortices will successfully merge. On the other hand, destructive interference results in vortex bursting.	
	\blue{The different regimes where coherent structures alternate with uncorrelated eddies at the trailing edge act as an amplitude modulation of the signal. The magnitude of the radiated acoustic waves depends on the coherence level of the vortical structures at the trailing edge, measured by \red{two-point, one-time autocovariance} of spanwise pressure fluctuations. Together with the autocovariance, a time-frequency analysis using continuous wavelet transforms show that the flow has two different time scales. One is related to the lowest tonal frequency, which drives the acoustic feedback loop and depends not only on the coherent structures but also on the uncorrelated turbulent packets. 
	The other is a faster time scale related to the passage of multiple coherent structures at the trailing edge. It has been found that the intermittent dynamics of this faster time scale impacts the magnitude of tonal peaks with the possibility of changing the instantaneous dominant tonal frequency.}
	

	The feedback loop mechanism is investigated by the time integration of the linearized Navier-Stokes equations using the mean flow perturbation technique. The system response to an impulsive actuation at the leading edge displays a wavepacket advected towards the trailing edge with subsequent radiation of acoustic waves due to a scattering mechanism. The wavepacket is composed of multiple frequencies in the eigenspectrum and its magnitude is maximum when the phases of the unstable frequencies from the eigenspectrum are aligned. \green{Visualization of the linearized solution shows that entropy disturbances are triggered by the secondary diffraction and propagate downstream along the boundary layer. These disturbances play an important role since they interact with upstream-propagating waves at the region of maximum forcing. Thus, it is likely that the feedback loop closes at the most sensitive region, described by the forcing modes, in order to satisfy the low frequency period of the wavepacket cycle.}

\red{		
	The spanwise pressure \red{covariance} is analyzed along the airfoil chord and shows an influence of the feedback mechanism, where specific patterns repeat in time. Intermittency due to transition to turbulence may shift the phase and magnitude of all frequencies, disrupting the feedback loop. Despite this, the flow reaches a new equilibrium and reestablish the periodic merging/bursting process, indicating that the periodicity of multiple stability modes may be the self-sustaining mechanism for multiple tones and flow transition.
}	
	
	\section{Acknowledgements}
	T.R. and W.W. acknowledge Fun\-da\-\c{c}\~{a}o de Amparo \`{a} Pesquisa do Estado de S\~{a}o Paulo, FAPESP, for supporting the present work under research grants No.\ 2013/08293-7, 2018/11835-0, 2019/20437-0 and 2021/06448-0, and Conselho Nacional de Desenvolvimento Cient\'{i}fico e Tecnol\'{o}gico, CNPq, for supporting this research under grants No.\ 407842/2018-7 and 304335/2018-5.
	K.T. thanks the support from the US Office of Naval Research under grant N00014-19-1-2460.
	The authors thank CENAPAD-SP (Project 551), UCLA cluster Hoffman2 and LNCC-Cluster SDumont (Project SimTurb) for providing the computational resources used in this study. We also thank C.-A. Yeh, C. Skene and J. H. M. Ribeiro for insightful discussions on stability and resolvent analyses.

\bibliographystyle{jfm}
\bibliography{bibfile}

\end{document}